\documentclass[%
 reprint,
 superscriptaddress, 
 amsmath,amssymb,
 aps,
 pra
]{revtex4-2}

\usepackage{graphicx}
\usepackage{subcaption}
\usepackage{dcolumn}
\usepackage{bm}
\usepackage{chemformula}
\usepackage{microtype}
\usepackage{hyperref}

\usepackage{lipsum}
\graphicspath{{./figures/}}

\definecolor{mutedGreen}{HTML}{3A7D44}

\usepackage{amsmath,amsfonts,bm}

\def\eqref#1{equation~\ref{#1}}

\def\1{\bm{1}}

\def\ve{{\bm{e}}}

\def\vy{{\bm{y}}}

\def\mC{{\bm{C}}}

\def\mE{{\bm{E}}}
\def\mF{{\bm{F}}}

\def\mH{{\bm{H}}}

\def\mX{{\bm{X}}}

\DeclareMathAlphabet{\mathsfit}{\encodingdefault}{\sfdefault}{m}{sl}
\SetMathAlphabet{\mathsfit}{bold}{\encodingdefault}{\sfdefault}{bx}{n}

\def\gE{{\mathcal{E}}}

\def\gG{{\mathcal{G}}}

\def\gV{{\mathcal{V}}}

\def\sR{{\mathbb{R}}}

\def\sY{{\mathbb{Y}}}

\begin{document}

\preprint{APS/123-QED}


\title{Inverse Design of Amorphous Materials with Targeted Properties}

\author{Jonas A.\ Finkler}
  \thanks{These authors contributed equally to this work} %
  \affiliation{Department of Chemistry and Bioscience, Aalborg University, Fredrik Bajers Vej 7H, 9220 Aalborg Øst, Denmark}

\author{Yan Lin}%
  \thanks{These authors contributed equally to this work} %
  \affiliation{Department of Computer Science, Aalborg University, Selma Lagerløfs Vej 300, 9220 Aalborg Øst, Denmark}

\author{Tao Du}
  \affiliation{Department of Applied Physics, The Hong Kong Polytechnic University, Kowloon, Hong Kong, 999077, China}

\author{Jilin Hu}%
  \email{hujilin@cs.aau.dk}
  \affiliation{Department of Computer Science, Aalborg University, Selma Lagerløfs Vej 300, 9220 Aalborg Øst, Denmark}
  \affiliation{School of Data Science and Engineering, Department of Computer Science, East China Normal University, Shanghai 200051, China}

\author{Morten M\@. Smedskjaer}
  \email{mos@bio.aau.dk}
  \affiliation{Department of Chemistry and Bioscience, Aalborg University, Fredrik Bajers Vej 7H, 9220 Aalborg Øst, Denmark}

\date{\today}

\begin{abstract}

  Disordered (amorphous) materials, such as glasses, are emerging as promising candidates for applications within energy storage, nonlinear optics, and catalysis. Their lack of long-range order and complex short- and medium-range orderings, which depend on composition as well as thermal and pressure history, offer a vast materials design space. 
  To this end, relying on machine learning methods instead of trial and error is promising, and among these, inverse design has emerged as a tool for generating materials with desired properties. 
  Although inverse design methods based on diffusion models have shown success for crystalline materials and molecules, similar methods targeting amorphous materials remain less developed, mainly because of the limited availability of large-scale datasets and the requirement for larger simulation cells.
  In this work, we propose and validate an inverse design method for amorphous materials, introducing AMDEN (Amorphous Material DEnoising Network), a diffusion model-based framework that generates structures of amorphous materials.
  First, we demonstrate the inherent challenges for diffusion models to generate relaxed structures. 
  These low-energy configurations are typically obtained through a thermal motion-driven random search-like process that cannot be replicated by standard denoising procedures.
  We therefore introduce an energy-based AMDEN variant that implements Hamiltonian Monte Carlo refinement for generating these relaxed structures. 
  We further introduce several amorphous material datasets with diverse properties and compositions to evaluate our framework and support future development.
\end{abstract}

\maketitle


\section{Introduction}

Amorphous materials are characterized by the absence of long-range atomic order, distinguishing them from crystalline materials with their periodic atomic arrangements.
These materials exhibit unique mechanical, thermal, and electrical properties that make them valuable candidates for diverse applications, including batteries, non-linear optics, and catalysis~\cite{liu2024amorphous}.

Inverse design has emerged as a promising approach for generating amorphous materials with targeted properties. 
In contrast to conventional trial-and-error methods, inverse design reverses the process by beginning with desired properties and systematically determining the necessary atomic configurations to achieve them. 
Relying on machine learning methods, inverse design has the potential to significantly accelerate the exploration of the materials design space~\cite{liu2024amorphous, zunger2018inverse}. 

Recent advances in machine learning and deep learning have accelerated efforts in the inverse design of crystalline materials and molecules through conditioned generation models. 
These models function essentially as learnable posterior probability models that capture the distribution of material structures given specified properties as conditions.
Early approaches mainly used two frameworks: variational auto-encoders (VAE)~\citep{DBLP:journals/corr/KingmaW13} and generative adversarial networks (GAN)~\cite{goodfellow2020generative}.
However, both VAE-based methods~\cite{gebauer2019symmetry,hoffmann2019data,noh2019inverse,court20203} and GAN-based methods~\cite{long2021constrained} demonstrated limitations in generating high-quality material structures due to insufficient expressivity~\cite{li2018limitations,DBLP:conf/iclr/DaunhawerSCPV22} and training instability~\cite{lucas2019don,xu2020understanding,becker2022instability}.
More recently, diffusion models~\cite{ho2020denoising} have shown superior performance~\cite{wu2022diffusion,DBLP:conf/iclr/XieFGBJ22,zeni2025generative} in this domain.
These models generate valid structures by progressively refining random noise into coherent structures.

Despite significant progress in inverse design for crystalline materials and the availability of advanced technical frameworks, the inverse design of amorphous materials remains underdeveloped~\cite{pham2023spectroscopy,lei2024grand}.
This gap stems primarily from the scarcity of diverse large-scale datasets of amorphous materials, which are essential for training effective deep learning-based generation models~\cite{horton2025accelerated}. 
This limitation is directly related to the inherent randomness of amorphous materials and their requirement for large simulation cells, with several hundred atoms, to accurately capture their properties.

The disordered nature of amorphous materials precludes the existence of a unique and definite structural representation; instead, only samples of an underlying structural distribution can be obtained.
This challenge is further complicated by the fact that atomic structures depend not only on composition but also on their thermal and pressure history.
As experimental characterization of atomic positions in amorphous materials, such as glasses, is typically still out of reach, molecular dynamics (MD) simulated melt-quenching with relatively low cooling rate is required to obtain realistic glass structures. In turn, this necessitates lengthy simulations that increase computational costs substantially.
Additionally, large simulation cells are required to minimize finite size effects and properly resolve medium-range order (typically 5-20 Å length scale), further increasing computational demands.
By contrast, crystalline materials can be consistently characterized by atomic positions in a minimal unit cell.
While exact positions and unit cell dimensions depend on the modeling methods employed (e.g., choice of exchange-correlation functional), re-relaxing atomic positions is usually sufficient and relatively affordable to obtain consistent structures, even when original data comes from different sources with inconsistent density functional theory (DFT) parameters. 
As an alternative to DFT, classical force fields can be used to drive the melt-quench simulations at a more affordable computational cost. 
While classical force fields have been successfully used for predicting materials properties~\cite{liu2022challenges}, they are usually parametrized for a specific materials class and range of elemental compositions limiting their applicability for general inverse design tasks.
Universal machine learned force fields~\cite{batatia2023foundation, yang2024mattersim} are emerging as a promising alternative, covering a wide range of elemental compositions at near DFT accuracy at a much lower computational cost. 
However, these methods can be unreliable for the complex high energy structures encountered during the high temperature phase of the melt-quench and further validation is needed~\cite{deng2025systematic}. 

Some efforts have been made to develop inverse design methods for amorphous materials.
For instance, machine learning based high-throughput screening of materials has been explored \citep{wang2021inverse, merchant2023scaling,li2025conditional}, yet these approaches still rely on conventional trial-and-error methods.
Further, while a generative framework for predicting compositions of glass materials given desired properties has been developed \citep{zhou2023generative}, compositions alone provide an incomplete picture of amorphous materials and thus limit the inverse design process.
There are a few efforts on generating atomic configurations of amorphous materials, with methods based on GAN~\cite{comin2019deep,xu2023generative,yong2024dismai} and VAE~\cite{chen2025physical,kilgour2020generating}. As mentioned before, their generation quality is limited by the limitations of the underlying GAN and VAE frameworks.
Diffusion models have also been applied to generate structures of graphite materials with specific x-ray absorption near edge structure ~\citep{pham2023spectroscopy}, with a follow-up work for crystalline phases and grain boundaries~\cite{lei2024grand}.
However, these approaches focus on relatively narrow types of amorphous materials and properties.
Recently, a diffusion generative framework was developed and applied to reproduce structural features and properties of amorphous \ch{SiO2} across different cooling rates~\cite{yang2025generative}.
These authors noted that standard denoising procedures are not sufficient to accurately generate high-quality amorphous samples and modifications to the method, such as the inclusion of ``extra noise'' and a final relaxation using a short MD trajectory, were included. 

In this work, we propose and validate a comprehensive inverse design framework for amorphous materials.
Our framework, named \textbf{AMDEN} (\underline{A}morphous \underline{M}aterial \underline{DE}noising \underline{N}etwork), is a diffusion model-based framework that generates structures of multi-element amorphous materials with desired properties.
At its core, AMDEN incorporates a material diffusion process that estimates the distribution of amorphous material samples with desired properties as prior information. A material sample can then be generated from this distribution by gradually removing noise from a random sample, guided by a learnable score function with equivariant structure. AMDEN also introduces a ghost atom mechanism that enables it to control material density during generation.

To effectively train and validate AMDEN and to support future development in inverse design of amorphous materials, we generate several amorphous material datasets with diverse properties and compositions.
First, a multi-element glass dataset covering a wide range of compositions to test AMDEN's inverse design capabilities. 
Second, a dataset with a fixed composition of \ch{SiO2}, to test AMDEN's ability to modify material properties purely based on the atomic structure, without relying on composition.
Third, three datasets of pure amorphous silicon with differing thermal histories to demonstrate the inherent challenges for diffusion models to generate relaxed structures.
The standard implementation of AMDEN is not able to generate low-energy structures, which are obtained when the material is quenched at a low cooling rate.  
We therefore developed an energy-based variant of the score function, which incorporates Hamiltonian Monte Carlo refinement into the material diffusion process.
This modified AMDEN implementation is able to generate samples that match the reference data closely in terms on energy and structure.
Since the score function is still learned and conditioned on the desired properties, its incorporation does not compromise the inverse design capability of AMDEN.



\section{Results}

\subsection{AMDEN model}

\begin{figure*}[t]
  \centering
  \includegraphics[width=\linewidth]{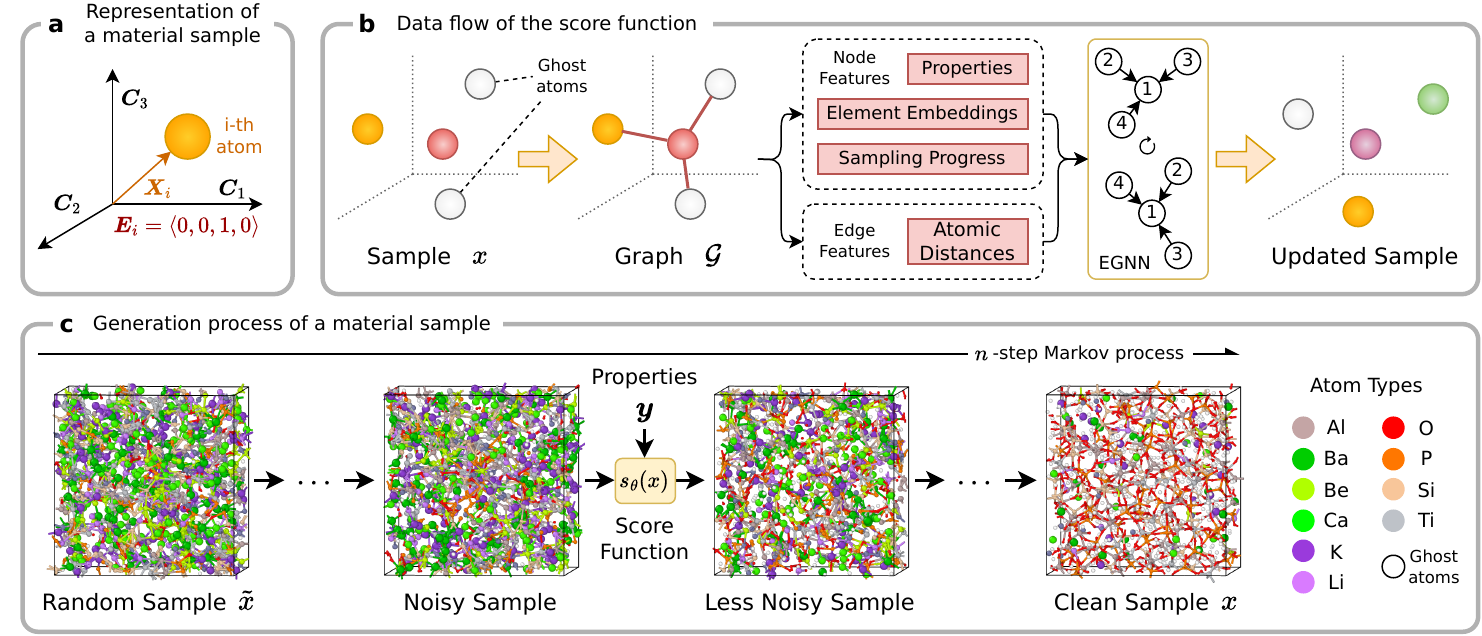}
  \caption{
    \textbf{Pipeline of the AMDEN model.} \textbf{a}, Material sample $x$ is represented by the cell $\mC$, atomic positions $\mX$ and a one-hot element embedding $\mE$. 
    \textbf{b}, Score function $s_\theta(x)$ is predicted by an equivariant graph neural network (EGNN), which takes interatomic distances as edge features and element embeddings, the diffusion time $t$ and target properties $\vy$ as node features.
    \textbf{c}, Reverse-time stochastic differential equation is solved to transform an initially completely random sample into a valid materials sample.
  }
  \label{fig:amden}
\end{figure*}


AMDEN introduces a materials diffusion process that aims to model the probabilistic distribution $p(x)$ of amorphous material samples $x$, both with and without a set of desired properties $\vy$ as prior information. We can then sample from the learned distribution $p_\theta(x)$, trained to match $p(x)$, to generate amorphous material samples that follow the same distribution as existing data or exhibit desired properties.

An illustration of the AMDEN model and the generation process is provided in Fig.~\ref{fig:amden}.
We begin by describing each amorphous material sample $x$ as the positions and elements of atoms located in a three-dimensional cell, formally represented as a tuple $x=(\mC, \mX, \mE)$. Here,
$\mC \in \sR^{3\times 3}$ denotes the three lattice vectors of the cell,
$\mX\in \sR^{n\times 3}$ represents the positions of atoms, with $n$ being the number of atoms and the $i$-th row of this matrix being the coordinate of the $i$-th atom. $\mE \in \sR^{n\times d}$ indicates the element embeddings, where we use one-hot encodings of atoms' atomic numbers, with the embedding dimension $d$ being the total number of elements under consideration.
The set of desired properties $\vy$ is a vector of $m$ property values, where each value denotes the desired degree or magnitude of that property.

In reality, the exact form of $p(x)$ is unknown. We build our material diffusion process based on stochastic differential equation (SDE) and score-matching~\cite{song2020score}.
SDE with score-matching has proven to be an effective framework for estimating complex distributions~\cite{blattmann2023stable,DBLP:conf/iclr/PodellELBDMPR24}. Compared to similar frameworks, such as ordinary differential equations (ODEs) with flow-matching~\cite{DBLP:conf/iclr/LipmanCBNL23}, the stochastic nature of SDEs is a better fit for amorphous materials since it provides more freedom for the exploration of optimal structures.
Specifically, we implement a learnable score function $s_\theta(x)$ to estimate the gradient of the log probability of $p(x)$ as $s_\theta \approx \nabla_x \ln p(x)$, where $\theta$ is the set of learnable neural network parameters.
To sample from $p(x)$, the material diffusion process starts from a noisy sample $\tilde x=(\mC, \tilde{\mX}, \tilde{\mE})$ where the positions and elements are random noise. It then solves a reverse-time SDE through an $n$-step Markov process guided by the score function $s_\theta(x)$, producing a noise-free sample $x$. Note that the cell $\mC$ remains intact throughout the process.

An amorphous material possesses several invariances, including: 1) \textit{permutation invariance,} changing the order of the atoms in $\mX$ and $\mE$ does not change the sample;
2) \textit{translation invariance,} translating the positions $\mX$ by an arbitrary vector does not change the sample; and
3) \textit{rotation and mirror invariance,} rotating or mirroring the cell $\mC$ and the positions $\mX$ do not change the sample.
To preserve these invariances in the material diffusion process, the score function $s_\theta(x)$ employs an equivariant architecture with the equivariant graph neural network (EGNN)~\cite{DBLP:conf/icml/SatorrasHW21} as its backbone.
EGNN processes both node and edge features of a graph structure, and updates node positions and features based on extracted correlations between nodes.

To prepare the input for the EGNN, we construct a graph $\gG=(\gV, \gE)$ of the atoms in each amorphous material sample, where each atom is a node in the node set $\gV$ and each edge in the edge set $\gE$ connects pairs of atoms with distances less than a cutoff radius $r_{\text{cut}}$. The distances are computed with periodic boundary conditions.
The nodes contain information about element embeddings $\mE$, desired properties $\vy$, and the sampling progress of the materials diffusion process. The edge features are derived from the distances between atoms.
Since all inputs to the EGNN backbone are invariant to the above three types of transformations, and the updates to node positions and features in EGNN are equivariant, the score function $s_\theta(x)$ properly preserves invariances of amorphous materials.

A unique feature of AMDEN is its ability to control material density during inference through a \textit{ghost atom} mechanism.
This approach enables generating materials with specific density targets, addressing the fundamental limitation that diffusion models cannot alter the number of atoms in generated structures.
In our context, diffusion models manipulate samples by adding or removing noise from atomic positions and element embeddings, but cannot add or remove atoms to change sample densities.
Existing solutions include voxel-based approaches~\cite{lei2024grand} that enable diffusion models to decide whether an atom exists in each small cell. However, voxel-based approaches cannot preserve the equivariance of amorphous materials.
In contrast, the ghost atom approach maintains a fixed total number of atoms and preserves AMDEN's equivariant architecture. It introduces a special type of atom that is removed from the final generated samples.
Given a sample $x$ and a target maximum density $\rho_{\text{target}}$, we first calculate the desired number of atoms based on $\rho_{\text{target}}$ and cell volume as $n_{\text{target}} = \lfloor \rho_{\text{target}} \cdot \text{Volume}(\mC) \rfloor$.
If $n_{\text{target}}$ exceeds the number of actual atoms in the generated structure, we supplement with ghost atoms. A number of ghost atoms, $n_{\text{ghost}} = \max(0, n_{\text{target}} - n_{\text{actual}})$, are then randomly positioned within the cell.
During training and denoising, ghost atoms are treated like normal atoms but are assigned a special chemical element class. 
The model can thus adjust the density of the sample by increasing or decreasing the fraction of atoms that are assigned the ghost atom type. 
As a final step after denoising, ghost atoms are removed from the sample. 

\subsection{Inverse material design}
To assess the inverse design capabilities of AMDEN, we created the multi-element glass (MEG) dataset using classical MD simulations (see Methods section for details). It features a large variety of compositions, containing eleven different elements. We then trained AMDEN on the MEG dataset and conditioned the model on the Young's modulus $E$ and the molar concentration of lithium $C_\text{Li}$. These were chosen as they are important properties for an emerging application of glass materials as solid electrolytes for batteries~\cite{ding2024amorphous,kalnaus2023solid}. 
During inference, the target lithium concentration was then set to $C_\text{Li}=0.15$ while the Young's modulus $E$ was set to values ranging from $20$\, to $160$\,GPa.

\begin{figure}[h]
  \centering
  \includegraphics[width=\linewidth]{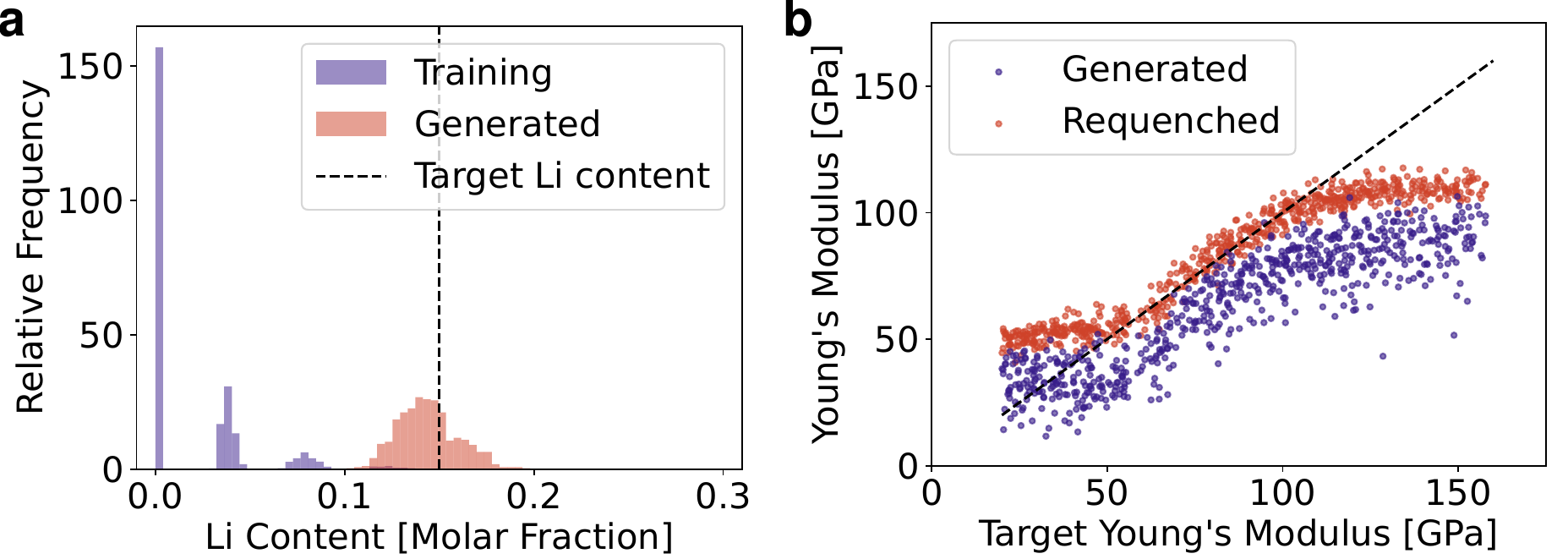}
  \caption{
    \textbf{Inverse design inference results.}
    \textbf{a}, Inferred distribution of molar \ch{Li} content for the generated and training structures and \textbf{b}, Young's modulus of the generated and requenched structures compared to their respective target values. The dashed line in panel b indicates $x=y$.
    The requenched structures have been obtained by performing the full MEG melt-quench workflow with the compositions of the inferred structures.
  }
  \label{fig:bmp}
\end{figure}

The inference results are shown in Fig.~\ref{fig:bmp}.
As shown in Fig.~\ref{fig:bmp}a, the generated \ch{Li} concentration is in good agreement with the targeted value (within few percent), despite only a very small fraction of the training data containing a similar ratio of \ch{Li} atoms. The Young's modulus of the generated structures is shown in Fig.~\ref{fig:bmp}b. While a decent correlation between the target and generated modulus is observed, the generated modulus falls short of the target, especially for larger values where it tapers off to about $100$\,GPa. 
The Young's modulus depends on both the material's elemental composition and the structural arrangement of the atoms (i.e., bond strength and bond density), which in practice are influenced by the thermal (and pressure) history of the material. 
There are thus two failure modes, which can explain the discrepancy between the target and obtained moduli in our inference data. 
Either, the model generates the correct composition but fails to accurately recreate the structure that would be obtained from the melt-quench procedure used for the training data generation, or the model already fails to generate the correct compositions necessary to obtain the targeted Young's moduli.
We thus re-ran melt-quench simulations following the same workflow as in the MEG dataset for all compositions generated by AMDEN to obtain a ground truth for the Young's modulus as a function of composition. 
The Young's moduli of the resulting samples are also shown in Fig.~\ref{fig:bmp}b, revealing a significantly improved correlation with the targeted values except for the extrapolative regime of low Young's modulus where the error in the generated structures appears to counteract the extrapolation error in terms of composition. 
The moduli of the requenched samples are still below the target for high values above 110\,GPa.
In this regime, the model is extrapolating outside the training data, and generating samples with such high moduli might simply not be physically realizable, especially with the additional constraint of rather high \ch{Li} content, which tends to reduce network connectivity and thus stiffness~\cite{rouxel2007elastic}.
Nevertheless, the generated samples extend the property range toward lower Young's modulus values beyond the training distribution, indicating that some degree of extrapolation is achieved.

The improved agreement between generated and target property after requenching the samples indicates that the problem lies in the structure of the generated samples.
In other words, the requenching experiment shows that AMDEN correctly identifies compositions that yield the targeted properties, but the atomic structures produced by standard denoising are not yet accurate.
This structural gap motivates the development of HMC denoising (Section~\ref{sec:relaxation-denoising}), which brings the generated Young's moduli closer to the target values without requenching (Fig.~\ref{fig:Si}f).
Element-resolved partial radial distribution functions and cumulative coordination numbers for representative cation--oxygen pairs further confirm that HMC denoising more closely aligns the first-shell environments with the training data, while standard denoising yields broader and lower first-shell peaks (Supplemental Section~4 and Figs.~S6 and~S7).

\subsection{Challenges in generating relaxed structures}
To gain a better understanding of the structural differences between samples obtained from MD melt-quench simulations and samples generated by AMDEN, we trained unconditional AMDEN models on each of the three amorphous \ch{Si} dataset variants.
The three datasets are identical in terms of composition and samples size but differ in the samples' thermal history. 
While samples of the \textit{melt} variant are obtained from a melt at 2500 \,K, the \textit{quench} variant is obtained after an almost instantaneous quench to 300 \,K and the \textit{anneal} variant is obtained after quenching the melted structures to 300 \,K at 1 \,K/ps.
For inference, the unit cells of the training samples were used and the number of atoms was kept fixed at 256.
To compare the structures of the generated samples with the respective training data, we computed radial distribution functions (RDFs), bond angle distributions and structure factors for the original data and after a local geometry optimization.
\begin{figure*}[!ht]
  \centering
  \includegraphics[width=\textwidth]{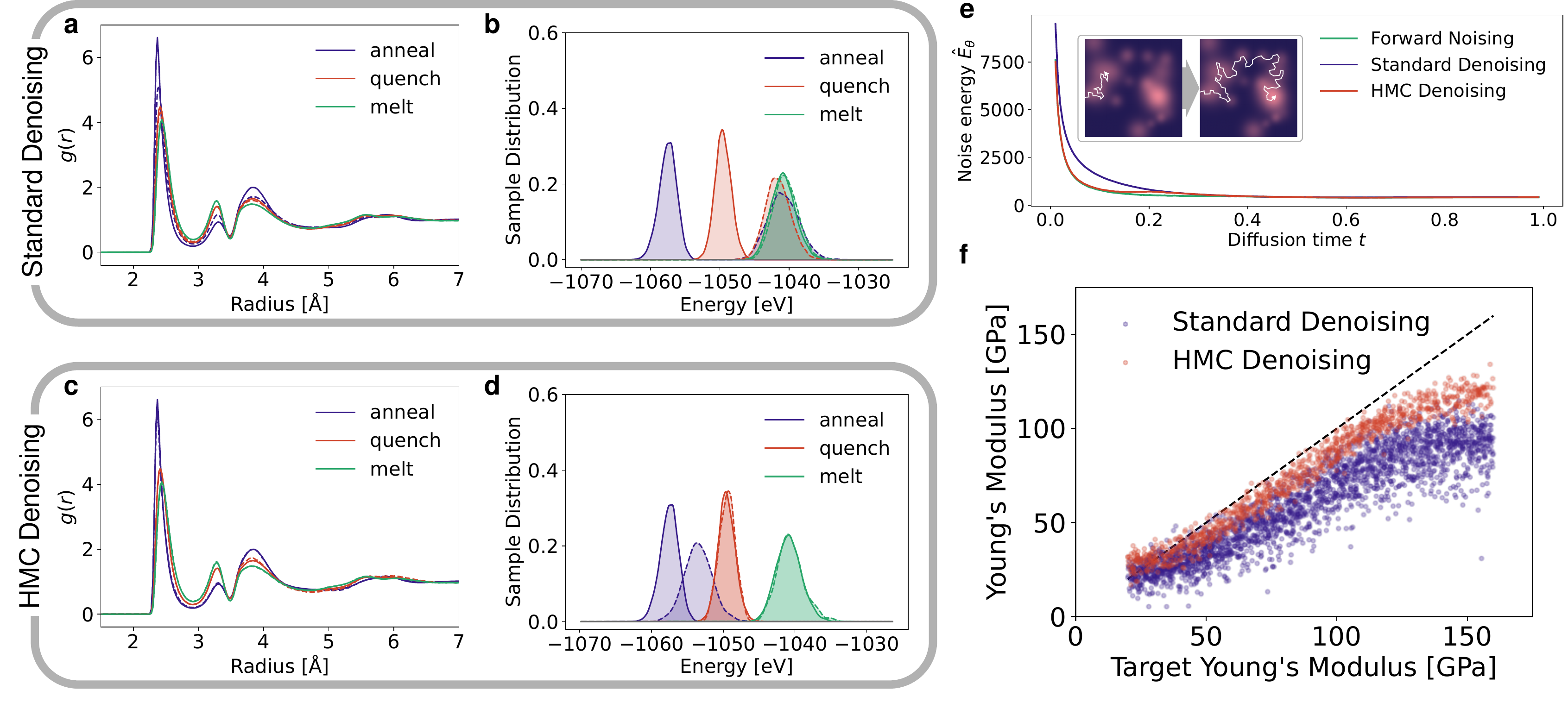}
  \caption{
    \textbf{Analysis of amorphous material samples generated by AMDEN with and without HMC denoising.}
    \textbf{a}, Radial distribution function (RDF) and \textbf{b}, potential energy distribution after geometry optimization for all three amorphous \ch{Si} datasets (solid lines) and samples generated by AMDEN with the standard denoising procedure (dashed lines).
    The generated \textit{quench} and \textit{anneal} samples differ clearly from the reference data. 
    Using HMC denoising, the RDFs in panel \textbf{c} match the training data closely, and the energy distributions in panel \textbf{d} are lowered, featuring closer agreement with the training data.
    \textbf{e}, Noise energy $\hat{E}_\theta$ predicted by the model trained on the \textit{anneal} \ch{Si} dataset plotted against diffusion time $t$.
    The \textit{forward noising} trajectory serves as the ground truth. For low $t$ values, the \textit{standard denoising} trajectory has a higher energy, while the \textit{HMC denoising} trajectory recovers to the lower values of the forward trajectory. 
    Inset illustrates the capability of HMC denoising to overcome barriers and find structures with lower $\hat{E}_\theta$.
    \textbf{f}, Young's moduli of samples generated by AMDEN trained on the MEG dataset plotted against the target value used for conditioning. The dashed line represents the parity line.
    While samples generated using the standard denoising process fall short of the targeted values, the HMC denoising process is able to generate samples with moduli close to the target values.
  }
  \label{fig:Si}
\end{figure*}

Fig.~\ref{fig:Si}a shows the RDFs after a local geometry optimization, while the other structural features and analysis of the unoptimized structures are provided in the Supplemental Fig.~S4.
Coordination number distributions, shown in Supplemental Fig.~S4i,j, and Voronoi volume distributions, shown in Supplemental Fig.~S4k, further confirm good short-range order agreement between the generated and training samples.
Local geometry optimizations were used to remove the influence of small residual noise and thermal motions and focus on the inherent structures instead. 
They were performed using the Stillinger-Weber potential~\cite{stillinger1985computer} which was also used for generating the training data. 
As expected, peaks in the RDF of Si become higher and narrower as the extent of relaxation increases from \textit{melt} to \textit{quench} and \textit{anneal} datasets.
Interestingly, the discrepancy between the generated and reference samples follows the same trend, with an almost perfect agreement being observed for the \textit{melt} data, while the largest difference can be seen in the \textit{anneal} data.
Considering the distribution of the potential energy of the geometry optimized samples shown in Fig.~\ref{fig:Si}b, we observe that the energy of the training samples decreases with increasing extent of relaxation, while samples generated by AMDEN have an almost identical energy distribution for all three datasets. 
AMDEN thus appears unable to generate the low-energy structures that are reached through relaxation processes. 
Despite the matching energy distributions, the atomic geometries generated by the model trained on the \textit{anneal} dataset are still structurally different from the samples obtained from the melt, as seen in Fig.~\ref{fig:Si}a and Supplemental Fig.~S4.
The generated \textit{quench} samples also match the training reference closely in terms of radial and angular distributions but differ in terms of energy.
These observations highlight the fact that structural features alone are not sufficient to assess the quality of generated samples. 

\subsection{Relaxation denoising}
\label{sec:relaxation-denoising}
To address the above-mentioned challenges, we therefore developed a new variant of AMDEN, which incorporates relaxation into the reverse diffusion process.
Instead of directly predicting the score function, this variant predicts a so-called noise energy $\hat{E}_\theta(x)$, from which the score $s_\theta(x)$ is calculated as 
\begin{equation}
  s_\theta(x) = -\frac{1}{k_B T}\nabla_x \hat{E}_\theta(x).
\end{equation}
Here, $k_\text{B} T$ are normalization constants and set to $1$. 
We note that $\hat{E}_\theta(x)$ is usually not a potential energy. 
Only in the case of an unconditional model, trained on Boltzmann distributed samples, will $\hat{E}_\theta(x)$ recover the potential energy of the system, given that $k_\text{B}$ and $T$ are set to the Boltzmann constant and the equilibrium temperature of the training data, respectively.
In the case of glasses, the distribution
\begin{equation}
  p(x) = \exp\left(\frac{-\hat{E}_\theta(x)}{k_\text{B} T}\right)
\end{equation}
is the out-of-equilibrium distribution of the system, which is not Boltzmann distributed and $\hat{E}_\theta \neq E_\text{pot}$.

Fig.~\ref{fig:Si}e shows $\hat{E}_\theta$ as predicted by a model trained on the \textit{anneal} \ch{Si} dataset. 
The values are plotted against the diffusion coordinate $t$ and obtained by averaging over ten forward noising trajectories and ten denoising trajectories labeled as \textit{forward} and \textit{std denoising}, respectively. 
The forward trajectories serve as the ground truth and ideally there should be no discernible difference in the denoising trajectories. 
However, as seen in the plot, both curves diverge around $t=0.4$ and the model ascribes a higher $\hat{E}_\theta$ and thus lower probability to the samples obtained from the standard denoising procedure. 
We therefore implement a modified variant of the denoising process, labeled \textit{HMC denoising} in the figure, which incorporates Hamiltonian Monte Carlo~\cite{duane1987hybrid} (HMC) steps on $\hat{E}_\theta$ during the denoising to equilibrate the structures during the denoising and sample from the true target distribution $p(x)$. 
As shown in Fig.~\ref{fig:Si}e, the HMC denoising process is able to recover noise-energies matching the forward trajectory by overcoming barriers of higher $\hat{E}_\theta$ as illustrated in the figure inset. 
Radial distribution functions shown in Fig.~\ref{fig:Si}e and bond angle distributions shown in Supplemental Fig.~S5 confirm that the samples obtained through HMC denoising are also structurally similar to those of the training data, while the standard denoising samples deviate significantly from the training data as shown in Fig.~\ref{fig:Si}a.
Coordination number distributions in Supplemental Fig.~S5i,j and Voronoi volume distributions in Supplemental Fig.~S5k similarly show improved agreement with HMC denoising.
Moreover, potential energies of the generated \textit{quench} and \textit{anneal} samples are reduced using HMC denoising, recovering the expected trend of lower potential energies given lower cooling rates as shown in Fig.~\ref{fig:Si}d.
This is remarkable as neither the potential energy nor forces are seen by the model during training, underlining the high quality of the generated structures.

\subsection{Material design through structure tuning}
The MEG results above show that AMDEN can perform inverse design by adjusting the composition to achieve targeted properties. In the following, we demonstrate that AMDEN can also perform purely structural inverse design on a dataset where composition is fixed.

In the previously presented MEG dataset, samples were generated following a fixed melt--quench procedure for a wide range of compositions.
Therefore, the variation in properties between samples can be explained purely by their composition. 
We should therefore expect that a generative model trained on this data relies purely on composition to obtain any given target property and then aims to recreate the structure that would be obtained from the same melt--quench procedure used for the training data generation. 
This is advantageous, as it allows for the synthesis of the generated materials via a comparable melt--quench procedure in an experimental setting. 

However, to learn the relationship between the atomic structure and the amorphous materials properties, a different training dataset is required that contains samples with the same composition but different atomic structures and resulting properties. 
For example, this can be achieved by varying the melt-quench procedure, e.g., by changing the cooling rate~\cite{li2017cooling} or varying the pressure~\cite{svenson2014composition}.
Samples generated by a model trained on such data can still be expected to be synthesizeable by varying the parameters of the melt-quench procedure. As an alternative, we here rely on the natural disorder of the atomic structure in amorphous materials. 
That is, although amorphous materials are homogeneous on a macroscopic scale, their structure varies on a local atomic scale. 
We can therefore use training samples containing relatively few atoms, in which certain structural features, such as ring sizes, network connectivity, or bond angles, deviate from the hypothetical average in an infinitely large sample.
This will allow the diffusion model to learn the relationship between those structural features and the materials properties. 
With AMDEN, we can then generate much larger samples with structural features that could not be obtained from any traditional MD melt--quench procedure. 

\begin{figure}[h]
  \centering
  \includegraphics[width=\linewidth]{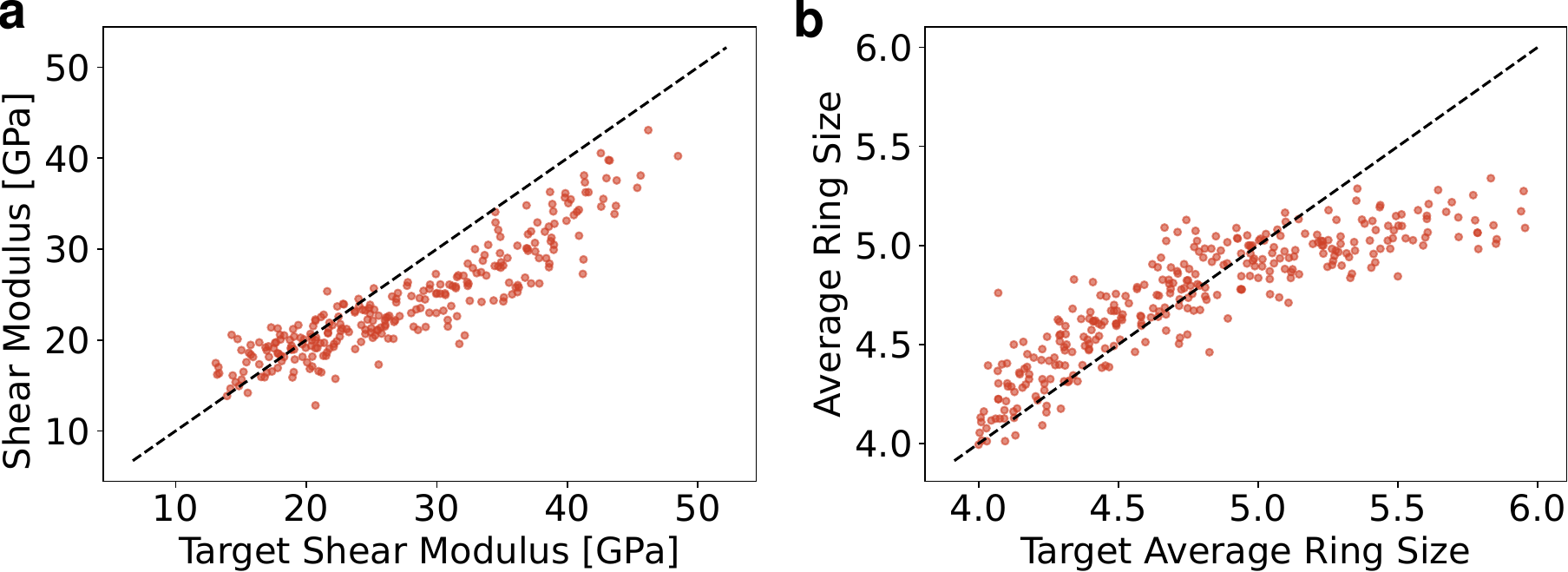}
  \caption{
    \textbf{Inference results for the amorphous silica dataset.}
    Comparison between generated and target \textbf{a} shear modulus and \textbf{b} average ring size. The dashed lines indicate parity lines in both panels.
    The models were trained on small samples to learn the structure--property relationship from the natural variation in local structure, while inference was performed on larger samples. 
  }
  \label{fig:SiO2}
\end{figure}

To illustrate this approach, we train AMDEN on the amorphous silica dataset, which consists of relatively small unit cells containing 80 to 250 atoms, while we used larger unit cells  for inference (350 to 500 atoms).
Since all samples share the same composition, the model has to rely on the atomic structure to modify the properties of the generated samples, providing a direct test of structural inverse design independent of composition.
We note that no requenching or post-processing MD simulation is applied; all results below are evaluated directly on the AMDEN-generated structures.
Inference results for our model trained on the silica dataset, conditioned on the shear modulus $G$ are shown in Fig.~\ref{fig:SiO2}a, revealing good correlation between the target and the predicted property.
The model is thus capable of adjusting the shear modulus of the generated sample by purely acting on the atomic structure, even beyond the range observed in the training data (23.2\,GPa to 32.5\,GPa, 5$^\text{th}$ and 95$^\text{th}$ percentile, respectively) despite extrapolating to a larger cell size. 

Besides directly conditioning on macroscopic materials properties, microscopic structural features can also be used in AMDEN. 
This enables new exciting opportunities to explore structure--property relationships. 
To demonstrate this, we conditioned the model on the average ring size (i.e., a medium-range order structural feature)~\cite{guttman1990ring}, where the ring size is given by the number of \ch{Si} atoms included in a ring.
A correlation plot between target and generated average ring size is shown in Fig.~\ref{fig:SiO2}b, showing a good correlation, especially for low target values, while the generated average ring size appears to plateau at around $5.25$.
Nevertheless, this is beyond the range observed by the model in the training data (4.59 to 5.18, 5$^\text{th}$ and 95$^\text{th}$ percentile, respectively).
We further verified that these enlarged generated cells remain physically realistic at the medium-range level by performing independent reference melt-quench MD simulations using the BKS potential~\cite{vanbeest1990force}, with an additional finite-size check using the Tersoff potential~\cite{munetoh2007interatomic} at three cubic supercell sizes (Supplemental Section~7 and Figs.~S11 and~S12).

Finally, we note that even though the amorphous silica dataset has a fixed composition, we used a model that predicts the element of each atom to enable the use of ghost atoms to adjust the density of the predicted structures. 
This is necessary as the shear modulus is closely linked to the density of the samples. 
The generated material samples do therefore not strictly follow the precise composition of \ch{SiO2} and slight deviations might occur due to the randomness of the denoising procedure. 
Considering formal charges of $+2\,e$ and $-1$\,e for \ch{Si} and \ch{O} atoms, respectively, the mean absolute errors in total formal charge per atom are $0.0025$\,e and $0.0096$\,e for the shear modulus and average ring size conditioning, respectively.
Of the 5\,000 generated samples, 275 (5.5\%) and 320 (6.4\%) are perfectly stoichiometrically balanced for the shear modulus and average ring size conditioning, respectively.
In Fig.~\ref{fig:SiO2}, only perfectly stoichiometrically balanced samples are included.
Correlation plots also including non-stoichiometric samples are shown in Supplemental Fig.~S3.
While stoichiometric deviations do not strongly correlate with the properties, samples conditioned on property values in the extrapolative regime are less likely to be stoichiometrically balanced (Supplemental Fig.~S9 and Table~S1).

\section{Discussion}
We have here introduced AMDEN, a diffusion model-based framework for the inverse design of amorphous materials.
Based on an equivariant graph neural network architecture, AMDEN generates atomic positions and elements and adjusts the density of the generated samples through ghost atoms. 
AMDEN operates in two complementary inverse design modes depending on the training data.
By training AMDEN on the MEG dataset, we highlight its capabilities for composition-based inverse design.
Our results show that AMDEN is able to adjust the Young's modulus of the generated samples while keeping their \ch{Li} content within a few percent of the targeted value.
However, the Young's moduli of the generated samples are underpredicted relative to the target values and differ from samples obtained by melt-quenching identical compositions, indicating inaccuracies in the atomic structure of the samples.
The requenching experiment confirms that AMDEN correctly identifies target compositions, isolating the structural inaccuracy as the source of the discrepancy.
With HMC denoising, this structural gap is substantially reduced, bringing the generated Young's moduli close to the target values without requenching (Fig.~\ref{fig:Si}f).
To investigate the origin of this discrepancy, we trained AMDEN on three amorphous \ch{Si} datasets with differing thermal histories.
When trained on structures sampled from the melt, the AMDEN-generated structures show almost perfect agreement in terms of radial distribution functions, angular distributions, coordination number distributions, and potential energy with the training data.
Samples generated by training on annealed samples, however, have a higher potential energy than the reference data and exhibit significant structural differences.
It thus appears that AMDEN is not able to generate lower energy samples obtained by slow quenching. 

During relaxation the potential energy of a glassy system is gradually lowered, as the system explores configuration space, driven by thermal motion. 
A key feature of glassy potential energy landscapes (PEL), preventing crystallization even on long timescales accessed by experiments, is the lack of a so-called funnel structure of the PEL~\cite{raza2015computer}. 
In systems with funnel structures, pathways to low-energy minima or even the global minimum of the PEL are present, which lead through a sequence of catchment basins with continuously decreasing energy and connected by low-energy barriers. 
This PEL structure thus allows for incremental improvements in lowering the system's energy. 
In glassy systems, however, this is not possible, as many funnels are present with varying depth, separated by high barriers ~\cite{niblett2017pathways}. 
Low-energy structures can therefore only be identified using a time-consuming random search like approach. 
This is problematic for diffusion models, which rely on the assumption that samples can be generated by incrementally improving the previous state in the denoising process.
Without HMC denoising, AMDEN thus fails to generate low-energy samples obtained upon slow annealing (in Fig.~\ref{fig:Si}b).

In image generating diffusion models, several phases can be identified during denoising~\cite{biroli2024dynamical}. 
After an initial phase of basically pure Brownian motion, speciation occurs, where the model commits to a single class from the training set, which is not changed anymore during later stages of the denoising. In the final phase, the denoising trajectory collapses onto a single sample. 
Assuming a perfect model, predicting the exact empirical score, the training samples would be reproduced. 
However, this is usually not desirable and in our case, when extrapolating to larger cell sizes, not possible. 
In practice, regularization is therefore used to obtain a smoothed score function~\cite{biroli2024dynamical,DBLP:conf/mm/ZengYZCCZY24}.
For image generating models, this enables interpolation between training samples, which is here prevented by the ruggedness of the PEL in models generating samples of glassy systems. 

We thus hypothesize that generating relaxed configurations beyond the training data is an inherent limitation of diffusion generative models, similar to their inability to sample spin glasses below a critical temperature~\cite{ghio2024sampling} and not the result of an insufficiently expressive model as suggested by Comin and Lewis~\cite{comin2019deep}.
Similarly, StriderNET, a reinforcement learning based approach for generating low-energy atomic structures, was shown to outperform classical local optimization methods but still failed to reliably predict the global minimum of the PEL~\cite{bihani2023stridernet}.
As observed by Lei \textit{et al.}~\cite{lei2024grand}, diffusion models can also fail to generate crystalline structures by getting trapped in local minima when the crystalline order is removed from the structure initializing the reverse diffusion process. 
The authors circumvent the problem by introducing a voxel-based diffusion model. 
In this approach, the orientation of the crystal lattice is implicitly provided to the model through the voxel grid breaking equivariance. 
It is thus unclear how voxel-based methods could be used to generate relaxed structures of amorphous materials.
Recently, it has been shown that a diffusion generative model was able to reproduce structural features and properties of amorphous \ch{SiO2} across a range of cooling rates~\cite{yang2025generative}.
The authors report that ``extra noise'' is required during the denoising process to escape local minima in the learned score function.
Additionally, short MD simulations are used to further refine the final structures and remove outlier environments.
Overall, these results match our observation that unmodified denoising procedures are insufficient to produce high quality amorphous structures that reproduce macroscopic properties.



In contrast, the amorphous \ch{SiO2} dataset demonstrates a purely structure-based design mode: since all samples share the same composition, the model must modulate properties through atomic arrangement alone.
We have demonstrated AMDEN's ability to learn the relationship between atomic structure and shear modulus from samples with few atoms using the \ch{SiO2} dataset.
This relationship can then be extrapolated to produce larger samples with a shear modulus that cannot be obtained by the melt--quench procedure used for the training data generation.
Similarly, we have shows that AMDEN can be directly conditioned on the average ring size of the generated samples, a structural feature which can only be controlled indirectly through traditional simulation procedures. 
This approach thus offers a new route for generating so-called ``forbidden glasses''~\cite{mauro2009forbidden}, i.e., glasses with structural features inaccessible through traditional melt--quench procedures, offering new opportunities to study structure--property relationships in disordered materials.

As with any data-driven generative model, AMDEN is most reliable when generating samples within the convex hull of its training data.
On the MEG dataset, the generated Young's modulus plateaus around 100\,GPa when targeting values above 110\,GPa, indicating limited extrapolation capability on the high end.
This is partly attributable to physical constraints, as very high Young's moduli may not be achievable at the targeted \ch{Li} content in oxide glasses.
Nevertheless, AMDEN demonstrates meaningful extrapolation toward lower property values, generating samples with Young's moduli well below the training range (Supplemental Fig.~S8a,b).
Similarly, on the \ch{SiO2} dataset, generated samples extend beyond the training distribution toward lower shear modulus and average ring size values (Supplemental Fig.~S8c,d), and the model extrapolates to larger cell sizes not seen during training.
We therefore characterize AMDEN as effective for targeted generation within and near the training distribution, with limited but non-trivial extrapolation capabilities that depend on the dataset and property of interest.
However, the fraction of stoichiometrically balanced samples decreases substantially in the extrapolative regime for the \ch{SiO2} dataset (Supplemental Fig.~S9), and the requench success rate for the MEG dataset drops in the same regime (Supplemental Fig.~S10), which should be taken into account when interpreting the filtered results.

While AMDEN is an important step towards the inverse design of amorphous materials, many challenges remain. 
Our proposed HMC denoising process is able to circumvent the limitation of the traditional approach in generating relaxed structures, but it comes with a significant computational cost.
During training, backpropagation has to be performed twice, i.e., first to derive the score function from the noise energy, and second to update the weights of the denoiser network.
Similarly, inference cost is increased as many evaluations of the model are required for each HMC update.
Potential solutions include shrinking the model size through neural network quantization techniques~\cite{DBLP:conf/cvpr/JacobKCZTHAK18} or employing shallower architectures, but these approaches may compromise model expressivity and generation performance.
Another factor hindering computational efficiency is that the model is not guaranteed to generate stoichiometrically balanced samples due to the stochastic nature of generative models.
As reported in Supplemental Section~6 and Table~S1, the valid fraction ranges from about 1--2\% for the MEG dataset to about 5--6\% for \ch{SiO2}, which may necessitate generating a larger pool of candidates when strict stoichiometric balance is required.
However, recent constrained sampling techniques such as physics-constrained flow matching~\cite{utkarsh2025physics} can enforce stoichiometric balance at inference time with negligible computational overhead and without retraining, and any remaining residual can be corrected by discrete search algorithms that find the minimum-cost set of element swaps.
Despite the increased per-sample cost of HMC denoising, generating a single MEG sample (${\sim}800$ atoms) takes approximately 1\,min on an NVIDIA A100 GPU, compared to 2.5--3.5\,hours for the corresponding LAMMPS melt--quench on a single CPU core.
More broadly, the primary advantage of the inverse design approach lies in accelerating the design loop, i.e., AMDEN directly generates structures conditioned on target properties, avoiding exhaustive trial-and-error exploration of the design space.

Finally, the biggest challenge that needs to be addressed is a lack of high-quality training data.
Recently, significant progress has been made in the generation and curation of large datasets suitable for training generative models for crystalline materials~\cite{jain2013commentary, saal2013materials, barroso2024open}.
Unfortunately, the situation for amorphous materials is currently much less developed.
This discrepancy can at least in parts be ascribed to the higher computation resources required for simulating amorphous materials on the DFT level. 
Larger unit cells are required to represent the medium-range order and many MD time steps are needed to adequately equilibrate the structures during a melt--quench simulation. 
Additionally, the dependence of the materials structure on the thermal history limits the compatibility between datasets from different sources (e.g., MD simulations vs. experiments) and promotes application specific but incompatible simulation workflows.
However, with increasing interest in glasses and other amorphous materials, efforts are being made to establish consistent simulation workflows that enable the generation of large datasets of amorphous materials~\cite{zheng2024ab}.

More diverse and extensive training data will enable exciting applications of AMDEN and help to design new materials and allow generative models to become an established tool in materials science research, extending the applicability of current models and ideas beyond the limitations of current datasets.
With DFT-level training data, functional properties such as optical spectra could also be targeted for inverse design, leveraging ML surrogate models for optical property prediction~\cite{hung2024universal}.
For instance, generative models could become an important tool to decipher the atomic structure of samples from experimental data as demonstrated by Kwon \textit{et al.}~\cite{pham2023spectroscopy} on amorphous carbon. As illustrated by our \ch{SiO2} example, AMDEN can also expand the search space for new materials beyond the limitations of traditional synthesis methods. 
While more research and new synthesis methods are needed to create ``forbidden'' structures predicted by AMDEN, \textit{in silico} studies can help to shed light on unexplored structure--property relationships in the meantime.

%

\section{Methods}

\subsection{Datasets}
The datasets used in this work were obtained from simulation workflows implemented using the Large-scale Atomic/Molecular Massively Parallel Simulator (LAMMPS)~\cite{lammps} and the atomic simulation environment (ASE)~\cite{larsen2017atomic}.
Geometry optimizations were performed using a stabilized quasi Newton method, adapted for periodic boundary conditions~\cite{gubler2023efficient}.
Elastic moduli were calculated using finite differences of the stress tensor.
More detailed information about the dataset generation and the calculation of properties are provided in the supplementary material.

\subsubsection{Multi Element Glass dataset}
The Multi Element Glass (MEG) Dataset consists of 9,027  glass samples, containing 11 different elements.
Initial structures were generated from varying compositions of the glass formers \ch{SiO2} and \ch{P2O5}, and the modifiers \ch{Al2O3} \ch{Li2O}, \ch{BeO}, \ch{K2O}, \ch{CaO}, \ch{TiO2}, \ch{BaO} and \ch{ZnO}.
The samples were then obtained using a simulation workflow based on the Bertain--Menziani--Pedone (BMP)-shrm classical interatomic potential~\cite{bertani2021improved}. 
The workflow determines a suitable initial temperature for the melt--quench procedure for each composition to ensure proper melting of the sample while avoiding evaporation. 
Samples are then quenched at $5$\,K/ps to $300$\,K and equilibrated.

\subsubsection{Amorphous \ch{Si} dataset}
Three amorphous silicon datasets were generated using the Stillinger-Weber potential~\cite{stillinger1985computer}.
Each of the three datasets consisted of 10,000 samples containing 256 \ch{Si} atoms. 
All samples were created using the same simulation workflow but using a different random seed ensuring a unique atomic structure.
Samples in the \textit{melt} dataset were obtained from the melt at 2500\,K, while samples in the \textit{quench} dataset were obtained after a rapid quench to 300\,K.
While the thermostat temperature was lowered instantaneously, a short amount of time was needed for the system to reach the new temperature, which is why a small amount of relaxation is also included in this dataset.
Samples in the \textit{anneal} dataset were slowly quenched to 300\,K at a cooling rate of 1\,K/ps.

\subsubsection{Amorphous \ch{SiO2} dataset}
Samples of the amorphous silica (\ch{SiO2}) dataset vary in size, with the number of atoms being uniformly distributed between 80 and 250 atoms.
Simulations were performed using the Tersoff potential parameterized by Munetoh \textit{et al.}~\cite{munetoh2007interatomic}.
Samples were initially melted at 3500\,K and the immediately quenched using a local geometry optimization to avoid relaxation effects. 
Finally, samples were equilibrated at 300\,K for 10\,ps.

\subsection{AMDEN}

\subsubsection{Materials diffusion process}
\label{sec:model-diffusion-process}

Following the SDE framework~\cite{song2020score}, our materials diffusion process is composed of two subprocesses: a forward diffusion process and a reverse denoising diffusion process. The forward diffusion process provides ground truth for noisy samples at intermediate steps $x_t$ and noise components to supervise the score function $s_\theta$. 

\textbf{The forward diffusion process} gradually transformed a clean material sample $x$ into random noise $\tilde x$ through a sequence of time step $t \in [t_{\text{min}}, t_{\text{max}}]$.
Formally, given a clean sample $x=(\mC, \mX, \mE)$, the positions and element embeddings of the noisy sample $x_t$ at time step $t$ were calculated as,
\begin{equation}
\begin{aligned}
{\mX}_t &= \alpha_{\mX}(t) \cdot \mX + \sigma_{\mX}(t) \cdot \boldsymbol{\epsilon}_{\mX}, \\
{\mE}_t &= \alpha_{\mE}(t) \cdot \mE + \sigma_{\mE}(t) \cdot \boldsymbol{\epsilon}_{\mE},
\end{aligned}
\label{eq:forward-diffusion-process}
\end{equation}
where $\boldsymbol{\epsilon}_{\mX}$ and $\boldsymbol{\epsilon}_{\mE}$ are noise components sampled from $\mathcal{N}(0, \mathbf{I})$, and $\alpha$ and $\sigma$ are time-dependent scaling factors specific to each noise schedule.

For atomic positions, we employed a variance exploding (VE) schedule~\cite{song2019generative},
\begin{equation}
\alpha_{\mX}(t) = 1, \quad 
\sigma_{\mX}(t) = \sigma_{\text{max}}^\mX \cdot t,
\end{equation}
where $\sigma_{\text{max}}^\mX$ was set to $1.7$\,\AA{} for models trained on the MEG dataset and $1.5$\,\AA{} for all other models.
For element embeddings, we implemented a variance preserving (VP) schedule~\cite{song2019generative} with a cosine progression,
\begin{equation}
\alpha_{\mE}(t) = \cos\left(\frac{\pi}{2}t\right), \quad
\sigma_{\mE}(t) = \sin\left(\frac{\pi}{2}t\right) \cdot \sigma_{\text{max}}^\mE,
\end{equation}
where $\sigma_{\text{max}}^\mE$ was set to $1.5$.
Combined with the periodic boundary condition, the two schedules determine the positions $\mX_{t_\text{max}}$ and element embeddings $\mE_{t_\text{max}}$ of the sample $x_{t_\text{max}}$.
At the last step of the forward diffusion process, the atomic position and element embeddings follow a uniform distribution within the cell $\mC$ and standard Gaussian distribution $\mathcal{N}(0, \mathbf{I})$, respectively. 
This created a starting point for the reverse denoising diffusion process that can be sampled trivially.

\textbf{The reverse denoising process} gradually transforms random noise into a material sample through iterative denoising manipulated by the learnable score function $s_\theta$.
To progress the process at $t$ by $\Delta t$, the Euler-Maruyama method was applied to solve the reverse SDE,
\begin{equation}
  \begin{aligned}
  \mX_{t-\Delta t} &= \mX_t - \left( f_{\mX}(t, \mX_t) \cdot \Delta t + g_{\mX}(t) \cdot \mathbf{z}_{\mX} \cdot \sqrt{\Delta t} \right), \\
  \mE_{t-\Delta t} &= \mE_t - \left(f_{\mE}(t, \mE_t) \cdot \Delta t + g_{\mE}(t) \cdot \mathbf{z}_{\mE} \cdot \sqrt{\Delta t} \right),
  \end{aligned}
\end{equation}
where $\mathbf{z}_{\mX}$ and $\mathbf{z}_{\mE}$ are noises sampled independently from a standard normal distribution, $f$ and $g$ are coefficients defined as,
\begin{equation}
  \begin{aligned}
  f_{\mX}(t,\mX_t) &= - g_\mX^2(t) \cdot s_\theta^{\mX}, \\
  g_{\mX}(t) &= \sqrt{2 {\sigma_{\text{max}}^\mX}^2 \cdot t}, \\
  f_{\mE}(t,\mE_t) &= -\frac{\pi}{2}\tan\left(\frac{\pi}{2}t\right) \cdot \mE_t - g_\mE^2(t) \cdot s_\theta^{\mE}, \\
  g_{\mE}(t) &= \sqrt{\pi \cdot \alpha_{\mE}(t) \cdot \sigma_{\mE}(t) \cdot \sigma_{\text{max}} + \pi \cdot \tan\left(\frac{\pi}{2}t\right) \cdot \sigma_{\mE}(t)^2}.
  \end{aligned}
\end{equation}

In these equations, $s_\theta^{\mX}$ and $s_\theta^{\mE}$ are position and element components of the score function, respectively.

The complete reverse denoising process started from a noise sample $\tilde x=(\mC, \mX_{t_\text{max}}, \mE_{t_\text{max}})$ where $\mX_{t_\text{max}}$ was sampled from uniform random distributed in $\mC$ and $\mE_{t_\text{max}}$ was sampled from $\mathcal{N}(0, \mathbf{I})$, aligning with the sample $x_{t_\text{max}}$ produced at the last step of the forward diffusion process.
The reverse process progressed sequentially from $t_{\text{max}}=0.99$ to $t_{\text{min}}=0.01$ with $n=200$ evenly spaced intervals and a final step to $t=0$.

\subsubsection{Equivariant graph neural network}
The EGNN serves as the equivariant backbone of the score function.
The graph $\gG=(\gV, \gE)$ provided to EGNN is a graph of the atoms in each sample $x$, calculated with a cutoff radius $r_{\text{cut}}=6.5$\,\AA{}.
The cutoff radius was chosen to ensure that all bonded and strongly interacting atoms share a direct edge connection while still keeping the graph sufficiently sparse.
Our EGNN implementation was composed of $L=4$ EGNN layers.
The $l$-th layer takes as input:
1) Node features $\mH^{(l)} \in \sR^{n \times d_h}$ containing information of the corresponding atoms;
2) Positional coordinates $\mX^{(l)} \in \sR^{n \times k \times 3}$ of the atoms, where $k=8$ is the number of vector channels~\cite{levy2023using}; and
3) Edge set $\gE$ of the graph $\gG$.

For the initial layer, the positions $\mX^{(0)}$ were replicated original positions $\mX$ for $k$ channels.
The edge attributes $\ve_{ij}$ were derived from the distance embedding,
\begin{equation}
\ve_{ij} = \tanh\left(\frac{\|\mX_i - \mX_j - \mathbf{o}_{ij}\|^2}{r_{\text{cut}}^2}\right) \cdot 2 - 1,
\end{equation}
where $\mathbf{o}_{ij}$ is the offset vector accounting for periodic boundary conditions.

Each layer updated the node features and positional coordinates, incorporating self-attention~\cite{DBLP:conf/nips/VaswaniSPUJGKP17} with a hidden dimension of $128$ as,
\begin{equation}
  \begin{aligned}
    \mathbf{m}_{ij}^{(l)} &= \phi_e^{(l)}(\mH_i^{(l-1)}, \mH_j^{(l-1)}, \ve_{ij}), \\
    \alpha_{ij}^{(l)} &= \sigma(\text{MLP}_{\text{att}}(\mathbf{m}_{ij}^{(l)})), \\
    \hat{\mathbf{m}}_{ij}^{(l)} &= \alpha_{ij}^{(l)} \cdot \mathbf{m}_{ij}^{(l)}, \\
    \mH_i^{(l)} &= \mH_i^{(l-1)} + \phi_H^{(l)}\left(\mH_i^{(l-1)}, \sum_{j \in N(i)} \frac{f_\text{cut}(d_{ik}^{(0)}) \cdot \hat{\mathbf{m}}_{ij}^{(l)}}{n_\text{norm}}\right), \\
    \mathbf{\Phi}_{ij}^{(l)} &= \text{MLP}_{\text{coord}}([\mH_i^{(l)}, \mH_j^{(l)}, \ve_{ij}]) \in \sR^{k \times k}, \\
    \mathbf{d}_{ij}^{(l)} &= \mX_i^{(l-1)} - \mX_j^{(l-1)} - \mathbf{o}_{ij}, \\
    \mX_i^{(l)}{'} &= \sum_{j \in N(i)} \frac{1}{n_\text{norm}} \cdot \mathbf{\Phi}_{ij}^{(l)} \cdot \mathbf{d}_{ij}^{(l)}, \\
    \mX_i^{(l)} &= \mX_i^{(l-1)} + \mX_i^{(l)}{'},
  \end{aligned}
\end{equation}
where $N(i)$ represents the neighbors of atom $i$, derived from the edge set $\gE$ and
$\sigma$ is the sigmoid activation function for self-attention.
$n_\text{norm}$ is a normalization factor (typically proportional to the average number of neighbors) to ensure numerical stability, which we set to $40$.
$\phi_e^{(l)}$, $\phi_H^{(l)}$ are implemented as multi-layer perceptrons (MLPs) with SiLU activation functions and layer normalization. 
$\mathbf{\Phi}_{ij}^{(l)}$ is a learned transformation matrix that maps between the $k$ vector channels.
A smooth cutoff function is used to prevent discontinuities when atoms leave or enter the cutoff radius, which is defined as follows. 
\begin{equation}
  f_\text{cut}(r) = 2 \tanh\left(1 - \frac{\min(r, r_\text{cut})}{r_\text{cut}}\right)^2.
\end{equation}

In our implementation, the learnable functions were structured as follows,
\begin{equation}
  \begin{aligned}
    \phi_e^{(l)}(\mH_i, \mH_j, \ve_{ij}) &= \text{MLP}_{\text{edge}}([\mH_i, \mH_j, \ve_{ij}]), \\
    \phi_H^{(l)}(\mH_i, \mathbf{m}_{\text{agg}}) &= \text{MLP}_{\text{node}}([\mH_i, \mathbf{m}_{\text{agg}}]).
  \end{aligned}
\end{equation}

At the last layer, EGNN outputs $\mH^{(L)}$ and $\mX^{(L)}$ as the final node features and positional coordinates, respectively.

\subsubsection{Score function}
The score function extended the EGNN framework to predict noise components in both positions and element embeddings during the reverse denoising diffusion process, optionally conditioned on desired properties.
The score function $s_\theta$ mapped a noisy material sample $x_t=(\mC, \mX_t, \mE_t)$, diffusion step $t$, and desired properties $\vy$ to position and element components,
\begin{equation}
s_\theta: (\mC, \mX_t, \mE_t, t, \vy) \mapsto (s_\theta^{\mX}, s_\theta^{\mE}),
\end{equation}
which we implemented with the following three main components.

First, the property embedding layer transformed material properties into a continuous embedding space.
For each property $y_i$ in the property set $\sY$, we implemented,
\begin{equation}
\mathbf{h}_{y_i} = \text{LayerNorm}(\text{Linear}(y_i)),
\label{eq:property_embedding}
\end{equation}
where $\text{Linear}$ is a fully-connected linear layer that maps the property $y_i$ into a $d_{\text{emb}}=8$-dimensional embedding vector.

Second, the feature assembly layer combined time embeddings, element representations, and property embeddings.
The initial node features provided to the EGNN backbone were assembled by concatenating,
1) diffusion time step $t$;
2) element embeddings: $\mathbf{e}_i \in \mathbb{R}^{d_e}$; and
3) property embeddings: $\mathbf{h}_{y_1}, \mathbf{h}_{y_2}, \ldots, \mathbf{h}_{y_m}$ (for each property in $\sY$).
Formally, the initial node features $\mH_i^{(0)}$ for atom $i$ were given by,
\begin{equation}
\mH_i^{(0)} = [t, \mathbf{e}_i, \mathbf{h}_{y_1}, \mathbf{h}_{y_2}, \ldots, \mathbf{h}_{y_m}].
\end{equation}

Third, the EGNN backbone processed the assembled features while preserving geometric equivariance.
The assembled node features, along with the raw noisy material sample $x_t$, were processed through the EGNN backbone. We calculated the predicted noise components for positions and elements as,
\begin{equation}
  \epsilon_\theta^{\mX} = \mX - \mX^{(L,0)},\quad \epsilon_\theta^{\mE}=\mH^{(L)},
\end{equation}
where for positions we used the deviation between the original positions and the first channel of output positions, and for elements we directly used the output node features. The score was then calculated as $s_\theta=-\epsilon_\theta/\sigma(t)$ for both positions and elements.

We further incorporate classifier-free guidance (CFG)~\cite{ho2022classifier} to enhance the calculation of conditioned scores. Denoting the unconditioned noise as $\tilde{\epsilon}_\theta^{\mX}$ and $\tilde{\epsilon}_\theta^{\mE}$, the final noise under CFG is calculated as,
\begin{equation}
  \begin{aligned}
    {\epsilon'}_\theta^{\mX} &= (1 + w_\text{cond}) \epsilon_\theta^{\mX} - w_\text{cond} \tilde{\epsilon}_\theta^{\mX}, \\
    {\epsilon'}_\theta^{\mE} &= (1 + w_\text{cond}) \epsilon_\theta^{\mE} - w_\text{cond} \tilde{\epsilon}_\theta^{\mE},
  \end{aligned}
\end{equation}
where $w_\text{cond}$ is the condition weight. The unconditioned noise can be calculated by using a dedicated unconditioned model trained without the properties $\vy$, or by leveraging the independent condition guidance (ICG)~\cite{sadat2024no}--feeding the model with random properties sampled from the distribution of the training set. For the MEG dataset, we use the ICG technique and set $w_\text{cond}=0.25$. For the \ch{SiO2} dataset, we use a dedicated unconditioned model and set $w_\text{cond}=1.0$.


The training of the score function was done by supervising the reverse denoising process at randomly sampled time steps. Specifically, for each training sample, we:
1) randomly generated a diffusion time step $t \sim \mathcal{u}(t_{\text{min}}, t_{\text{max}})$;
2) applied the forward diffusion process to obtain the noisy sample and ground truth noise components;
3) fed the noisy sample into the score function to predict the noise components; and
4) computed the loss as the mean squared error between the prediction and ground truth.

Formally, the training objective was,
\begin{equation}
\mathcal{L}(\theta) = \mathbb{E}_{x,t} \Big[ \| \boldsymbol{\epsilon}_{\mX} -  \epsilon_\theta^{\mX} \|^2 + \lambda \| \boldsymbol{\epsilon}_{\mE} - \epsilon_\theta^{\mE} \|^2 \Big],
\end{equation}
where $\lambda=0.5$ is a weighting factor balancing the importance of position and element noise prediction.
The training was performed with the Adam optimizer~\cite{DBLP:journals/corr/KingmaB14} for 400, 500, and 1000 epochs with batch sizes of 2, 8, and 8 and learning rates of $10^{-3}$, $10^{-3}$, and $5\cdot 10^{-4}$ for the MEG, Si, and SiO$_2$ datasets, respectively.

\subsubsection{Hamiltonian Monte Carlo refinement}

\textbf{The energy-based score function} was introduced to support the prediction of noise energy utilized in Section~\ref{sec:relaxation-denoising}.
The key difference between the standard score function and the energy-based variant lies in how noise components are predicted.
While the standard score function directly predicted position and element noise components, the energy-based variant reformulated the noise prediction problem using an energy function $\hat{E}_\theta$ and then computed noise as derivatives of this energy.
It should be noted that $\hat{E}_\theta$ does not necessarily relate to any physical energy, such as the potential energy.
Only in the special case, when the model was trained on Boltzmann distributed samples and $t=0$, the model would learn the potential energy of the system, up to an additive constant.

The standard score function $s_\theta$ was defined through the targeted distribution $p(x)$ as,
\begin{equation}
  s_\theta(x) \approx \nabla_x \ln p(x).
\end{equation}

In the energy-based variant, $p(x)$ was predicted through $\hat{E}_\theta$ as,
\begin{equation}
  p(x) = \frac{1}{Z} \exp\left(\frac{-\hat{E}_\theta(x)}{k_\text{B} \times T}\right),
\end{equation}
where $Z$ is the unknown partition function normalizing $p(x)$, $k_\text{B}$ is the Boltzmann constant, and $T$ is the temperature. 
Since $\hat{E}_\theta$ does not match the potential energy $k_\text{B}$ and $T$ are arbitrary normalization constants and chosen to be $k_\text{B} \times T = 1$,
we thus obtain 
\begin{equation}
  s_\theta(x) = -\frac{1}{k_\text{B} \times T} \nabla_x \hat{E}_\theta(x).
\end{equation}

Specifically, the score function was reconfigured to output a scalar atomic energy value for each atom, instead of directly outputting noise vectors.
For each atom $i$, the noise energy was constructed as a combination of:
1) position-based energy~\cite{salimans2021should}: $e^{\mX}_i = \frac{1}{2}\Vert\mX_i - \mX_i^{(L)}\Vert^2$
, derived from squared distances between original positions and EGNN-output positions; and
2) atomic energy: $e^{\text{atom}}_i = \mH_i^{(L)}$, EGNN-output node features.

The total system noise energy was given by,
\begin{equation}
  \hat{E}_\theta(x) = \frac{1}{\sigma(t)}\sum_{i=1}^{n}\left(e^{\mX}_i + \gamma \cdot e^{\text{atom}}_i\right),
\end{equation}
where $\gamma$ is a learnable scale factor balancing the contribution of atomic energies.

Then, the position and element components of the score function were computed through automatic differentiation,
\begin{equation}
\begin{aligned}
  s_\theta^\mX &= -\frac{1}{k_\text{B} T} \nabla_\mX \hat{E}_\theta(x), \\
  s_\theta^\mE &= -\frac{1}{k_\text{B} T} \nabla_\mE \hat{E}_\theta(x).
\end{aligned}
\end{equation}

We note that the energy-based score function was computationally heavier compared to the standard variant, thus its training was performed with smaller batch sizes: $1$, $4$, and $4$ for MEG, Si, and \ch{SiO2} datasets, respectively.
For the same reason it is trained for fewer--150--epochs on the MEG dataset.

\textbf{Hamiltonian Monte Carlo denoising.}
As discussed in previous sections, the standard denoising procedure outlined in Section~\ref{sec:model-diffusion-process} was unable to produce samples of relaxed structures. This was because the model failed to steer the denoising trajectory towards relaxed samples in the early stages of denoising.
However, analysis of energy-based score function shows that the model is able to assign a higher probability to relaxed samples at later stages of the denoising process.
This analysis allowed us to refine the generated samples by equilibrating on the predicted distribution $p(x)$ during the denoising process using Hamiltonian Monte Carlo (HMC).

A series of HMC iterations were therefore performed between the traditional denoising iterations.
The application of HMC iterations can be limited to a range of diffusion time steps $t$ within a specified range $[t_{\text{min}}^{\text{HMC}}, t_{\text{max}}^{\text{HMC}}]$ to reduce the computational cost. We set $t_{\text{min}}^{\text{HMC}}=0.0$ and $t_{\text{max}}^{\text{HMC}}=0.5$, respectively. Exactly one HMC iteration is performed before each diffusion denoising step within the range.

In each HMC iteration, atomic coordinates were updated using the following steps: 
\begin{enumerate}
  \item Initializing momenta from a Maxwell--Boltzmann distribution: $\mathbf{p} \sim \mathcal{N}(0, k_\text{B} T \mathbf{M})$, where $\mathbf{M}$ is the diagonal mass matrix chosen as $\mathbf{M}=\mathbf{I}$;
  \item Computing the initial energy $\hat{E}_\theta$ and total energy $E_{\text{tot}} = \hat{E}_\theta + \frac{1}{2} \Vert \mathbf{p} \Vert^2$;
  \item Evolving the system using velocity Verlet integration with $15$ steps,
   \begin{equation}
   \begin{aligned}
   \mX &\leftarrow \mX + \mathbf{p} \cdot dt + \frac{1}{2} \mF \cdot dt^2, \\
     \mF &\leftarrow -\nabla_{\mX} \hat{E}_\theta(\mX), \\
   \mathbf{p} &\leftarrow \mathbf{p} + \frac{1}{2}(\mF + \mF_{\text{last}}) \cdot dt;
   \end{aligned}
   \end{equation}
\item Computing the final energy and accepting or rejecting the final structure according to the Metropolis--Hastings criterion with probability, 
   \begin{equation}
     \alpha = \min\left[1, \exp\left(\frac{-(E_{\text{tot}}^{\text{final}} - E_{\text{tot}}^{\text{initial}})}{k_\text{B}T}\right)\right].
   \end{equation}
\end{enumerate}

To ensure a consistent acceptance rate of roughly 0.5, the timestep $dt$ was adjusted using $dt = \sigma_t dt^0$, where $dt^0$ is a constant.
This accounted for the fact that $\hat{E}_\theta$ becomes less smooth for lower $\sigma_t$.
We set $dt^0 = 0.2$ for the MEG dataset and $dt^0 = 0.4$ for both Si and \ch{SiO2} datasets.
We set the number of diffusion steps to $n=2000$ when performing HMC refinement to ensure there are enough HMC iterations for finding lower energy structures.
We also set $w_\text{cond}=0$, i.e., we do not use CFG when performing HMC refinement to avoid adding further computational burden.

This sampling approach draws on an analogy with physical relaxation: just as thermal motion enables a glassy system to overcome barriers on the potential energy landscape, the HMC steps enable the denoising trajectory to cross barriers on the $\hat{E}_\theta$ landscape.
We emphasize that $\hat{E}_\theta$ encodes the learned probability distribution of the training data, not the physical potential energy (see Section~\ref{sec:relaxation-denoising}).
Nevertheless, this approach enabled the generation of more stable and realistic material structures by exploring the probability landscape at each diffusion step.
On the other hand, since the energy was provided by the denoiser network conditioned on properties, incorporation of HMC sampling preserved the property-conditioned generation process.

\section{Data availability}
The source code of AMDEN and all datasets used for this work are available at \url{https://github.com/Logan-Lin/AMDEN-code}.

\section{Author contributions}
Conceptualization, M.M.S. and J.H.; Methodology, M.M.S., J.H, T.D., Y.L. and J.A.F.; Software, Y.L. and J.A.F.; Validation, Y.L. and J.A.F.; Formal analysis, Y.L. and J.A.F.; Investigation, Y.L. and J.A.F.; Resources, M.M.S. and J.H.; Data curation, Y.L. and J.A.F.; Writing---original draft, Y.L. and J.A.F.; Writing---review \& editing, Y.L., J.A.F., T.D., M.M.S. and J.H.; Visualization, Y.L. and J.A.F.; Supervision, M.M.S., J.H. and T.D.; Project administration, M.M.S. and J.H.; Funding acquisition, M.M.S. and J.H.

\section{Acknowledgements}
We thank N.M.A. Krishnan (IIT Delhi) for helpful discussions. 
This work was supported by a research grant (VIL57373) from VILLUM FONDEN. 
We also acknowledge the computational resources from EuroHPC Joint Undertaking (EHPC-EXT-2024E02-123) and Danish e-Infrastructure Cooperation (DeiC-AAU-N5-000006) for access to the LUMI supercomputer, hosted by CSC (Finland).

\section{Competing interests}
The authors declare no competing interests.

\bibliography{paper_abbreviated.bib}

\begin{thebibliography}{10}
\expandafter\ifx\csname url\endcsname\relax
  \def\url#1{\texttt{#1}}\fi
\expandafter\ifx\csname urlprefix\endcsname\relax\def\urlprefix{URL }\fi
\providecommand{\bibinfo}[2]{#2}
\providecommand{\eprint}[2][]{\url{#2}}

\bibitem{liu2024amorphous}
\bibinfo{author}{Liu, Y.}, \bibinfo{author}{Madanchi, A.},
  \bibinfo{author}{Anker, A.~S.}, \bibinfo{author}{Simine, L.} \&
  \bibinfo{author}{Deringer, V.~L.}
\newblock \bibinfo{title}{The amorphous state as a frontier in computational
  materials design}.
\newblock \emph{\bibinfo{journal}{Nat. Rev. Mater.}} \bibinfo{pages}{1--14}
  (\bibinfo{year}{2024}).

\bibitem{zunger2018inverse}
\bibinfo{author}{Zunger, A.}
\newblock \bibinfo{title}{Inverse design in search of materials with target
  functionalities}.
\newblock \emph{\bibinfo{journal}{Nat. Rev. Chem.}}
  \textbf{\bibinfo{volume}{2}}, \bibinfo{pages}{0121} (\bibinfo{year}{2018}).

\bibitem{DBLP:journals/corr/KingmaW13}
\bibinfo{author}{Kingma, D.~P.} \& \bibinfo{author}{Welling, M.}
\newblock \bibinfo{title}{Auto-encoding variational bayes}.
\newblock In \emph{\bibinfo{booktitle}{ICLR}} (\bibinfo{year}{2014}).

\bibitem{goodfellow2020generative}
\bibinfo{author}{Goodfellow, I.} \emph{et~al.}
\newblock \bibinfo{title}{Generative adversarial networks}.
\newblock \emph{\bibinfo{journal}{Commun. ACM}} \textbf{\bibinfo{volume}{63}},
  \bibinfo{pages}{139--144} (\bibinfo{year}{2020}).

\bibitem{gebauer2019symmetry}
\bibinfo{author}{Gebauer, N.}, \bibinfo{author}{Gastegger, M.} \&
  \bibinfo{author}{Sch{\"u}tt, K.}
\newblock \bibinfo{title}{Symmetry-adapted generation of 3d point sets for the
  targeted discovery of molecules}.
\newblock \emph{\bibinfo{journal}{Adv. Neural Inf. Process. Syst.}}
  \textbf{\bibinfo{volume}{32}} (\bibinfo{year}{2019}).

\bibitem{hoffmann2019data}
\bibinfo{author}{Hoffmann, J.} \emph{et~al.}
\newblock \bibinfo{title}{Data-driven approach to encoding and decoding 3-d
  crystal structures}.
\newblock \emph{\bibinfo{journal}{arXiv preprint arXiv:1909.00949}}
  (\bibinfo{year}{2019}).

\bibitem{noh2019inverse}
\bibinfo{author}{Noh, J.} \emph{et~al.}
\newblock \bibinfo{title}{Inverse design of solid-state materials via a
  continuous representation}.
\newblock \emph{\bibinfo{journal}{Matter}} \textbf{\bibinfo{volume}{1}},
  \bibinfo{pages}{1370--1384} (\bibinfo{year}{2019}).

\bibitem{court20203}
\bibinfo{author}{Court, C.~J.}, \bibinfo{author}{Yildirim, B.},
  \bibinfo{author}{Jain, A.} \& \bibinfo{author}{Cole, J.~M.}
\newblock \bibinfo{title}{3-d inorganic crystal structure generation and
  property prediction via representation learning}.
\newblock \emph{\bibinfo{journal}{J. Chem. Inf. Model.}}
  \textbf{\bibinfo{volume}{60}}, \bibinfo{pages}{4518--4535}
  (\bibinfo{year}{2020}).

\bibitem{long2021constrained}
\bibinfo{author}{Long, T.} \emph{et~al.}
\newblock \bibinfo{title}{Constrained crystals deep convolutional generative
  adversarial network for the inverse design of crystal structures}.
\newblock \emph{\bibinfo{journal}{npj Comput. Mater.}}
  \textbf{\bibinfo{volume}{7}}, \bibinfo{pages}{66} (\bibinfo{year}{2021}).

\bibitem{li2018limitations}
\bibinfo{author}{Li, J.}, \bibinfo{author}{Madry, A.},
  \bibinfo{author}{Peebles, J.} \& \bibinfo{author}{Schmidt, L.}
\newblock \bibinfo{title}{On the limitations of first-order approximation in
  gan dynamics}.
\newblock In \emph{\bibinfo{booktitle}{ICML}}, \bibinfo{pages}{3005--3013}
  (\bibinfo{organization}{PMLR}, \bibinfo{year}{2018}).

\bibitem{DBLP:conf/iclr/DaunhawerSCPV22}
\bibinfo{author}{Daunhawer, I.}, \bibinfo{author}{Sutter, T.~M.},
  \bibinfo{author}{Chin{-}Cheong, K.}, \bibinfo{author}{Palumbo, E.} \&
  \bibinfo{author}{Vogt, J.~E.}
\newblock \bibinfo{title}{On the limitations of multimodal vaes}.
\newblock In \emph{\bibinfo{booktitle}{ICLR}} (\bibinfo{year}{2022}).

\bibitem{lucas2019don}
\bibinfo{author}{Lucas, J.}, \bibinfo{author}{Tucker, G.},
  \bibinfo{author}{Grosse, R.~B.} \& \bibinfo{author}{Norouzi, M.}
\newblock \bibinfo{title}{Don't blame the elbo! a linear vae perspective on
  posterior collapse}.
\newblock \emph{\bibinfo{journal}{Adv. Neural Inf. Process. Syst.}}
  \textbf{\bibinfo{volume}{32}} (\bibinfo{year}{2019}).

\bibitem{xu2020understanding}
\bibinfo{author}{Xu, K.}, \bibinfo{author}{Li, C.}, \bibinfo{author}{Zhu, J.}
  \& \bibinfo{author}{Zhang, B.}
\newblock \bibinfo{title}{Understanding and stabilizing gans’ training
  dynamics using control theory}.
\newblock In \emph{\bibinfo{booktitle}{ICML}}, \bibinfo{pages}{10566--10575}
  (\bibinfo{organization}{PMLR}, \bibinfo{year}{2020}).

\bibitem{becker2022instability}
\bibinfo{author}{Becker, E.}, \bibinfo{author}{Pandit, P.},
  \bibinfo{author}{Rangan, S.} \& \bibinfo{author}{Fletcher, A.~K.}
\newblock \bibinfo{title}{Instability and local minima in gan training with
  kernel discriminators}.
\newblock \emph{\bibinfo{journal}{Adv. Neural Inf. Process. Syst.}}
  \textbf{\bibinfo{volume}{35}}, \bibinfo{pages}{20300--20312}
  (\bibinfo{year}{2022}).

\bibitem{ho2020denoising}
\bibinfo{author}{Ho, J.}, \bibinfo{author}{Jain, A.} \&
  \bibinfo{author}{Abbeel, P.}
\newblock \bibinfo{title}{Denoising diffusion probabilistic models}.
\newblock In \emph{\bibinfo{booktitle}{NeurIPS}}, vol.~\bibinfo{volume}{33},
  \bibinfo{pages}{6840--6851} (\bibinfo{year}{2020}).

\bibitem{wu2022diffusion}
\bibinfo{author}{Wu, L.}, \bibinfo{author}{Gong, C.}, \bibinfo{author}{Liu,
  X.}, \bibinfo{author}{Ye, M.} \& \bibinfo{author}{Liu, Q.}
\newblock \bibinfo{title}{Diffusion-based molecule generation with informative
  prior bridges}.
\newblock \emph{\bibinfo{journal}{Adv. Neural Inf. Process. Syst.}}
  \textbf{\bibinfo{volume}{35}}, \bibinfo{pages}{36533--36545}
  (\bibinfo{year}{2022}).

\bibitem{DBLP:conf/iclr/XieFGBJ22}
\bibinfo{author}{Xie, T.}, \bibinfo{author}{Fu, X.}, \bibinfo{author}{Ganea,
  O.}, \bibinfo{author}{Barzilay, R.} \& \bibinfo{author}{Jaakkola, T.~S.}
\newblock \bibinfo{title}{Crystal diffusion variational autoencoder for
  periodic material generation}.
\newblock In \emph{\bibinfo{booktitle}{ICLR}} (\bibinfo{year}{2022}).

\bibitem{zeni2025generative}
\bibinfo{author}{Zeni, C.} \emph{et~al.}
\newblock \bibinfo{title}{A generative model for inorganic materials design}.
\newblock \emph{\bibinfo{journal}{Nature}} \bibinfo{pages}{1--3}
  (\bibinfo{year}{2025}).

\bibitem{pham2023spectroscopy}
\bibinfo{author}{Kwon, H.} \emph{et~al.}
\newblock \bibinfo{title}{Spectroscopy-guided discovery of three-dimensional
  structures of disordered materials with diffusion models}.
\newblock \emph{\bibinfo{journal}{Mach. Learn.: Sci. Technol.}}
  \textbf{\bibinfo{volume}{5}}, \bibinfo{pages}{045037} (\bibinfo{year}{2024}).

\bibitem{lei2024grand}
\bibinfo{author}{Lei, B.} \emph{et~al.}
\newblock \bibinfo{title}{Grand canonical generative diffusion model for
  crystalline phases and grain boundaries}.
\newblock \emph{\bibinfo{journal}{arXiv preprint arXiv:2408.15601}}
  (\bibinfo{year}{2024}).

\bibitem{horton2025accelerated}
\bibinfo{author}{Horton, M.~K.} \emph{et~al.}
\newblock \bibinfo{title}{Accelerated data-driven materials science with the
  materials project}.
\newblock \emph{\bibinfo{journal}{Nat. Mater.}} \bibinfo{pages}{1--11}
  (\bibinfo{year}{2025}).

\bibitem{liu2022challenges}
\bibinfo{author}{Liu, H.} \emph{et~al.}
\newblock \bibinfo{title}{Challenges and opportunities in atomistic simulations
  of glasses: a review}.
\newblock \emph{\bibinfo{journal}{Comptes Rendus. G{\'e}oscience}}
  \textbf{\bibinfo{volume}{354}}, \bibinfo{pages}{35--77}
  (\bibinfo{year}{2022}).

\bibitem{batatia2023foundation}
\bibinfo{author}{Batatia, I.} \emph{et~al.}
\newblock \bibinfo{title}{A foundation model for atomistic materials
  chemistry}.
\newblock \emph{\bibinfo{journal}{arXiv preprint arXiv:2401.00096}}
  (\bibinfo{year}{2023}).

\bibitem{yang2024mattersim}
\bibinfo{author}{Yang, H.} \emph{et~al.}
\newblock \bibinfo{title}{Mattersim: A deep learning atomistic model across
  elements, temperatures and pressures}.
\newblock \emph{\bibinfo{journal}{arXiv preprint arXiv:2405.04967}}
  (\bibinfo{year}{2024}).

\bibitem{deng2025systematic}
\bibinfo{author}{Deng, B.} \emph{et~al.}
\newblock \bibinfo{title}{Systematic softening in universal machine learning
  interatomic potentials}.
\newblock \emph{\bibinfo{journal}{npj Comput. Mater.}}
  \textbf{\bibinfo{volume}{11}}, \bibinfo{pages}{9} (\bibinfo{year}{2025}).

\bibitem{wang2021inverse}
\bibinfo{author}{Wang, Q.} \& \bibinfo{author}{Zhang, L.}
\newblock \bibinfo{title}{Inverse design of glass structure with deep graph
  neural networks}.
\newblock \emph{\bibinfo{journal}{Nat. Commun.}} \textbf{\bibinfo{volume}{12}},
  \bibinfo{pages}{5359} (\bibinfo{year}{2021}).

\bibitem{merchant2023scaling}
\bibinfo{author}{Merchant, A.} \emph{et~al.}
\newblock \bibinfo{title}{Scaling deep learning for materials discovery}.
\newblock \emph{\bibinfo{journal}{Nature}} \bibinfo{pages}{1--6}
  (\bibinfo{year}{2023}).

\bibitem{li2025conditional}
\bibinfo{author}{Li, H.} \emph{et~al.}
\newblock \bibinfo{title}{Conditional generative modeling for amorphous
  multi-element materials}.
\newblock \emph{\bibinfo{journal}{arXiv preprint arXiv:2503.07043}}
  (\bibinfo{year}{2025}).

\bibitem{zhou2023generative}
\bibinfo{author}{Zhou, Z.}, \bibinfo{author}{Shang, Y.}, \bibinfo{author}{Liu,
  X.} \& \bibinfo{author}{Yang, Y.}
\newblock \bibinfo{title}{A generative deep learning framework for inverse
  design of compositionally complex bulk metallic glasses}.
\newblock \emph{\bibinfo{journal}{npj Comput. Mater.}}
  \textbf{\bibinfo{volume}{9}}, \bibinfo{pages}{15} (\bibinfo{year}{2023}).

\bibitem{comin2019deep}
\bibinfo{author}{Comin, M.} \& \bibinfo{author}{Lewis, L.~J.}
\newblock \bibinfo{title}{Deep-learning approach to the structure of amorphous
  silicon}.
\newblock \emph{\bibinfo{journal}{Phys. Rev. B}}
  \textbf{\bibinfo{volume}{100}}, \bibinfo{pages}{094107}
  (\bibinfo{year}{2019}).

\bibitem{xu2023generative}
\bibinfo{author}{Xu, X.} \& \bibinfo{author}{Hu, J.}
\newblock \bibinfo{title}{A generative adversarial networks (gan) based
  efficient sampling method for inverse design of metallic glasses}.
\newblock \emph{\bibinfo{journal}{J. Non-Cryst. Solids}}
  \textbf{\bibinfo{volume}{613}}, \bibinfo{pages}{122378}
  (\bibinfo{year}{2023}).

\bibitem{yong2024dismai}
\bibinfo{author}{Yong, A. X.~B.}, \bibinfo{author}{Su, T.} \&
  \bibinfo{author}{Ertekin, E.}
\newblock \bibinfo{title}{Dismai-bench: benchmarking and designing generative
  models using disordered materials and interfaces}.
\newblock \emph{\bibinfo{journal}{Digit. Discov.}}
  \textbf{\bibinfo{volume}{3}}, \bibinfo{pages}{1889--1909}
  (\bibinfo{year}{2024}).

\bibitem{chen2025physical}
\bibinfo{author}{Chen, Q.}, \bibinfo{author}{Annamareddy, A.},
  \bibinfo{author}{Li, Y.-F.}, \bibinfo{author}{Morgan, D.} \&
  \bibinfo{author}{Wang, B.}
\newblock \bibinfo{title}{Physical regularized hierarchical generative model
  for metallic glass structural generation and energy prediction}.
\newblock \emph{\bibinfo{journal}{arXiv preprint arXiv:2505.09977}}
  (\bibinfo{year}{2025}).

\bibitem{kilgour2020generating}
\bibinfo{author}{Kilgour, M.}, \bibinfo{author}{Gastellu, N.},
  \bibinfo{author}{Hui, D.~Y.}, \bibinfo{author}{Bengio, Y.} \&
  \bibinfo{author}{Simine, L.}
\newblock \bibinfo{title}{Generating multiscale amorphous molecular structures
  using deep learning: a study in 2d}.
\newblock \emph{\bibinfo{journal}{J. Phys. Chem. Lett.}}
  \textbf{\bibinfo{volume}{11}}, \bibinfo{pages}{8532--8537}
  (\bibinfo{year}{2020}).

\bibitem{yang2025generative}
\bibinfo{author}{Yang, K.} \& \bibinfo{author}{Schwalbe-Koda, D.}
\newblock \bibinfo{title}{A generative diffusion model for amorphous
  materials}.
\newblock \emph{\bibinfo{journal}{arXiv preprint arXiv:2507.05024}}
  (\bibinfo{year}{2025}).

\bibitem{song2020score}
\bibinfo{author}{Song, Y.} \emph{et~al.}
\newblock \bibinfo{title}{Score-based generative modeling through stochastic
  differential equations}.
\newblock \emph{\bibinfo{journal}{arXiv preprint arXiv:2011.13456}}
  (\bibinfo{year}{2020}).

\bibitem{blattmann2023stable}
\bibinfo{author}{Blattmann, A.} \emph{et~al.}
\newblock \bibinfo{title}{Stable video diffusion: Scaling latent video
  diffusion models to large datasets}.
\newblock \emph{\bibinfo{journal}{arXiv preprint arXiv:2311.15127}}
  (\bibinfo{year}{2023}).

\bibitem{DBLP:conf/iclr/PodellELBDMPR24}
\bibinfo{author}{Podell, D.} \emph{et~al.}
\newblock \bibinfo{title}{{SDXL:} improving latent diffusion models for
  high-resolution image synthesis}.
\newblock In \emph{\bibinfo{booktitle}{ICLR}} (\bibinfo{year}{2024}).

\bibitem{DBLP:conf/iclr/LipmanCBNL23}
\bibinfo{author}{Lipman, Y.}, \bibinfo{author}{Chen, R. T.~Q.},
  \bibinfo{author}{Ben{-}Hamu, H.}, \bibinfo{author}{Nickel, M.} \&
  \bibinfo{author}{Le, M.}
\newblock \bibinfo{title}{Flow matching for generative modeling}.
\newblock In \emph{\bibinfo{booktitle}{ICLR}} (\bibinfo{year}{2023}).

\bibitem{DBLP:conf/icml/SatorrasHW21}
\bibinfo{author}{Satorras, V.~G.}, \bibinfo{author}{Hoogeboom, E.} \&
  \bibinfo{author}{Welling, M.}
\newblock \bibinfo{title}{E(n) equivariant graph neural networks}.
\newblock In \emph{\bibinfo{booktitle}{ICML}}, vol. \bibinfo{volume}{139},
  \bibinfo{pages}{9323--9332} (\bibinfo{year}{2021}).

\bibitem{ding2024amorphous}
\bibinfo{author}{Ding, J.}, \bibinfo{author}{Ji, D.}, \bibinfo{author}{Yue, Y.}
  \& \bibinfo{author}{Smedskjaer, M.~M.}
\newblock \bibinfo{title}{Amorphous materials for lithium-ion and
  post-lithium-ion batteries}.
\newblock \emph{\bibinfo{journal}{Small}} \textbf{\bibinfo{volume}{20}},
  \bibinfo{pages}{2304270} (\bibinfo{year}{2024}).

\bibitem{kalnaus2023solid}
\bibinfo{author}{Kalnaus, S.}, \bibinfo{author}{Dudney, N.~J.},
  \bibinfo{author}{Westover, A.~S.}, \bibinfo{author}{Herbert, E.} \&
  \bibinfo{author}{Hackney, S.}
\newblock \bibinfo{title}{Solid-state batteries: The critical role of
  mechanics}.
\newblock \emph{\bibinfo{journal}{Science}} \textbf{\bibinfo{volume}{381}},
  \bibinfo{pages}{eabg5998} (\bibinfo{year}{2023}).

\bibitem{rouxel2007elastic}
\bibinfo{author}{Rouxel, T.}
\newblock \bibinfo{title}{Elastic properties and short-to medium-range order in
  glasses}.
\newblock \emph{\bibinfo{journal}{Journal of the American Ceramic Society}}
  \textbf{\bibinfo{volume}{90}}, \bibinfo{pages}{3019--3039}
  (\bibinfo{year}{2007}).

\bibitem{stillinger1985computer}
\bibinfo{author}{Stillinger, F.~H.} \& \bibinfo{author}{Weber, T.~A.}
\newblock \bibinfo{title}{Computer simulation of local order in condensed
  phases of silicon}.
\newblock \emph{\bibinfo{journal}{Phys. Rev. B}} \textbf{\bibinfo{volume}{31}},
  \bibinfo{pages}{5262} (\bibinfo{year}{1985}).

\bibitem{duane1987hybrid}
\bibinfo{author}{Duane, S.}, \bibinfo{author}{Kennedy, A.~D.},
  \bibinfo{author}{Pendleton, B.~J.} \& \bibinfo{author}{Roweth, D.}
\newblock \bibinfo{title}{Hybrid monte carlo}.
\newblock \emph{\bibinfo{journal}{Phys. Lett. B}}
  \textbf{\bibinfo{volume}{195}}, \bibinfo{pages}{216--222}
  (\bibinfo{year}{1987}).

\bibitem{li2017cooling}
\bibinfo{author}{Li, X.} \emph{et~al.}
\newblock \bibinfo{title}{Cooling rate effects in sodium silicate glasses:
  Bridging the gap between molecular dynamics simulations and experiments}.
\newblock \emph{\bibinfo{journal}{J. Chem. Phys.}}
  \textbf{\bibinfo{volume}{147}} (\bibinfo{year}{2017}).

\bibitem{svenson2014composition}
\bibinfo{author}{Svenson, M.~N.} \emph{et~al.}
\newblock \bibinfo{title}{Composition-structure-property relations of
  compressed borosilicate glasses}.
\newblock \emph{\bibinfo{journal}{Phys. Rev. Appl.}}
  \textbf{\bibinfo{volume}{2}}, \bibinfo{pages}{024006} (\bibinfo{year}{2014}).

\bibitem{guttman1990ring}
\bibinfo{author}{Guttman, L.}
\newblock \bibinfo{title}{Ring structure of the crystalline and amorphous forms
  of silicon dioxide}.
\newblock \emph{\bibinfo{journal}{J. Non-Cryst. Solids}}
  \textbf{\bibinfo{volume}{116}}, \bibinfo{pages}{145--147}
  (\bibinfo{year}{1990}).

\bibitem{vanbeest1990force}
\bibinfo{author}{van Beest, B. W.~H.}, \bibinfo{author}{Kramer, G.~J.} \&
  \bibinfo{author}{van Santen, R.~A.}
\newblock \bibinfo{title}{Force fields for silicas and aluminophosphates based
  on ab initio calculations}.
\newblock \emph{\bibinfo{journal}{Phys. Rev. Lett.}}
  \textbf{\bibinfo{volume}{64}}, \bibinfo{pages}{1955--1958}
  (\bibinfo{year}{1990}).

\bibitem{munetoh2007interatomic}
\bibinfo{author}{Munetoh, S.}, \bibinfo{author}{Motooka, T.},
  \bibinfo{author}{Moriguchi, K.} \& \bibinfo{author}{Shintani, A.}
\newblock \bibinfo{title}{Interatomic potential for si--o systems using tersoff
  parameterization}.
\newblock \emph{\bibinfo{journal}{Comput. Mater. Sci.}}
  \textbf{\bibinfo{volume}{39}}, \bibinfo{pages}{334--339}
  (\bibinfo{year}{2007}).

\bibitem{raza2015computer}
\bibinfo{author}{Raza, Z.}, \bibinfo{author}{Alling, B.} \&
  \bibinfo{author}{Abrikosov, I.~A.}
\newblock \bibinfo{title}{Computer simulations of glasses: the potential energy
  landscape}.
\newblock \emph{\bibinfo{journal}{J. Phys.: Condens. Matter}}
  \textbf{\bibinfo{volume}{27}}, \bibinfo{pages}{293201}
  (\bibinfo{year}{2015}).

\bibitem{niblett2017pathways}
\bibinfo{author}{Niblett, S.}, \bibinfo{author}{Biedermann, M.},
  \bibinfo{author}{Wales, D.} \& \bibinfo{author}{De~Souza, V.}
\newblock \bibinfo{title}{Pathways for diffusion in the potential energy
  landscape of the network glass former {SiO2}}.
\newblock \emph{\bibinfo{journal}{J. Chem. Phys.}}
  \textbf{\bibinfo{volume}{147}} (\bibinfo{year}{2017}).

\bibitem{biroli2024dynamical}
\bibinfo{author}{Biroli, G.}, \bibinfo{author}{Bonnaire, T.},
  \bibinfo{author}{De~Bortoli, V.} \& \bibinfo{author}{M{\'e}zard, M.}
\newblock \bibinfo{title}{Dynamical regimes of diffusion models}.
\newblock \emph{\bibinfo{journal}{Nat. Commun.}} \textbf{\bibinfo{volume}{15}},
  \bibinfo{pages}{9957} (\bibinfo{year}{2024}).

\bibitem{DBLP:conf/mm/ZengYZCCZY24}
\bibinfo{author}{Zeng, W.} \emph{et~al.}
\newblock \bibinfo{title}{Infusion: Preventing customized text-to-image
  diffusion from overfitting}.
\newblock In \emph{\bibinfo{booktitle}{{ACM} Multimedia}},
  \bibinfo{pages}{3568--3577} (\bibinfo{year}{2024}).

\bibitem{ghio2024sampling}
\bibinfo{author}{Ghio, D.}, \bibinfo{author}{Dandi, Y.},
  \bibinfo{author}{Krzakala, F.} \& \bibinfo{author}{Zdeborov{\'a}, L.}
\newblock \bibinfo{title}{Sampling with flows, diffusion, and autoregressive
  neural networks from a spin-glass perspective}.
\newblock \emph{\bibinfo{journal}{Proc. Natl. Acad. Sci. U.S.A.}}
  \textbf{\bibinfo{volume}{121}}, \bibinfo{pages}{e2311810121}
  (\bibinfo{year}{2024}).

\bibitem{bihani2023stridernet}
\bibinfo{author}{Bihani, V.}, \bibinfo{author}{Manchanda, S.},
  \bibinfo{author}{Sastry, S.}, \bibinfo{author}{Ranu, S.} \&
  \bibinfo{author}{Krishnan, N.~A.}
\newblock \bibinfo{title}{Stridernet: A graph reinforcement learning approach
  to optimize atomic structures on rough energy landscapes}.
\newblock In \emph{\bibinfo{booktitle}{ICML}}, \bibinfo{pages}{2431--2451}
  (\bibinfo{organization}{PMLR}, \bibinfo{year}{2023}).

\bibitem{mauro2009forbidden}
\bibinfo{author}{Mauro, J.~C.} \& \bibinfo{author}{Loucks, R.~J.}
\newblock \bibinfo{title}{Forbidden glasses and the failure of fictive
  temperature}.
\newblock \emph{\bibinfo{journal}{J. Non-Cryst. Solids}}
  \textbf{\bibinfo{volume}{355}}, \bibinfo{pages}{676--680}
  (\bibinfo{year}{2009}).

\bibitem{DBLP:conf/cvpr/JacobKCZTHAK18}
\bibinfo{author}{Jacob, B.} \emph{et~al.}
\newblock \bibinfo{title}{Quantization and training of neural networks for
  efficient integer-arithmetic-only inference}.
\newblock In \emph{\bibinfo{booktitle}{CVPR}}, \bibinfo{pages}{2704--2713}
  (\bibinfo{year}{2018}).

\bibitem{utkarsh2025physics}
\bibinfo{author}{Utkarsh, U.}, \bibinfo{author}{Cai, P.},
  \bibinfo{author}{Edelman, A.}, \bibinfo{author}{Gomez-Bombarelli, R.} \&
  \bibinfo{author}{Rackauckas, C.~V.}
\newblock \bibinfo{title}{Physics-constrained flow matching: Sampling
  generative models with hard constraints}.
\newblock In \emph{\bibinfo{booktitle}{NeurIPS}} (\bibinfo{year}{2025}).

\bibitem{jain2013commentary}
\bibinfo{author}{Jain, A.} \emph{et~al.}
\newblock \bibinfo{title}{Commentary: The materials project: A materials genome
  approach to accelerating materials innovation}.
\newblock \emph{\bibinfo{journal}{APL Mater.}} \textbf{\bibinfo{volume}{1}}
  (\bibinfo{year}{2013}).

\bibitem{saal2013materials}
\bibinfo{author}{Saal, J.~E.}, \bibinfo{author}{Kirklin, S.},
  \bibinfo{author}{Aykol, M.}, \bibinfo{author}{Meredig, B.} \&
  \bibinfo{author}{Wolverton, C.}
\newblock \bibinfo{title}{Materials design and discovery with high-throughput
  density functional theory: the open quantum materials database (oqmd)}.
\newblock \emph{\bibinfo{journal}{JOM}} \textbf{\bibinfo{volume}{65}},
  \bibinfo{pages}{1501--1509} (\bibinfo{year}{2013}).

\bibitem{barroso2024open}
\bibinfo{author}{Barroso-Luque, L.} \emph{et~al.}
\newblock \bibinfo{title}{Open materials 2024 (omat24) inorganic materials
  dataset and models}.
\newblock \emph{\bibinfo{journal}{arXiv preprint arXiv:2410.12771}}
  (\bibinfo{year}{2024}).

\bibitem{zheng2024ab}
\bibinfo{author}{Zheng, H.} \emph{et~al.}
\newblock \bibinfo{title}{The ab initio non-crystalline structure database:
  empowering machine learning to decode diffusivity}.
\newblock \emph{\bibinfo{journal}{npj Comput. Mater.}}
  \textbf{\bibinfo{volume}{10}}, \bibinfo{pages}{295} (\bibinfo{year}{2024}).

\bibitem{hung2024universal}
\bibinfo{author}{Hung, N.~T.}, \bibinfo{author}{Okabe, R.},
  \bibinfo{author}{Chotrattanapituk, A.} \& \bibinfo{author}{Li, M.}
\newblock \bibinfo{title}{Universal ensemble-embedding graph neural network for
  direct prediction of optical spectra from crystal structures}.
\newblock \emph{\bibinfo{journal}{Adv. Mater.}} \textbf{\bibinfo{volume}{36}},
  \bibinfo{pages}{2409175} (\bibinfo{year}{2024}).

\bibitem{lammps}
\bibinfo{author}{Thompson, A.~P.} \emph{et~al.}
\newblock \bibinfo{title}{{LAMMPS} - a flexible simulation tool for
  particle-based materials modeling at the atomic, meso, and continuum scales}.
\newblock \emph{\bibinfo{journal}{Comp. Phys. Comm.}}
  \textbf{\bibinfo{volume}{271}}, \bibinfo{pages}{108171}
  (\bibinfo{year}{2022}).

\bibitem{larsen2017atomic}
\bibinfo{author}{Larsen, A.~H.} \emph{et~al.}
\newblock \bibinfo{title}{The atomic simulation environment—a python library
  for working with atoms}.
\newblock \emph{\bibinfo{journal}{J. Phys.: Condens. Matter}}
  \textbf{\bibinfo{volume}{29}}, \bibinfo{pages}{273002}
  (\bibinfo{year}{2017}).

\bibitem{gubler2023efficient}
\bibinfo{author}{Gubler, M.}, \bibinfo{author}{Krummenacher, M.},
  \bibinfo{author}{Huber, H.} \& \bibinfo{author}{Goedecker, S.}
\newblock \bibinfo{title}{Efficient variable cell shape geometry optimization}.
\newblock \emph{\bibinfo{journal}{J. Comput. Phys. X}}
  \textbf{\bibinfo{volume}{17}}, \bibinfo{pages}{100131}
  (\bibinfo{year}{2023}).

\bibitem{bertani2021improved}
\bibinfo{author}{Bertani, M.}, \bibinfo{author}{Menziani, M.~C.} \&
  \bibinfo{author}{Pedone, A.}
\newblock \bibinfo{title}{Improved empirical force field for multicomponent
  oxide glasses and crystals}.
\newblock \emph{\bibinfo{journal}{Phys. Rev. Mater.}}
  \textbf{\bibinfo{volume}{5}}, \bibinfo{pages}{045602} (\bibinfo{year}{2021}).

\bibitem{song2019generative}
\bibinfo{author}{Song, Y.} \& \bibinfo{author}{Ermon, S.}
\newblock \bibinfo{title}{Generative modeling by estimating gradients of the
  data distribution}.
\newblock In \emph{\bibinfo{booktitle}{NeurIPS}}, vol.~\bibinfo{volume}{32},
  \bibinfo{pages}{11895--11907} (\bibinfo{year}{2019}).

\bibitem{levy2023using}
\bibinfo{author}{Levy, D.}, \bibinfo{author}{Kaba, S.-O.},
  \bibinfo{author}{Gonzales, C.}, \bibinfo{author}{Miret, S.} \&
  \bibinfo{author}{Ravanbakhsh, S.}
\newblock \bibinfo{title}{Using multiple vector channels improves e
  (n)-equivariant graph neural networks}.
\newblock \emph{\bibinfo{journal}{arXiv preprint arXiv:2309.03139}}
  (\bibinfo{year}{2023}).

\bibitem{DBLP:conf/nips/VaswaniSPUJGKP17}
\bibinfo{author}{Vaswani, A.} \emph{et~al.}
\newblock \bibinfo{title}{Attention is all you need}.
\newblock In \emph{\bibinfo{booktitle}{NeurIPS}}, \bibinfo{pages}{5998--6008}
  (\bibinfo{year}{2017}).

\bibitem{ho2022classifier}
\bibinfo{author}{Ho, J.} \& \bibinfo{author}{Salimans, T.}
\newblock \bibinfo{title}{Classifier-free diffusion guidance}.
\newblock \emph{\bibinfo{journal}{arXiv preprint arXiv:2207.12598}}
  (\bibinfo{year}{2022}).

\bibitem{sadat2024no}
\bibinfo{author}{Sadat, S.}, \bibinfo{author}{Kansy, M.},
  \bibinfo{author}{Hilliges, O.} \& \bibinfo{author}{Weber, R.~M.}
\newblock \bibinfo{title}{No training, no problem: Rethinking classifier-free
  guidance for diffusion models}.
\newblock \emph{\bibinfo{journal}{arXiv preprint arXiv:2407.02687}}
  (\bibinfo{year}{2024}).

\bibitem{DBLP:journals/corr/KingmaB14}
\bibinfo{author}{Kingma, D.~P.} \& \bibinfo{author}{Ba, J.}
\newblock \bibinfo{title}{Adam: {A} method for stochastic optimization}.
\newblock In \bibinfo{editor}{Bengio, Y.} \& \bibinfo{editor}{LeCun, Y.} (eds.)
  \emph{\bibinfo{booktitle}{ICLR}} (\bibinfo{year}{2015}).

\bibitem{salimans2021should}
\bibinfo{author}{Salimans, T.} \& \bibinfo{author}{Ho, J.}
\newblock \bibinfo{title}{Should ebms model the energy or the score?}
\newblock In \emph{\bibinfo{booktitle}{Energy Based Models Workshop-ICLR 2021}}
  (\bibinfo{year}{2021}).

\end{thebibliography}


\begin{thebibliography}{7}%
\makeatletter
\providecommand \@ifxundefined [1]{%
 \@ifx{#1\undefined}
}%
\providecommand \@ifnum [1]{%
 \ifnum #1\expandafter \@firstoftwo
 \else \expandafter \@secondoftwo
 \fi
}%
\providecommand \@ifx [1]{%
 \ifx #1\expandafter \@firstoftwo
 \else \expandafter \@secondoftwo
 \fi
}%
\providecommand \natexlab [1]{#1}%
\providecommand \enquote  [1]{``#1''}%
\providecommand \bibnamefont  [1]{#1}%
\providecommand \bibfnamefont [1]{#1}%
\providecommand \citenamefont [1]{#1}%
\providecommand \href@noop [0]{\@secondoftwo}%
\providecommand \href [0]{\begingroup \@sanitize@url \@href}%
\providecommand \@href[1]{\@@startlink{#1}\@@href}%
\providecommand \@@href[1]{\endgroup#1\@@endlink}%
\providecommand \@sanitize@url [0]{\catcode `\\12\catcode `\$12\catcode
  `\&12\catcode `\#12\catcode `\^12\catcode `\_12\catcode `\%12\relax}%
\providecommand \@@startlink[1]{}%
\providecommand \@@endlink[0]{}%
\providecommand \url  [0]{\begingroup\@sanitize@url \@url }%
\providecommand \@url [1]{\endgroup\@href {#1}{\urlprefix }}%
\providecommand \urlprefix  [0]{URL }%
\providecommand \Eprint [0]{\href }%
\providecommand \doibase [0]{http://dx.doi.org/}%
\providecommand \selectlanguage [0]{\@gobble}%
\providecommand \bibinfo  [0]{\@secondoftwo}%
\providecommand \bibfield  [0]{\@secondoftwo}%
\providecommand \translation [1]{[#1]}%
\providecommand \BibitemOpen [0]{}%
\providecommand \bibitemStop [0]{}%
\providecommand \bibitemNoStop [0]{.\EOS\space}%
\providecommand \EOS [0]{\spacefactor3000\relax}%
\providecommand \BibitemShut  [1]{\csname bibitem#1\endcsname}%
\let\auto@bib@innerbib\@empty
\bibitem [{\citenamefont {Thompson}\ \emph {et~al.}(2022)\citenamefont
  {Thompson}, \citenamefont {Aktulga}, \citenamefont {Berger}, \citenamefont
  {Bolintineanu}, \citenamefont {Brown}, \citenamefont {Crozier}, \citenamefont
  {in~'t Veld}, \citenamefont {Kohlmeyer}, \citenamefont {Moore}, \citenamefont
  {Nguyen}, \citenamefont {Shan}, \citenamefont {Stevens}, \citenamefont
  {Tranchida}, \citenamefont {Trott},\ and\ \citenamefont {Plimpton}}]{lammps}%
  \BibitemOpen
  \bibfield  {author} {\bibinfo {author} {\bibfnamefont {A.~P.}\ \bibnamefont
  {Thompson}}, \bibinfo {author} {\bibfnamefont {H.~M.}\ \bibnamefont
  {Aktulga}}, \bibinfo {author} {\bibfnamefont {R.}~\bibnamefont {Berger}},
  \bibinfo {author} {\bibfnamefont {D.~S.}\ \bibnamefont {Bolintineanu}},
  \bibinfo {author} {\bibfnamefont {W.~M.}\ \bibnamefont {Brown}}, \bibinfo
  {author} {\bibfnamefont {P.~S.}\ \bibnamefont {Crozier}}, \bibinfo {author}
  {\bibfnamefont {P.~J.}\ \bibnamefont {in~'t Veld}}, \bibinfo {author}
  {\bibfnamefont {A.}~\bibnamefont {Kohlmeyer}}, \bibinfo {author}
  {\bibfnamefont {S.~G.}\ \bibnamefont {Moore}}, \bibinfo {author}
  {\bibfnamefont {T.~D.}\ \bibnamefont {Nguyen}}, \bibinfo {author}
  {\bibfnamefont {R.}~\bibnamefont {Shan}}, \bibinfo {author} {\bibfnamefont
  {M.~J.}\ \bibnamefont {Stevens}}, \bibinfo {author} {\bibfnamefont
  {J.}~\bibnamefont {Tranchida}}, \bibinfo {author} {\bibfnamefont
  {C.}~\bibnamefont {Trott}}, \ and\ \bibinfo {author} {\bibfnamefont {S.~J.}\
  \bibnamefont {Plimpton}},\ }\href {\doibase 10.1016/j.cpc.2021.108171}
  {\bibfield  {journal} {\bibinfo  {journal} {Comp. Phys. Comm.}\ }\textbf
  {\bibinfo {volume} {271}},\ \bibinfo {pages} {108171} (\bibinfo {year}
  {2022})}\BibitemShut {NoStop}%
\bibitem [{\citenamefont {Bertani}\ \emph {et~al.}(2021)\citenamefont
  {Bertani}, \citenamefont {Menziani},\ and\ \citenamefont
  {Pedone}}]{bertani2021improved}%
  \BibitemOpen
  \bibfield  {author} {\bibinfo {author} {\bibfnamefont {M.}~\bibnamefont
  {Bertani}}, \bibinfo {author} {\bibfnamefont {M.~C.}\ \bibnamefont
  {Menziani}}, \ and\ \bibinfo {author} {\bibfnamefont {A.}~\bibnamefont
  {Pedone}},\ }\href@noop {} {\bibfield  {journal} {\bibinfo  {journal} {Phys.
  Rev. Mater.}\ }\textbf {\bibinfo {volume} {5}},\ \bibinfo {pages} {045602}
  (\bibinfo {year} {2021})}\BibitemShut {NoStop}%
\bibitem [{\citenamefont {Munetoh}\ \emph {et~al.}(2007)\citenamefont
  {Munetoh}, \citenamefont {Motooka}, \citenamefont {Moriguchi},\ and\
  \citenamefont {Shintani}}]{munetoh2007interatomic}%
  \BibitemOpen
  \bibfield  {author} {\bibinfo {author} {\bibfnamefont {S.}~\bibnamefont
  {Munetoh}}, \bibinfo {author} {\bibfnamefont {T.}~\bibnamefont {Motooka}},
  \bibinfo {author} {\bibfnamefont {K.}~\bibnamefont {Moriguchi}}, \ and\
  \bibinfo {author} {\bibfnamefont {A.}~\bibnamefont {Shintani}},\ }\href@noop
  {} {\bibfield  {journal} {\bibinfo  {journal} {Comput. Mater. Sci.}\ }\textbf
  {\bibinfo {volume} {39}},\ \bibinfo {pages} {334} (\bibinfo {year}
  {2007})}\BibitemShut {NoStop}%
\bibitem [{\citenamefont {Stillinger}\ and\ \citenamefont
  {Weber}(1985)}]{stillinger1985computer}%
  \BibitemOpen
  \bibfield  {author} {\bibinfo {author} {\bibfnamefont {F.~H.}\ \bibnamefont
  {Stillinger}}\ and\ \bibinfo {author} {\bibfnamefont {T.~A.}\ \bibnamefont
  {Weber}},\ }\href@noop {} {\bibfield  {journal} {\bibinfo  {journal} {Phys.
  Rev. B}\ }\textbf {\bibinfo {volume} {31}},\ \bibinfo {pages} {5262}
  (\bibinfo {year} {1985})}\BibitemShut {NoStop}%
\bibitem [{\citenamefont {Guttman}(1990)}]{guttman1990ring}%
  \BibitemOpen
  \bibfield  {author} {\bibinfo {author} {\bibfnamefont {L.}~\bibnamefont
  {Guttman}},\ }\href@noop {} {\bibfield  {journal} {\bibinfo  {journal} {J.
  Non-Cryst. Solids}\ }\textbf {\bibinfo {volume} {116}},\ \bibinfo {pages}
  {145} (\bibinfo {year} {1990})}\BibitemShut {NoStop}%
\bibitem [{\citenamefont {Gersten}\ and\ \citenamefont
  {Smith}(2001)}]{gersten2001physics}%
  \BibitemOpen
  \bibfield  {author} {\bibinfo {author} {\bibfnamefont {J.~I.}\ \bibnamefont
  {Gersten}}\ and\ \bibinfo {author} {\bibfnamefont {F.~W.}\ \bibnamefont
  {Smith}},\ }\href@noop {} {\emph {\bibinfo {title} {The Physics and Chemistry
  of Materials}}}\ (\bibinfo  {publisher} {John Wiley \& Sons, Inc.},\ \bibinfo
  {address} {New York, Chichester, Weinheim, Brisbane, Singapore, Toronto},\
  \bibinfo {year} {2001})\ \bibinfo {note} {a Wiley-Interscience publication.
  Joel I. Gersten and Frederick W. Smith, The City College of the City
  University of New York}\BibitemShut {NoStop}%
\bibitem [{\citenamefont {van Beest}\ \emph {et~al.}(1990)\citenamefont {van
  Beest}, \citenamefont {Kramer},\ and\ \citenamefont {van
  Santen}}]{vanbeest1990force}%
  \BibitemOpen
  \bibfield  {author} {\bibinfo {author} {\bibfnamefont {B.~W.~H.}\
  \bibnamefont {van Beest}}, \bibinfo {author} {\bibfnamefont {G.~J.}\
  \bibnamefont {Kramer}}, \ and\ \bibinfo {author} {\bibfnamefont {R.~A.}\
  \bibnamefont {van Santen}},\ }\href@noop {} {\bibfield  {journal} {\bibinfo
  {journal} {Phys. Rev. Lett.}\ }\textbf {\bibinfo {volume} {64}},\ \bibinfo
  {pages} {1955} (\bibinfo {year} {1990})}\BibitemShut {NoStop}%
\end{thebibliography}%

\end{document}


\maketitle

\section{Data Set Generation}
\subsection{Multi Element Glass Data Set}
We created the multi element glass (MEG) data set to test our model's performance on data including a larger variety of elements. 
As also mentioned in the Methods section, the data set consists of 9,027 samples, containing 11 different elements. 
Initial structures were generated from varying compositions of the glass formers \ch{SiO2} and \ch{P2O5}, and the modifiers \ch{Al2O3} \ch{Li2O}, \ch{BeO}, \ch{K2O}, \ch{CaO}, \ch{TiO2}, \ch{BaO} and \ch{ZnO}.

Structural samples and corresponding properties of the MEG data set were obtained using the workflow described below.
Simulations were performed using LAMMPS~\cite{lammps} software and the Bertain--Menziani--Pedone (BMP)-shrm potential~\cite{bertani2021improved}.
\begin{enumerate}
  \item Elemental compositions were generated to include different ratios of the three glass formers, up to four different modifiers with total concentration of $40\,\%$ relative to the glass former concentration. 
  \item Initial structures of the generated compositions, containing roughly 800 atoms, were created by randomly placing the atoms in a simulation cell with a volume $V = 3 \sum_i \frac{4}{3} \pi r_i^3$, with $r_i$ being the covalent radius of atom $i$. The atoms positions were then adjusted to ensure that no two atoms were closer than the sum of their respective covalent radii. Finally, a local geometry optimization was performed to optimize the atomic positions and cell dimensions. 
  \item To ensure proper melting while avoiding evaporation, an initial temperature for the melt-quench procedure needed to be determined for each composition. For this task, the initial cells were doubled in size along one dimension to form a vacuum region. A short molecular dynamics (MD) simulation was then performed in the NVT ensemble during which the temperature was increased up to 8000\,K for a duration of 100\,ps. The evaporation temperature $T_\text{evap}$ was then identified at the onset of pressure increase during the dynamics simulation. 
  \item Structural samples were obtained from a melt-quench simulation in the NPT ensemble, initialized at $T_\text{init}=\frac{3}{4}T_\text{evap}$. The samples were first melted for 400\,ps, then quenched to 300\,K at 5\,K/ps and finally equilibrated for 300\,ps. Out of 9,240 compositions, 213 samples were identified that did not melt properly during the initial phase of the simulation and were thus excluded from the final data set.
  \item Melt-quenched samples were then equilibrated at 50\,K for 100\,ps and subsequently heated to 500\,K over 500\,ps to extract heat capacities and thermal expansion coefficients. 
  \item Samples were also relaxed to compute the elasticity tensor using finite differences of the stress tensor. 
\end{enumerate}

\begin{figure}[!ht]
  \centering
  \begin{subfigure}[b]{0.32\linewidth}
    \includegraphics[width=\textwidth]{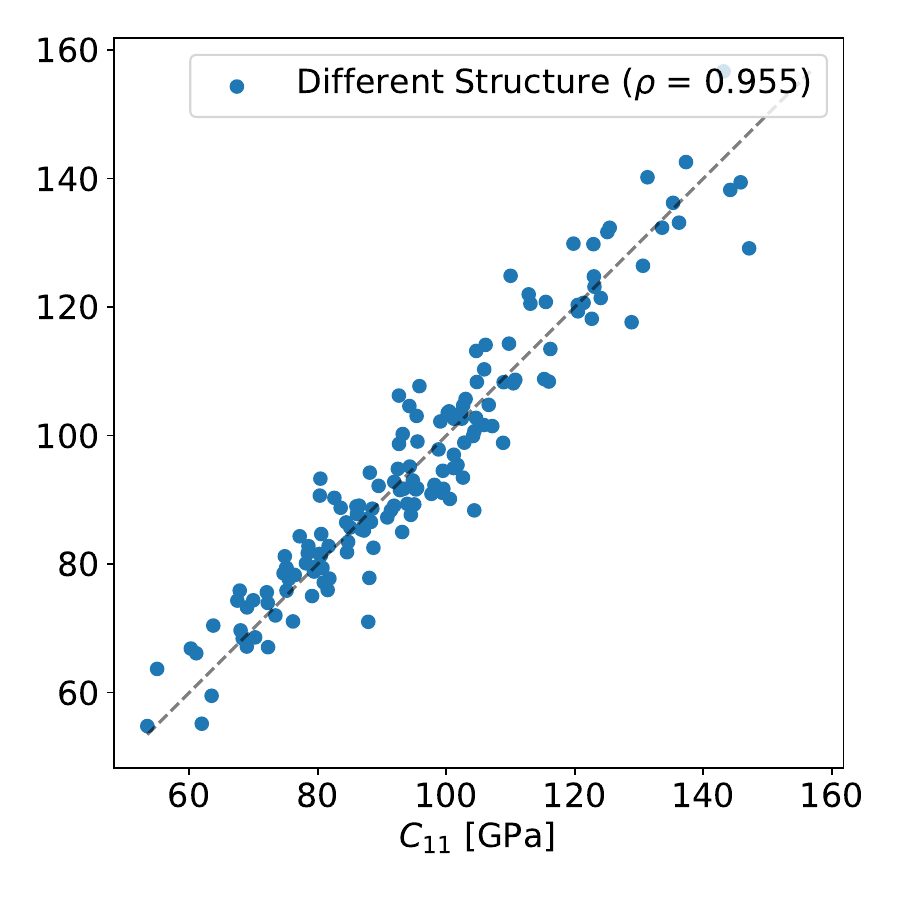}
    \caption{}
  \end{subfigure}
  \begin{subfigure}[b]{0.32\linewidth}
    \includegraphics[width=\textwidth]{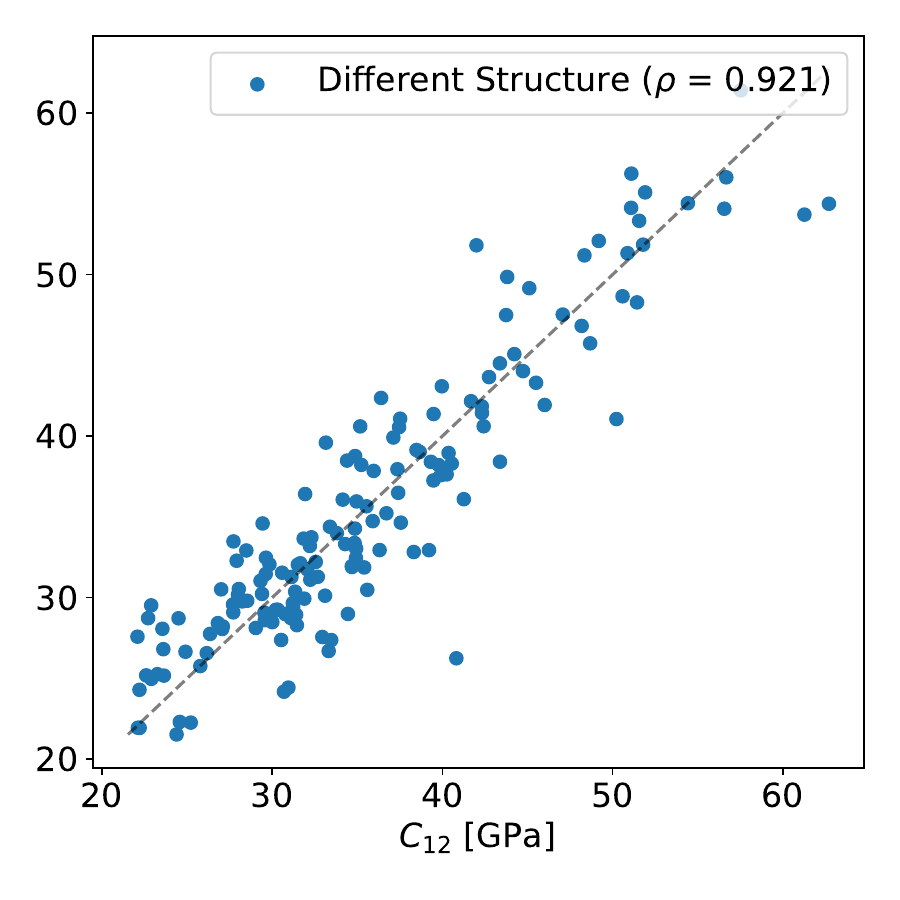}
    \caption{}
  \end{subfigure}
  \begin{subfigure}[b]{0.32\textwidth}
    \includegraphics[width=\textwidth]{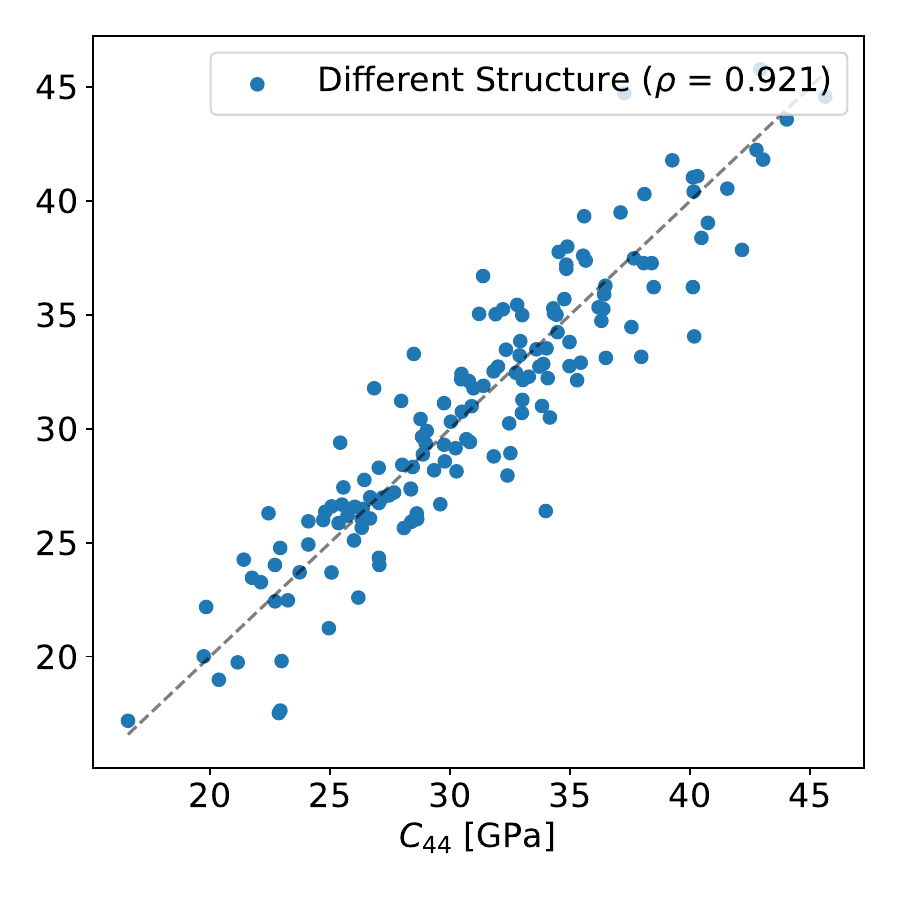}
    \caption{}
  \end{subfigure} 
  \newline
  \begin{subfigure}[b]{0.32\textwidth}
    \includegraphics[width=\textwidth]{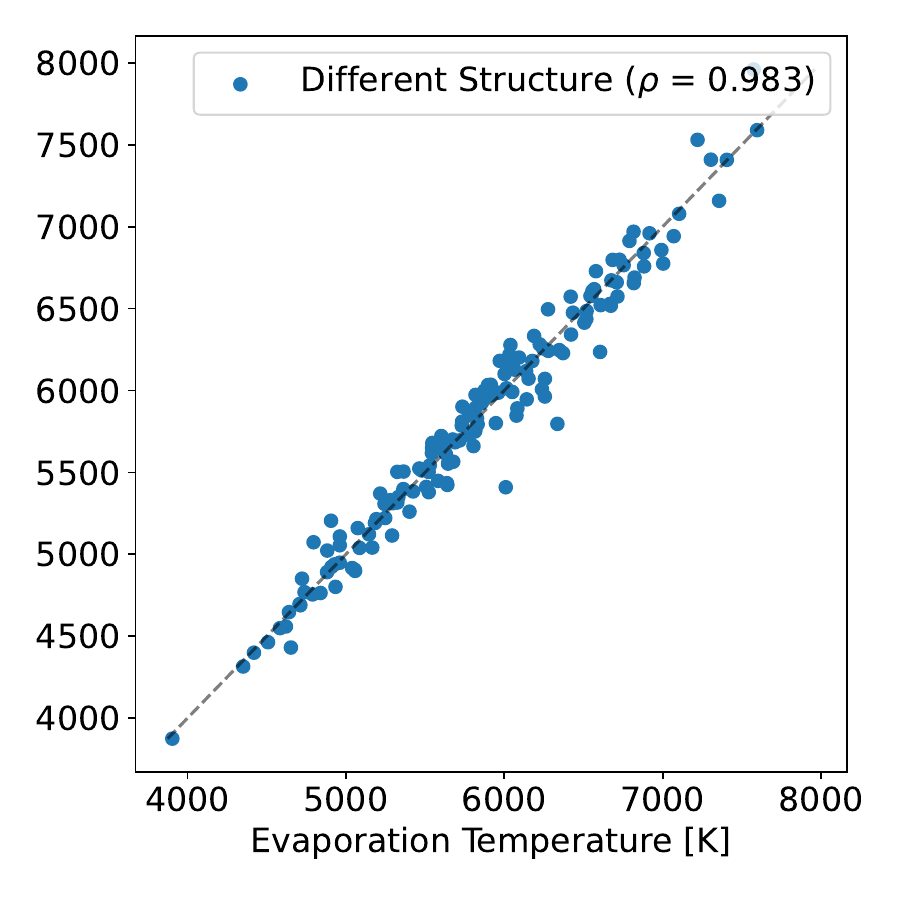}
    \caption{}
  \end{subfigure}
  \begin{subfigure}[b]{0.32\textwidth}
    \includegraphics[width=\textwidth]{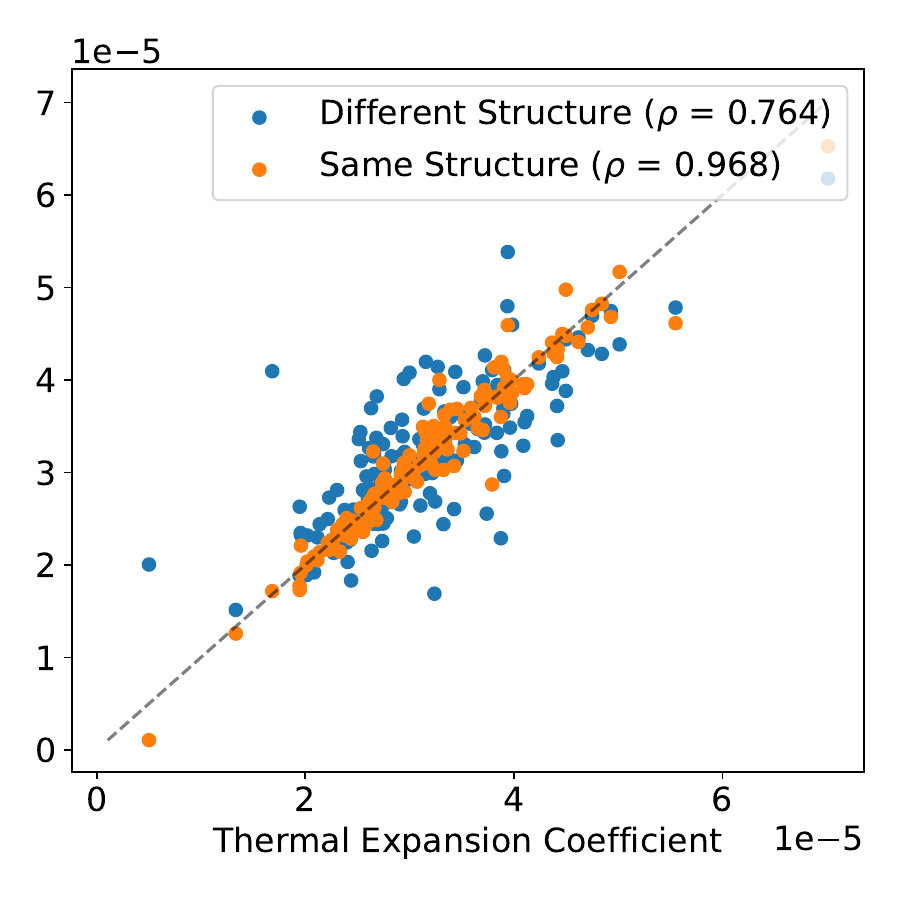}
    \caption{}
  \end{subfigure}
  \begin{subfigure}[b]{0.32\textwidth}
    \includegraphics[width=\textwidth]{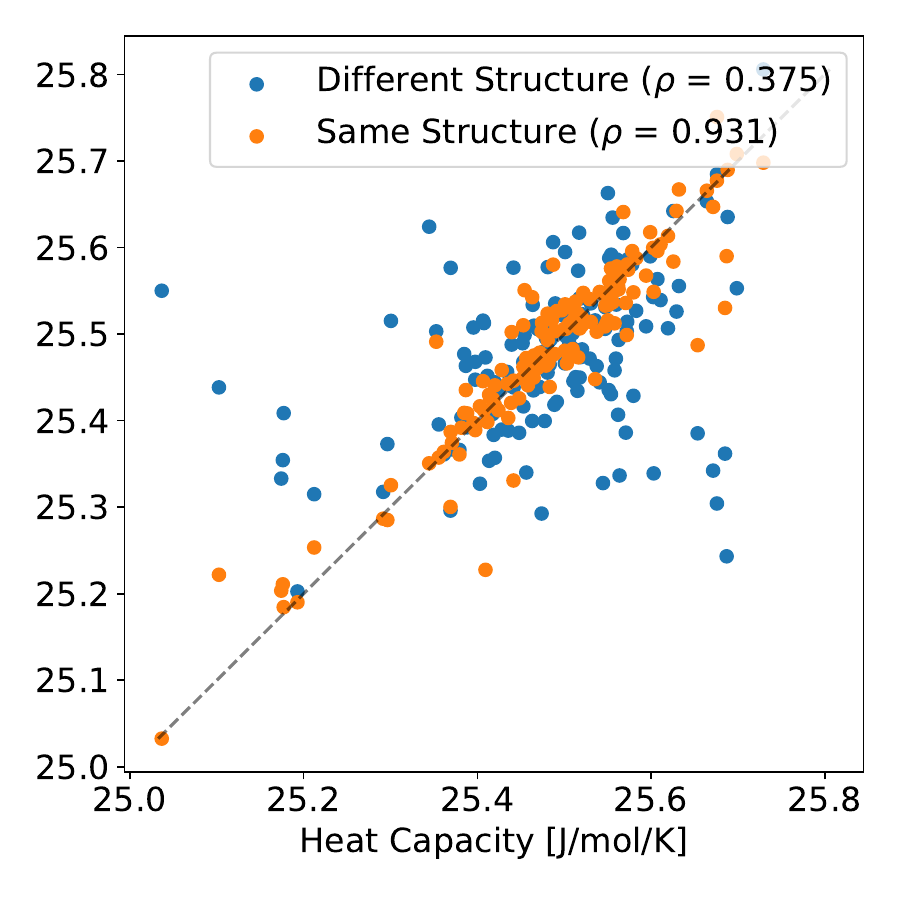}
    \caption{}
  \end{subfigure}
  \caption{Parity plots of the independent elastic constants $C_{11}$ (a), C$_{12}$ (b) and $C_{44}$ (c), the evaporation temperature $T_\text{evap}$ (d), the thermal expansion coefficient (e), and the molar heat capacity (f) for a random subset of samples of the MEG data set. The properties were computed from two independent runs of the workflow, initialized with the same compositions. Properties obtained from the heating simulation of the workflow were computed for a third time by re-running the heating simulation initialized with the same structural sample but using a different random seed. Pearson correlation coefficients $\rho$ are shown in the figure legends.
  }
  \label{fig:bmp_corr}
\end{figure}

Due to the finite number of atoms, some amount of uncertainty in the computed properties is expected. 
To assess these, we performed two independent runs of the workflow for a random subset of compositions, resulting in two sets of structural samples for which properties were calculated. 
For one set of samples, the final heating simulation of the workflow was then repeated with the same initial structure but using a different random seed to assess the variability introduced by the heating simulation.
Correlation plots of all properties and corresponding Pearson correlation coefficients are shown in Fig.~\ref{fig:bmp_corr}.
The elastic constants, which were deterministically computed from the structural samples, correlate well between the independent runs of the workflow, indicating a strong dependence on the composition. 
Similarly, the evaporation temperature shows a strong correlation between the independent runs.
The thermal expansion coefficient and the molar heat capacity show a weaker correlation between the independent runs but a good agreement between the two heating simulations. 
Overall, this indicates that the simulation workflows to obtain glass properties work reliably, with variability being attributed to differences between the structures of the samples.

Heat capacities were obtained as the gradient of a linear fit to the total energy versus temperature of the heating simulation in step 5 of the workflow.
Similarly, the thermal expansion coefficient ($\alpha$)  at room temperature was obtained from a linear fit $V(T)$ to the volume versus temperature of the heating simulation and calculated as
\begin{equation}
  \alpha = \frac{1}{V(T)} \left. \frac{\partial V(T)}{\partial T} \right|_{T=300\,\text{K}}.
\end{equation}
Elastic constants were obtained as described in Section~\ref{sec:mechanical_properties}.

\subsection{Amorphous Silica Data Set}
We developed the amorphous silica data set to investigate our model's performance acting exclusively on the structure of the sample without changing the material's composition. 
As such, all $6,000$ samples share the composition of pure silica, \ch{SiO2}.
To maximize the variation of properties between the samples generated with the same simulation workflow, relatively small unit cells were chosen with the number of atoms uniformly selected in the range of 80 to 250. 
Atoms were initially placed in a unit cell with a volume $V = 4 \sum_i \frac{4}{3} \pi r_i^3$, with $r_i$ being the covalent radius of the $i$-th atom, avoiding unphysical overlap between neighboring atoms.
A local structure relaxation was performed on the initial configuration followed by an MD simulation in the NPT ensemble at 3500\,K for 2000\,ps.
To limit the effects of relaxation, which we observed for our other data sets, we used an instantaneous quenching procedure by performing a local structure optimization and a subsequent equilibration at 300\,K for 10\,ps.
Elastic constants were computed from the relaxed final structures using finite differences of the stress tensor as described in Section~\ref{sec:mechanical_properties}.
All simulations were performed using LAMMPS~\cite{lammps} software and the Tersoff potential parameterized by Munetoh \textit{et al.}~\cite{munetoh2007interatomic}.

\subsection{Amorphous Silicon Data Sets}
We created three data sets (\textit{melt}, \textit{quench} and \textit{anneal}) of amorphous silicon to study the effects of relaxation on the generation performance. 
All three data sets were created using LAMMPS~\cite{lammps} software with the Stillinger--Weber potential~\cite{stillinger1985computer} and consisted of $10\,000$ samples each.
The simulations were initialized with a unit cell containing 256 atoms of crystalline silicon, but different thermal schedules were applied to obtain the final samples. 
All MD simulations were performed in the NPT ensemble at zero pressure. 

The \textit{melt} data set was generated by heating the crystalline silicon from 2500\,K to 3000\,K over 200\,ps, equilibrating the melt for 300\,ps, and then cooling it down again to 2500\,K at a rate of $10^{12}$\,K/s. 
The final samples were taken after equilibrating for another 300\,ps at 2500\,K.

The \textit{anneal} and \textit{quench} data sets were both initialized at 300\,K, heated to 2500\,K over 200\,ps, and equilibrated for 300\,ps.
Samples for the \textit{anneal} data set were then cooled down at a rate of $10^{12}$\,K/s to 300\,K and equilibrated for another 300\,ps, while the cooling step was omitted for samples for the \textit{quench} data set.
The structures of the \textit{anneal} data set were thus allowed to relax during the cooling period, while the \textit{quench} samples were obtained from an almost instantaneous quenching procedure.
However, a small amount of relaxation is still expected during the time period taken by the thermostat to adjust the temperature of the system to the lower target value.

Ring sizes were computed according to the definition of \citet{guttman1990ring} and reported as the number of \ch{Si} atoms in the ring.
Atoms where considered bonded if the distance between them was below the sum of their covalent radii ($r_\text{Si}=1.11$\,\AA, $r_\text{O}=0.66$\,\AA) multiplied with a factor of $1.3$.

\begin{figure}[ht]
  \centering
  \begin{subfigure}[b]{0.4\textwidth}
    \includegraphics[width=\textwidth]{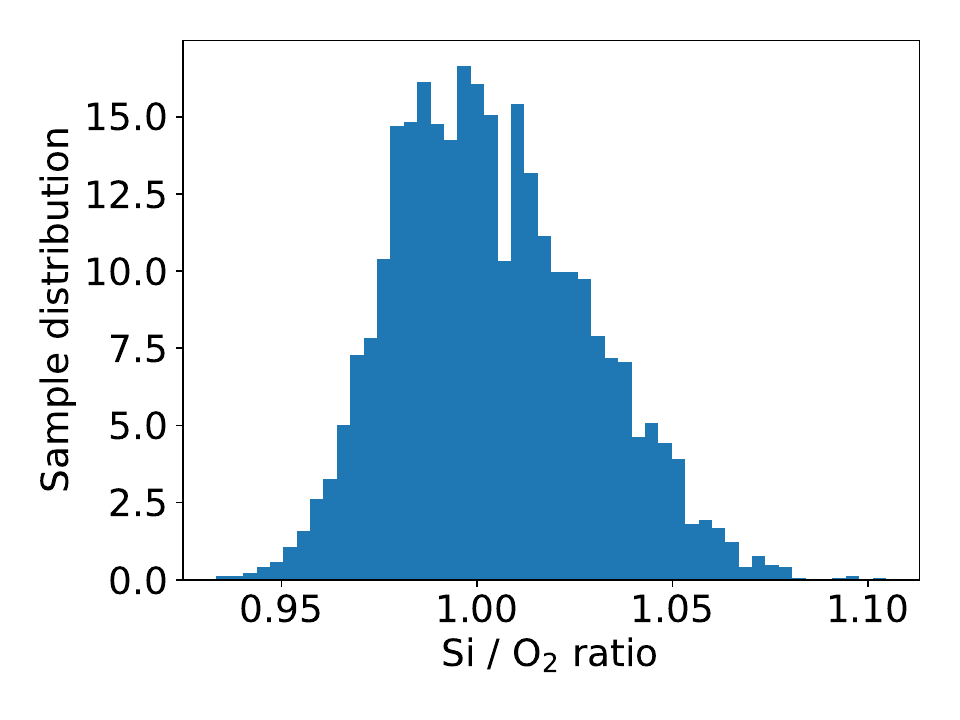}
    \caption{}
  \end{subfigure} 
  \begin{subfigure}[b]{0.4\textwidth}
    \includegraphics[width=\textwidth]{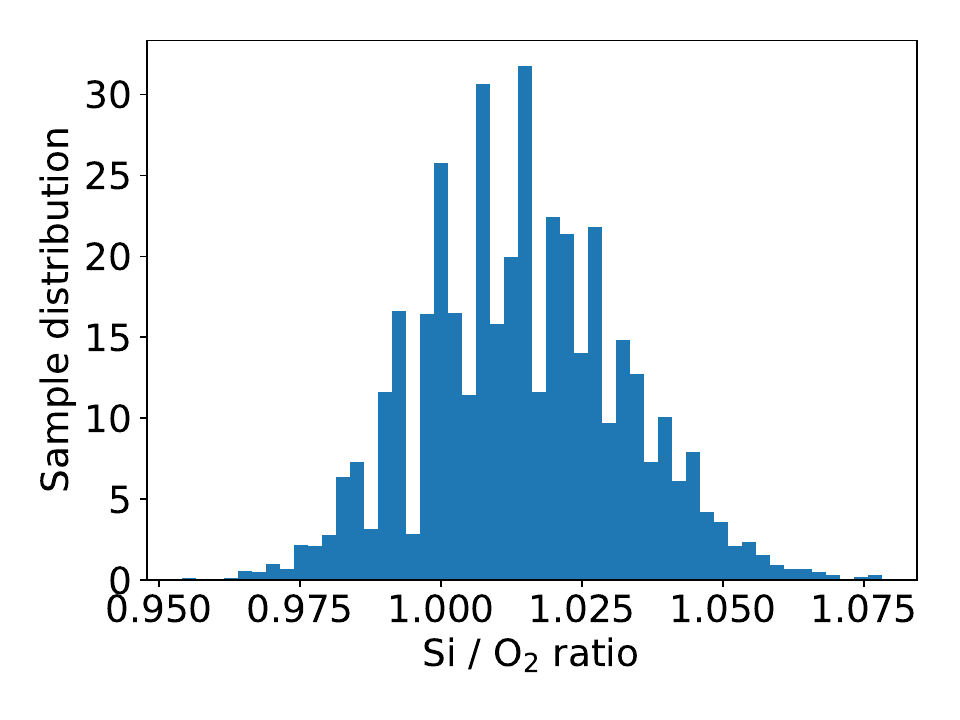}
    \caption{}
  \end{subfigure} 
  \caption{Histograms of the ratio between \ch{Si} and \ch{O2} content in the samples generated by AMDEN conditioned on shear modulus (a) and average ring size (b).
  }
  \label{fig:SiO2_ratio}
\end{figure}

\begin{figure}[ht]
  \centering
  \begin{subfigure}[b]{0.4\textwidth}
    \includegraphics[width=\textwidth]{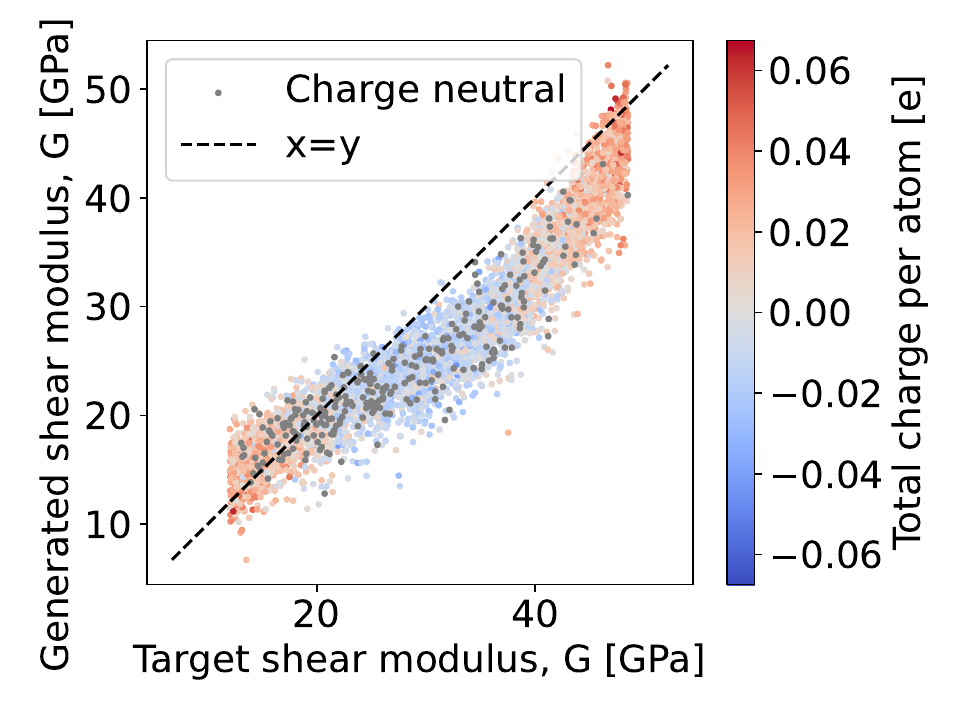}
    \caption{}
  \end{subfigure} 
  \begin{subfigure}[b]{0.4\textwidth}
    \includegraphics[width=\textwidth]{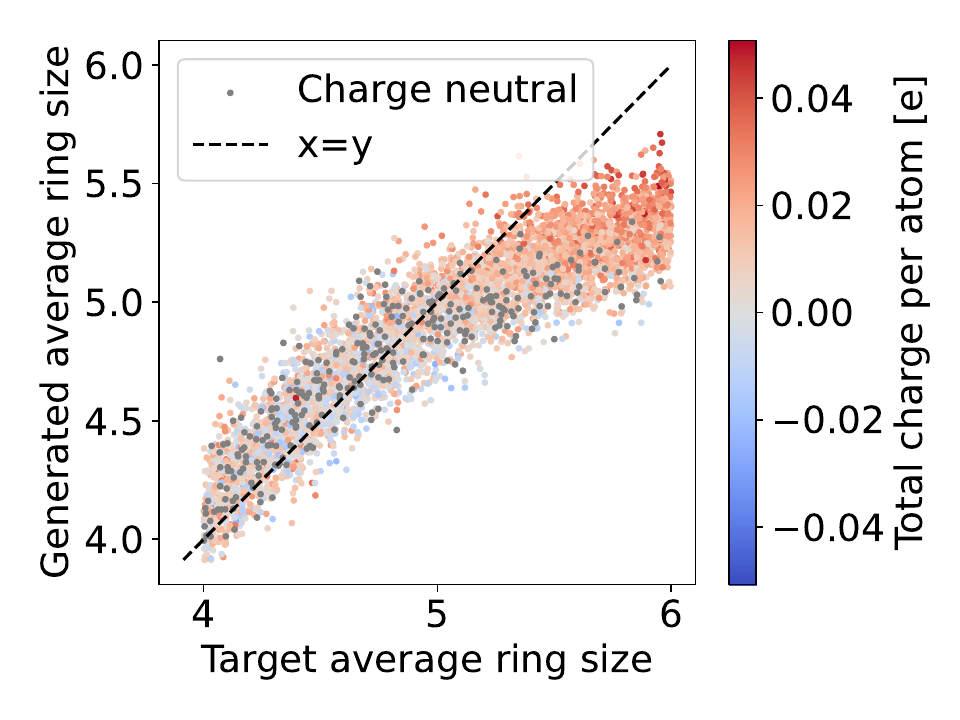}
    \caption{}
  \end{subfigure} 
    \caption{Parity plots between target and generated shear modulus (a) and average ring size (b). 
    The total system charge per atom is indicated by color and was calculated using formal charges of $+2$\,e and $-1$\,e for \ch{Si} and \ch{O} atoms, respectively.}
  \label{fig:SiO2_corr}
\end{figure}

Since ghost atoms were used for the generation of the reported \ch{SiO2} structures, stoichiometric balance of the generated samples is not strictly guaranteed.
Instead, the model learns to predict the correct ratio between \ch{Si} and \ch{O} atoms in the samples.
Histograms of the \ch{Si} to \ch{O2} ratios in the generated samples are shown in Fig.~\ref{fig:SiO2_ratio}. We note that the figures included in the main text only contain perfectly stoichiometrically balanced samples.
Parity plots of the shear modulus and the average ring size, including also the non-stoichiometric samples, are shown in Fig.~\ref{fig:SiO2_corr}.
The dependence of the valid fraction on the target property value is shown in Fig.~\ref{fig:valid_ratio_sio2}.

\section{Structural features of the amorphous \ch{Si} data set}
\begin{figure}[ht]
  \centering
  \begin{subfigure}[b]{0.24\textwidth}
    \includegraphics[width=\textwidth]{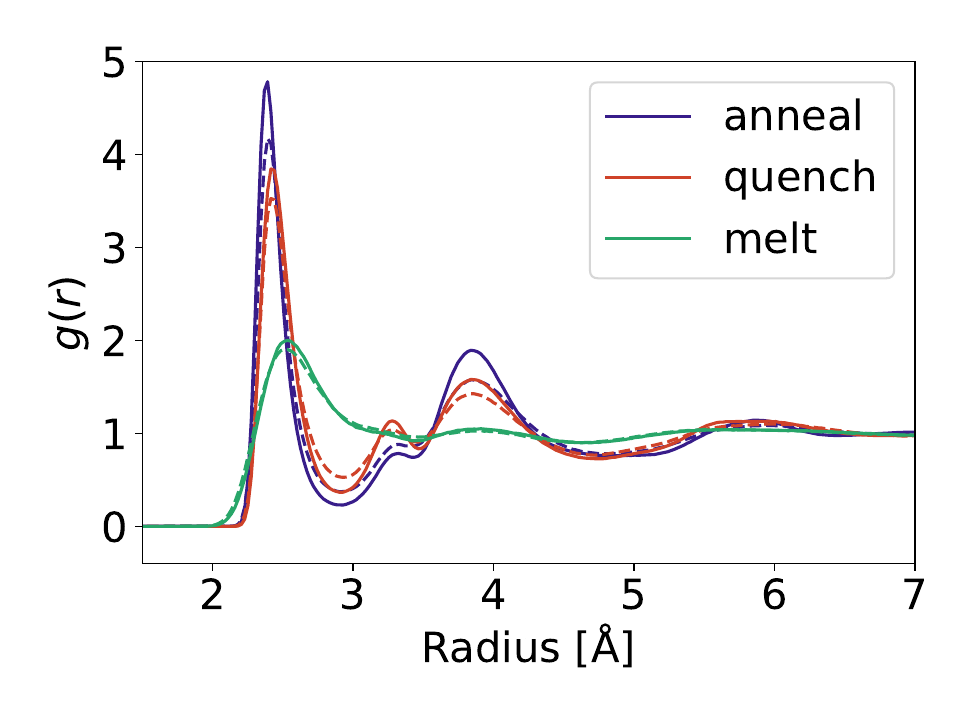}
    \caption{}
  \end{subfigure}
  \begin{subfigure}[b]{0.24\textwidth}
    \includegraphics[width=\textwidth]{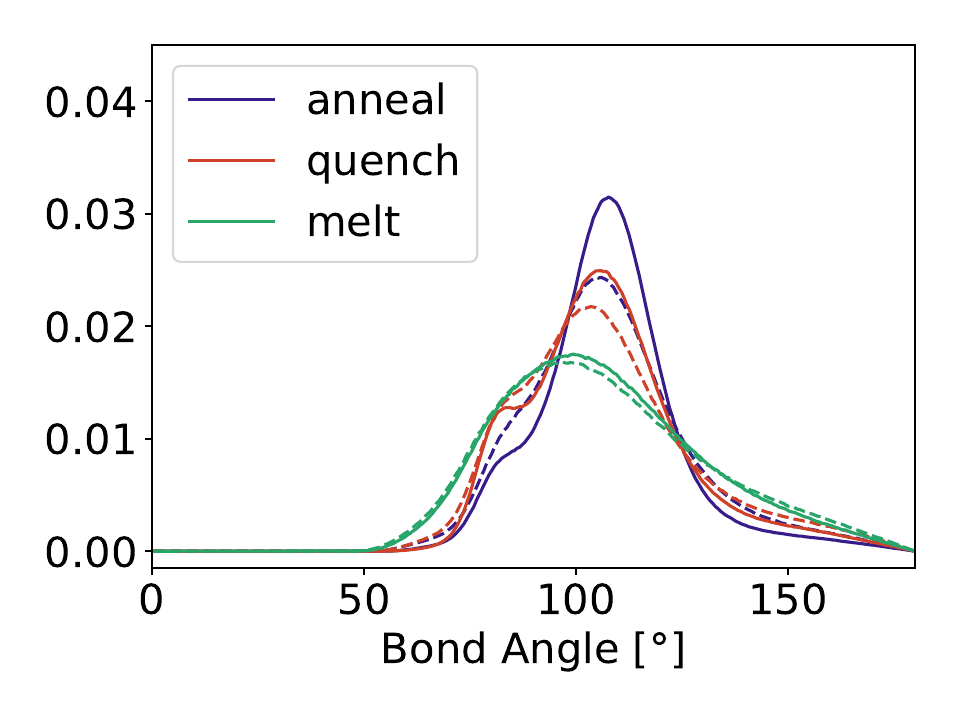}
    \caption{}
  \end{subfigure}
  \begin{subfigure}[b]{0.24\textwidth}
    \includegraphics[width=\textwidth]{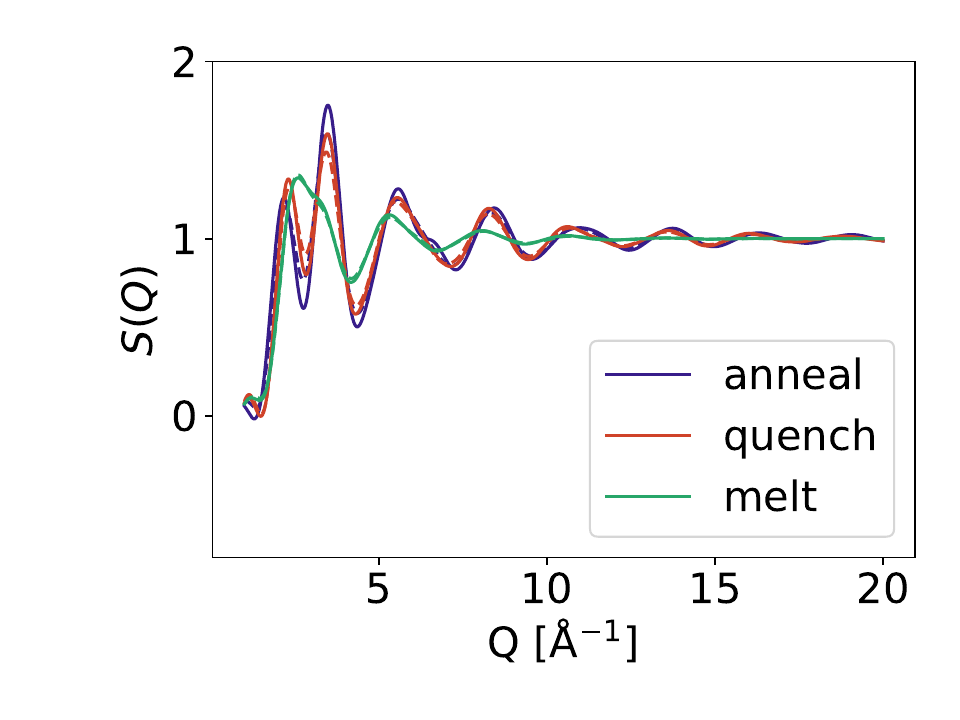}
    \caption{}
  \end{subfigure} 
  \begin{subfigure}[b]{0.24\textwidth}
    \includegraphics[width=\textwidth]{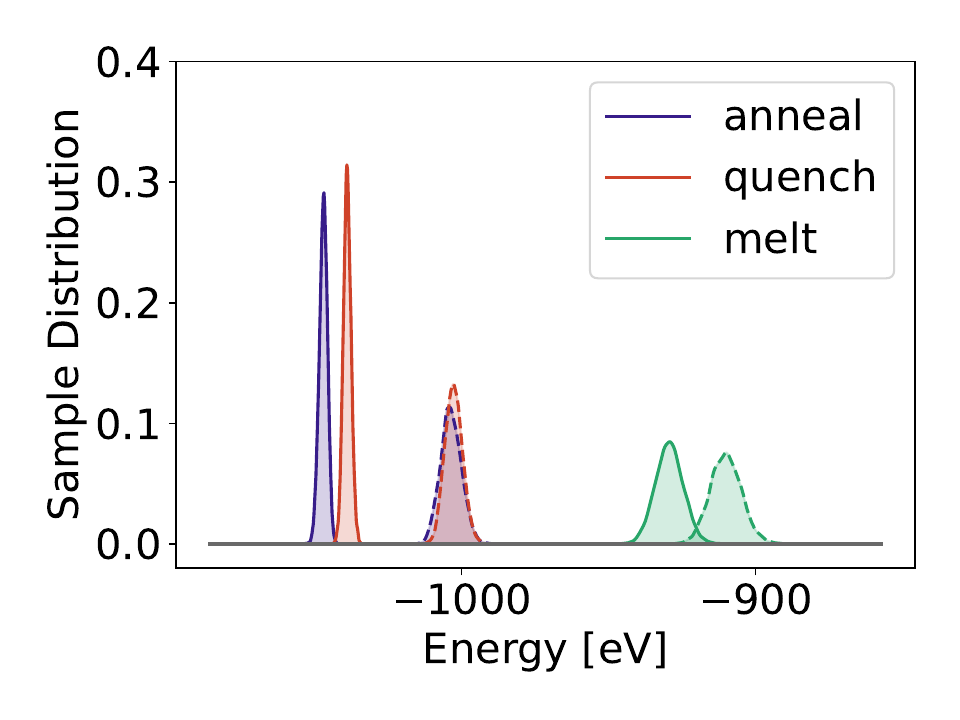}
    \caption{}
  \end{subfigure} 
  \newline
  \begin{subfigure}[b]{0.24\textwidth}
    \includegraphics[width=\textwidth]{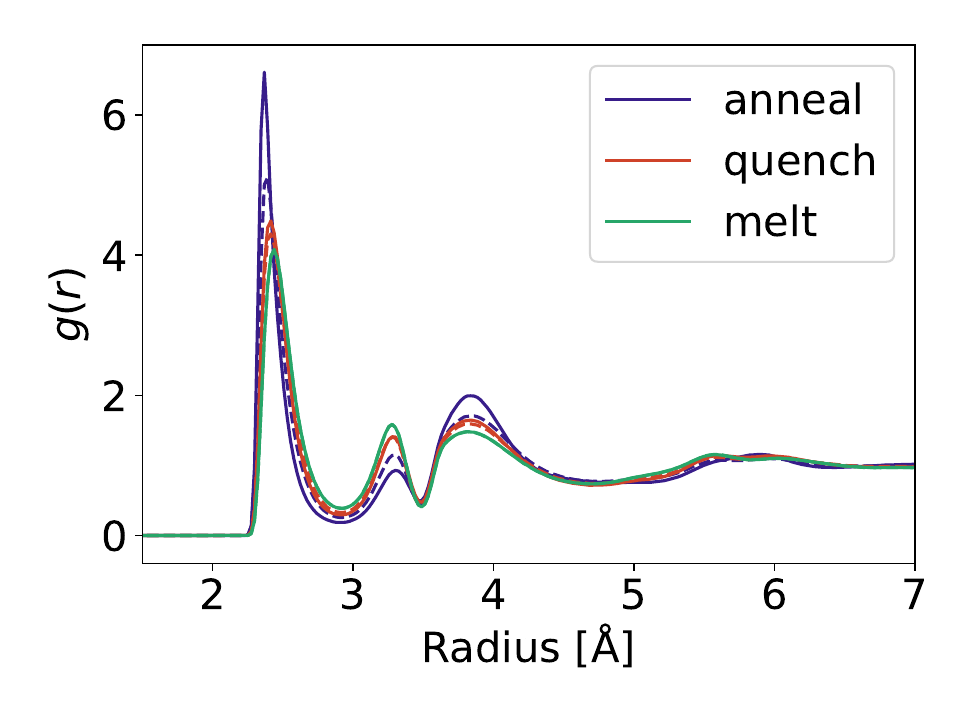}
    \caption{}
  \end{subfigure}
  \begin{subfigure}[b]{0.24\textwidth}
    \includegraphics[width=\textwidth]{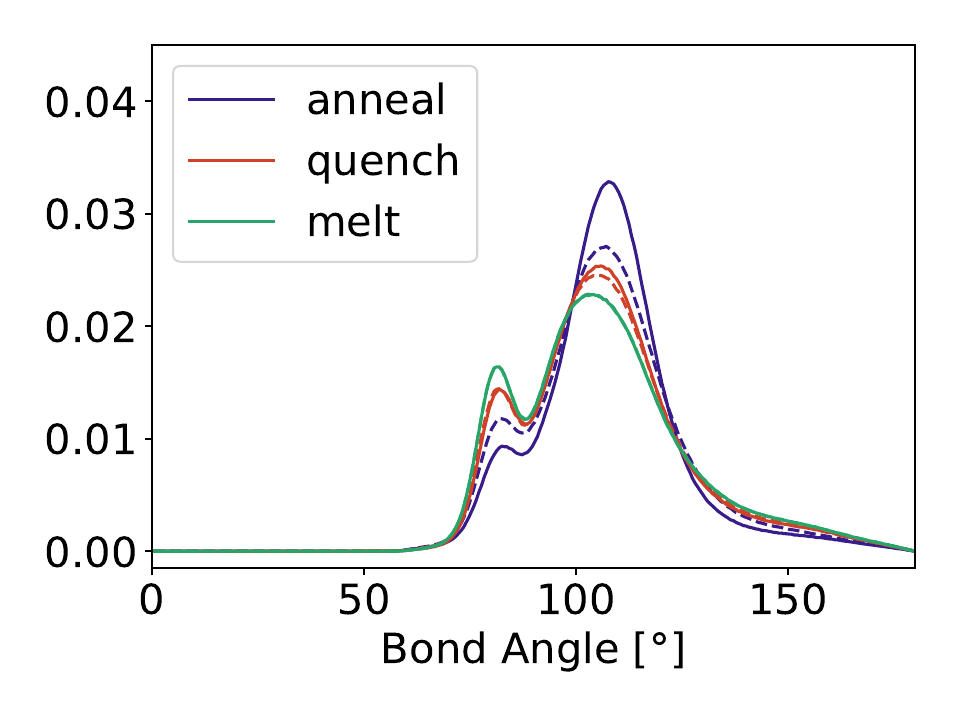}
    \caption{}
  \end{subfigure}
  \begin{subfigure}[b]{0.24\textwidth}
    \includegraphics[width=\textwidth]{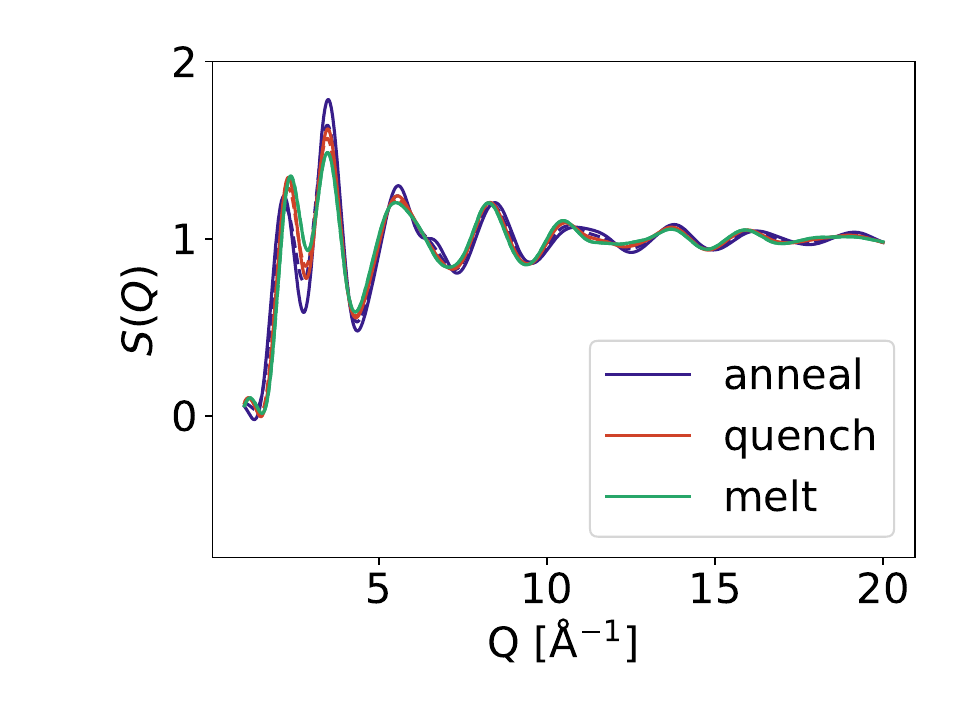}
    \caption{}
  \end{subfigure} 
  \begin{subfigure}[b]{0.24\textwidth}
    \includegraphics[width=\textwidth]{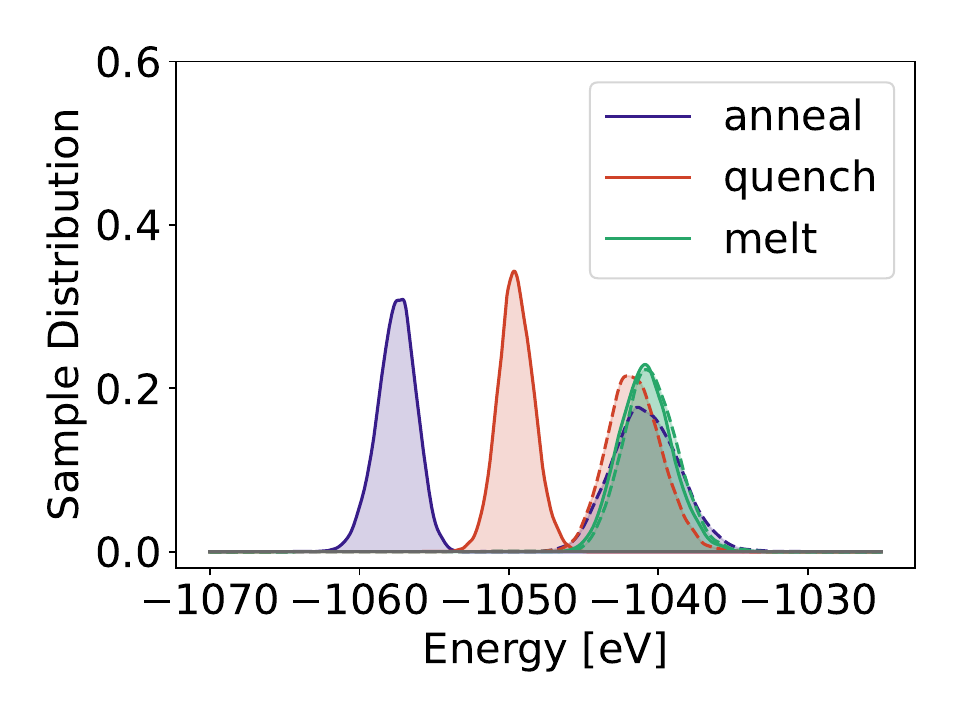}
    \caption{}
  \end{subfigure}
  \newline
  \begin{subfigure}[b]{0.24\textwidth}
    \includegraphics[width=\textwidth]{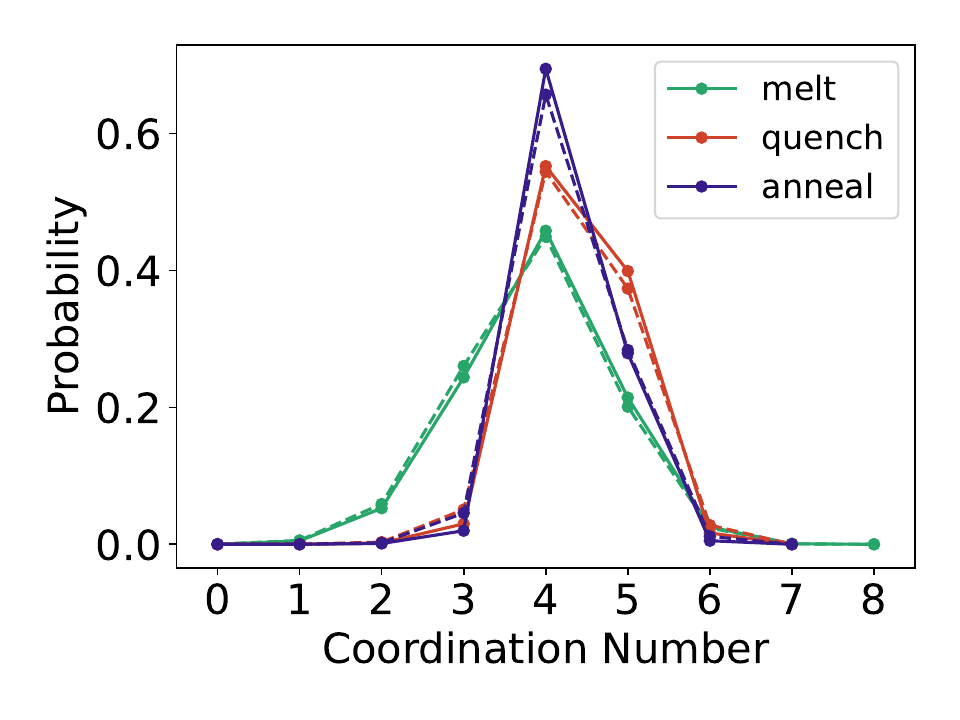}
    \caption{}
  \end{subfigure}
  \begin{subfigure}[b]{0.24\textwidth}
    \includegraphics[width=\textwidth]{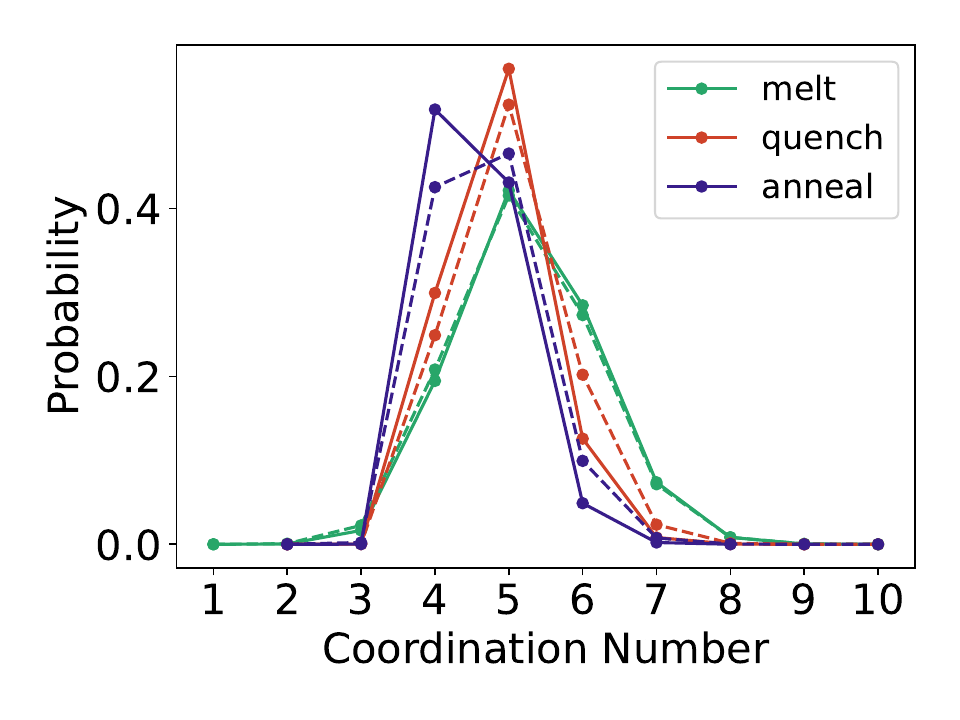}
    \caption{}
  \end{subfigure}
  \begin{subfigure}[b]{0.24\textwidth}
    \includegraphics[width=\textwidth]{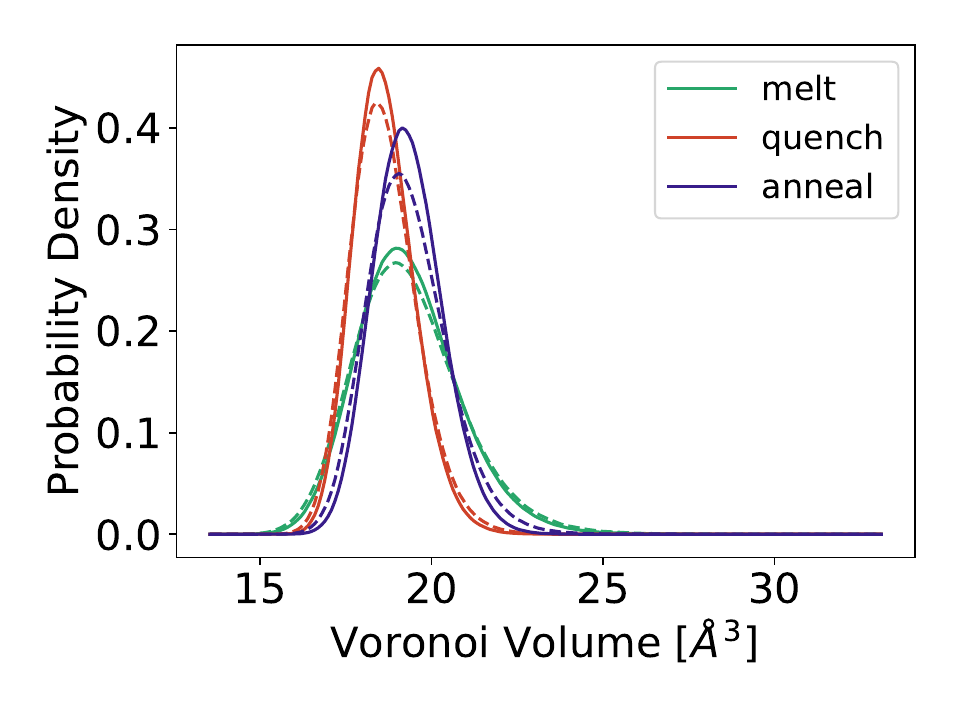}
    \caption{}
  \end{subfigure}
  \caption{Radial distribution function (a), bond angle distribution (b), structure factor (c), energies (d), coordination number distributions at cutoff radii of $2.8$\,\AA{} (i) and $3.0$\,\AA{} (j), and Voronoi volume distribution (k) of the generated structures compared to the training data. Panels (e), (f), (g) and (h) show the features in the same order after performing a local geometry optimization using the Tersoff potential used for generating the training data.
    Training data are shown by solid lines, while dashed lines are obtained from the AMDEN-generated samples.
  }
  \label{fig:Si_std}
\end{figure}

\begin{figure}[ht]
  \centering
  \begin{subfigure}[b]{0.24\textwidth}
    \includegraphics[width=\textwidth]{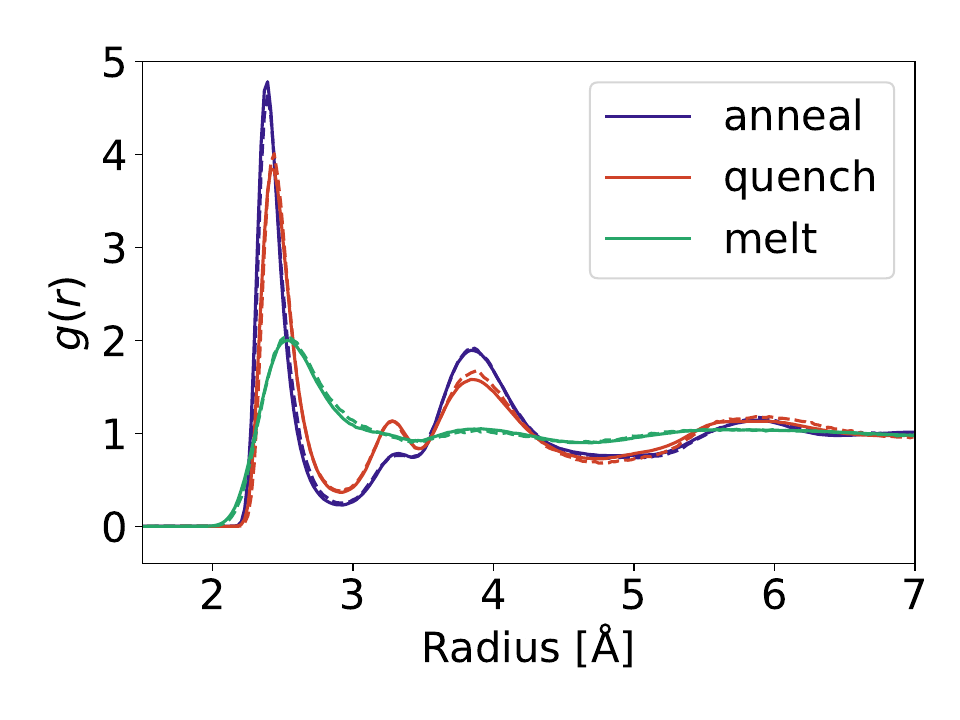}
    \caption{}
  \end{subfigure}
  \begin{subfigure}[b]{0.24\textwidth}
    \includegraphics[width=\textwidth]{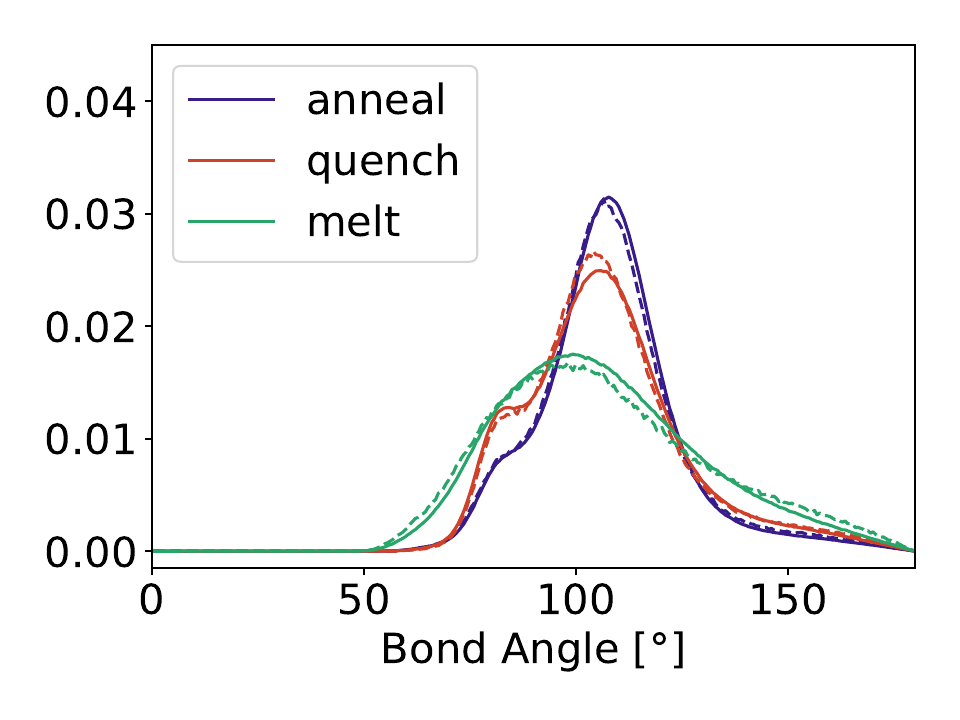}
    \caption{}
  \end{subfigure}
  \begin{subfigure}[b]{0.24\textwidth}
    \includegraphics[width=\textwidth]{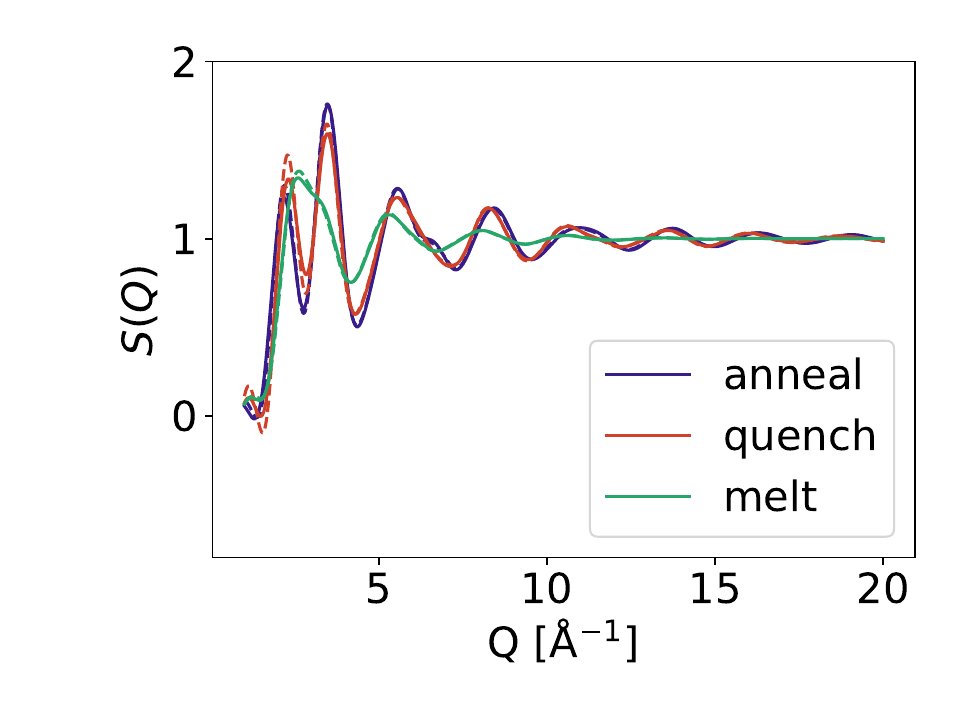}
    \caption{}
  \end{subfigure} 
  \begin{subfigure}[b]{0.24\textwidth}
    \includegraphics[width=\textwidth]{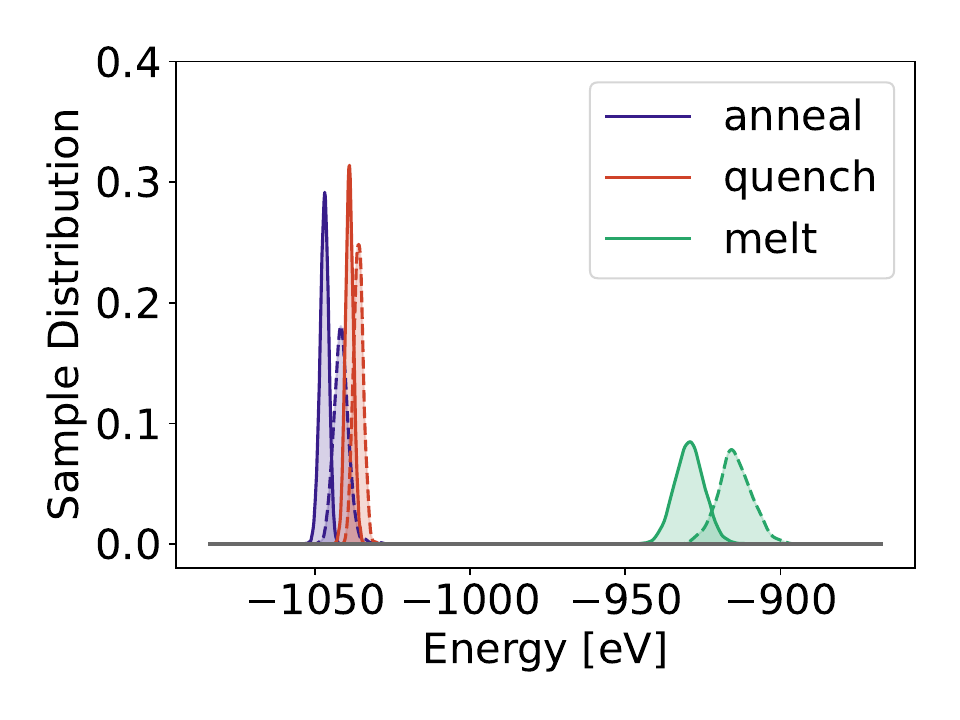}
    \caption{}
  \end{subfigure} 
  \newline
  \begin{subfigure}[b]{0.24\textwidth}
    \includegraphics[width=\textwidth]{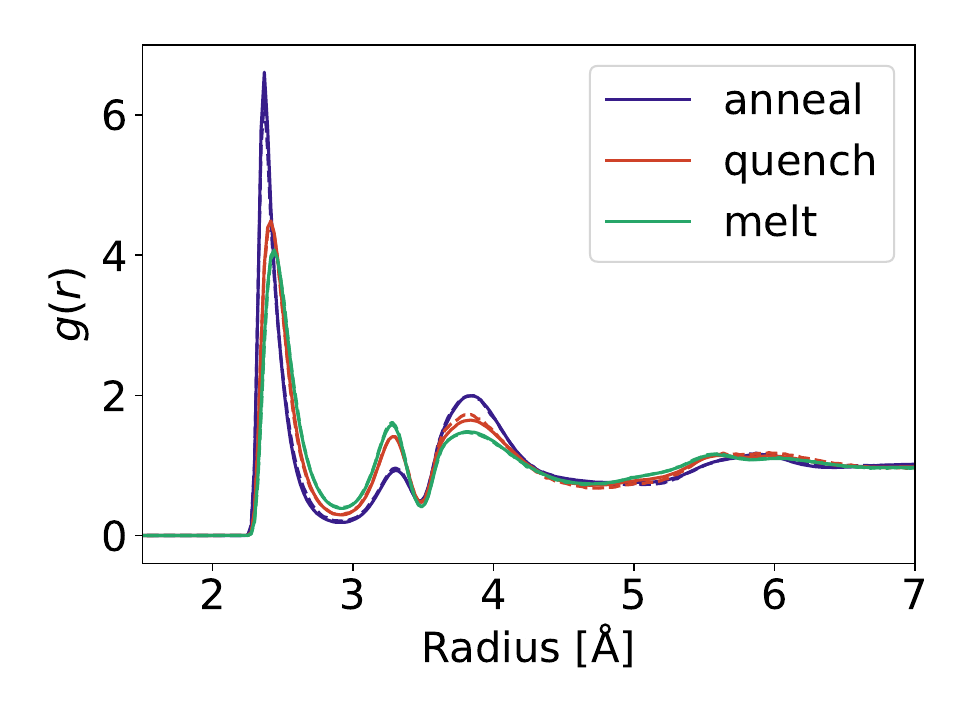}
    \caption{}
  \end{subfigure}
  \begin{subfigure}[b]{0.24\textwidth}
    \includegraphics[width=\textwidth]{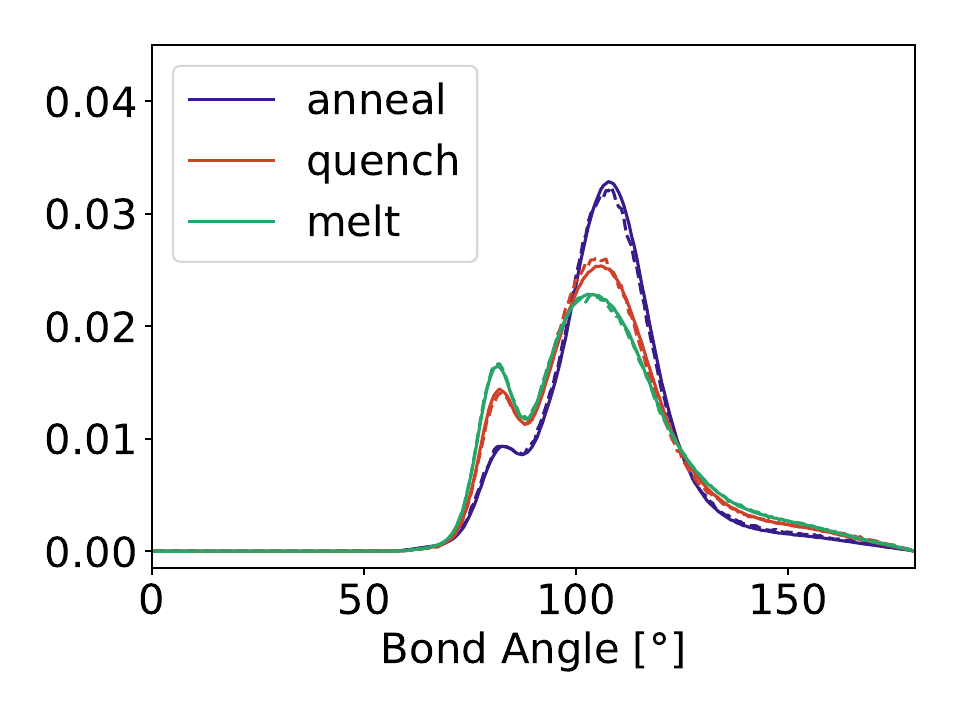}
    \caption{}
  \end{subfigure}
  \begin{subfigure}[b]{0.24\textwidth}
    \includegraphics[width=\textwidth]{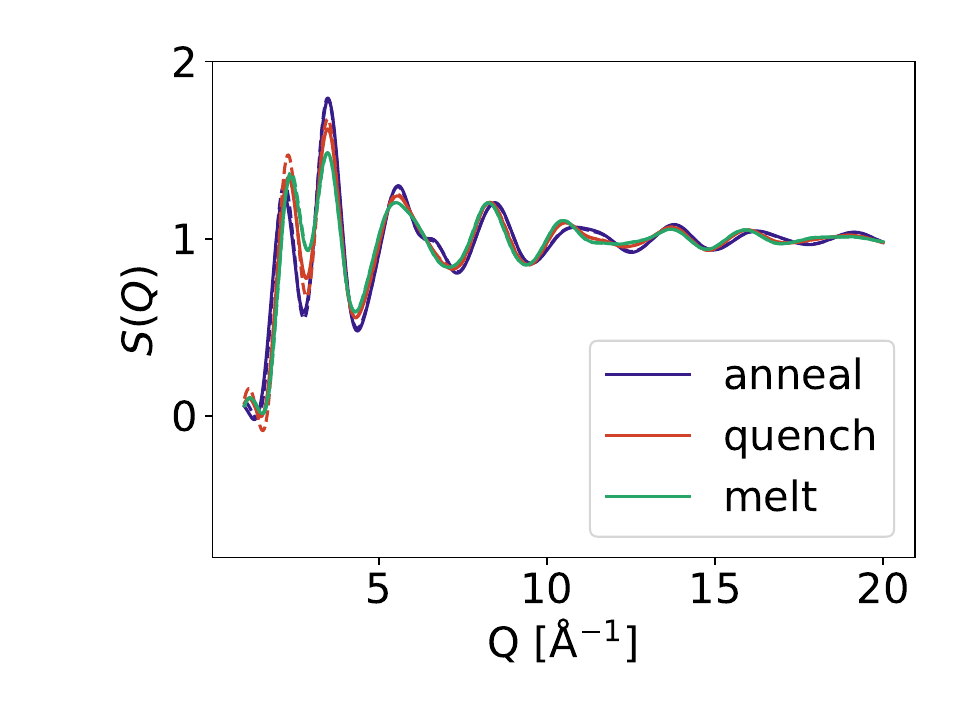}
    \caption{}
  \end{subfigure} 
  \begin{subfigure}[b]{0.24\textwidth}
    \includegraphics[width=\textwidth]{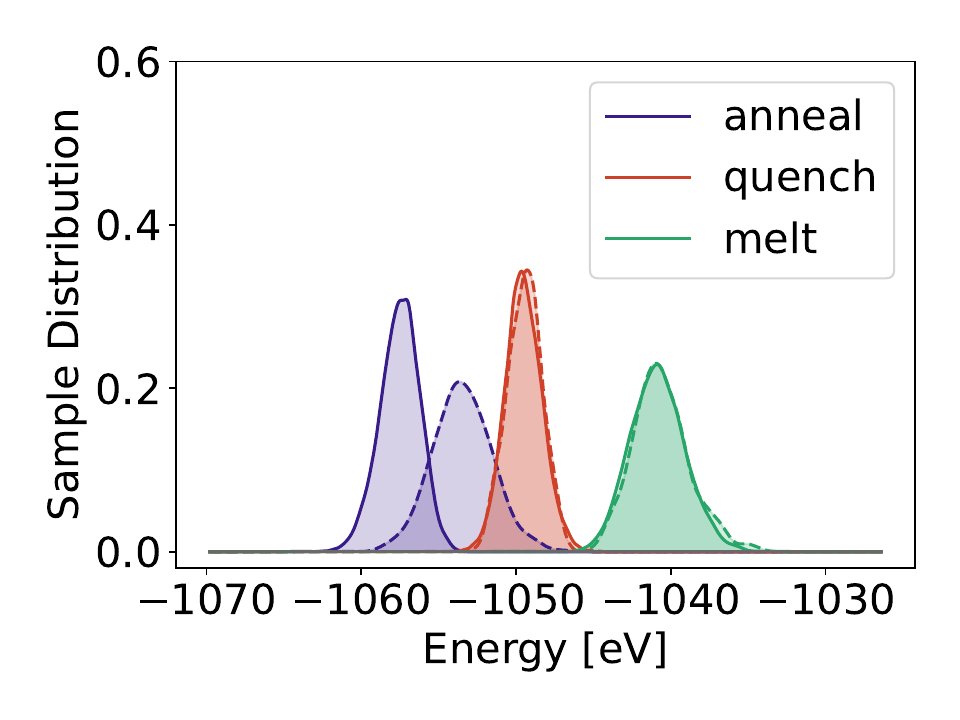}
    \caption{}
  \end{subfigure}
  \newline
  \begin{subfigure}[b]{0.24\textwidth}
    \includegraphics[width=\textwidth]{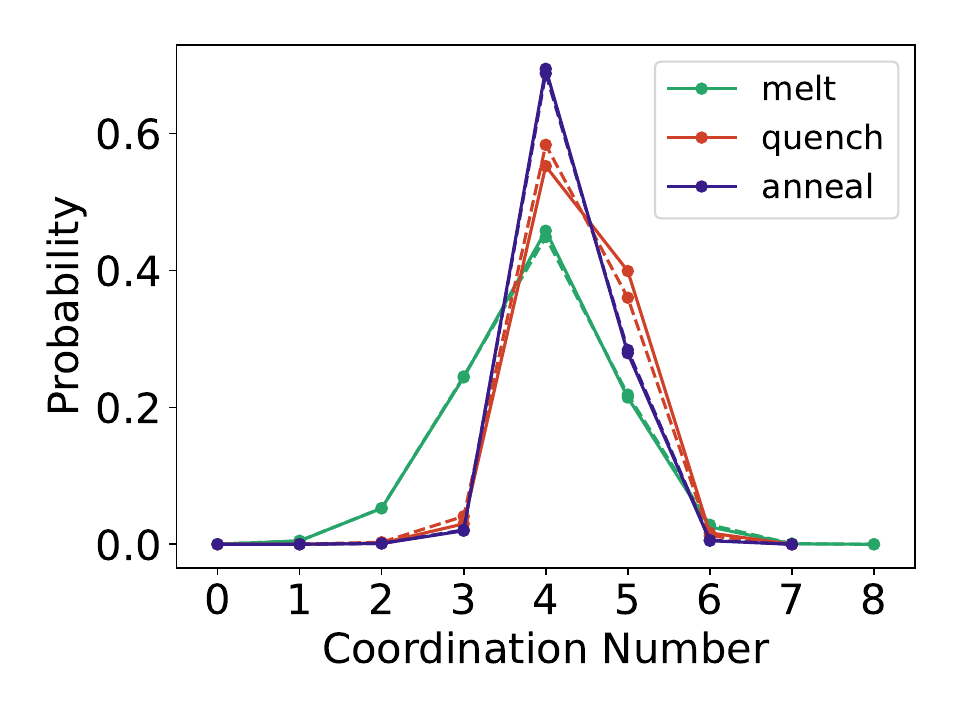}
    \caption{}
  \end{subfigure}
  \begin{subfigure}[b]{0.24\textwidth}
    \includegraphics[width=\textwidth]{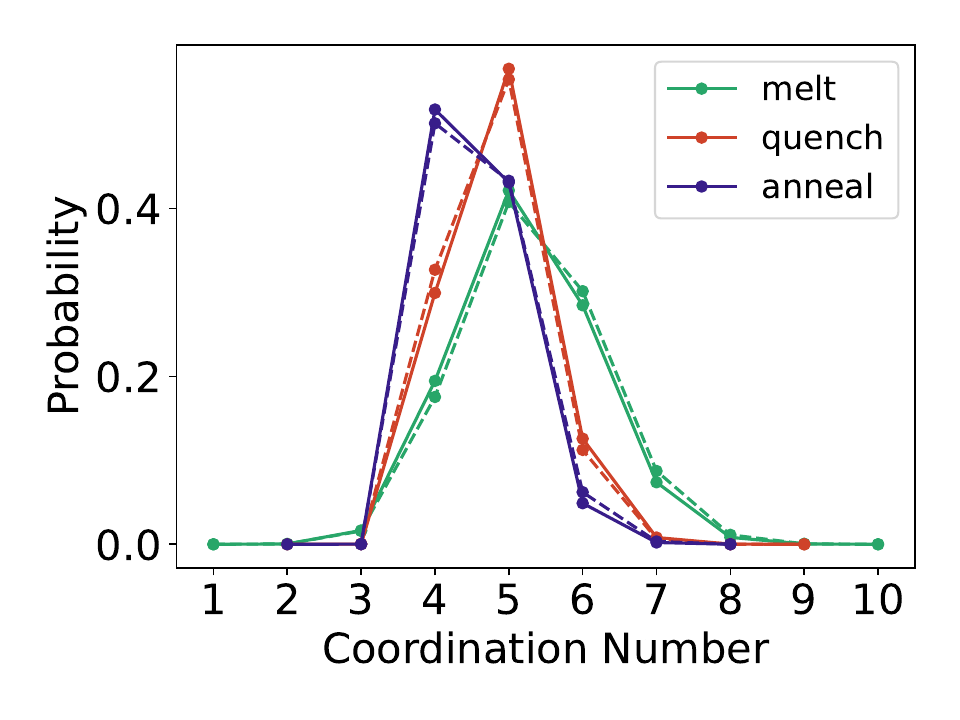}
    \caption{}
  \end{subfigure}
  \begin{subfigure}[b]{0.24\textwidth}
    \includegraphics[width=\textwidth]{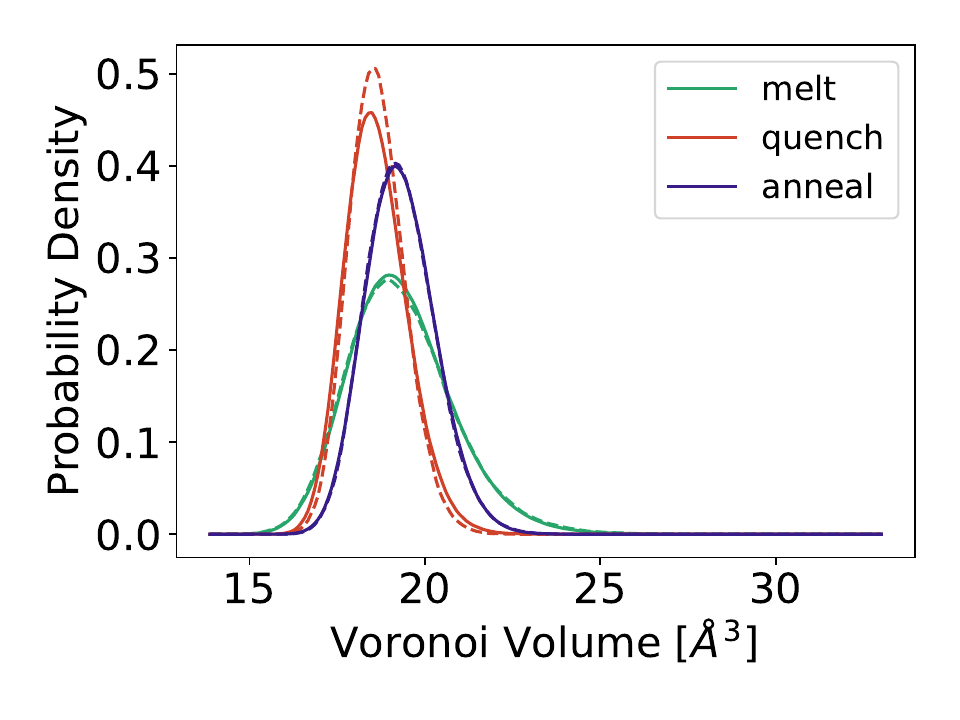}
    \caption{}
  \end{subfigure}
  \caption{Radial distribution function (a), bond angle distribution (b), structure factor (c), energies (d), coordination number distributions at cutoff radii of $2.8$\,\AA{} (i) and $3.0$\,\AA{} (j), and Voronoi volume distribution (k) of the structures generated using Hamiltonian Monte Carlo (HMC) denoising compared to the training data. Panels (e), (f), (g) and (h) show the features in the same order after performing a local geometry optimization using the Tersoff potential used for generating the training data.
    Training data are shown by solid lines, while dashed lines are obtained from the AMDEN-generated samples.
  }
  \label{fig:Si_hmc}
\end{figure}

To analyze the quality of the structures generated by AMDEN, we computed radial distribution functions, bond angle distributions, structure factors, the potential energy distribution, coordination number distributions, and Voronoi volume distributions of the generated and the training samples. 
Bond angle distributions were computed using a radial cutoff of $2.7$\,\AA.
All features were computed before and after performing a local geometry optimization of the structures.
Features obtained from the standard denoising procedure are shown in Fig.~\ref{fig:Si_std}, while those obtained from the Hamiltonian Monte Carlo (HMC) denoising procedure are shown in Fig.~\ref{fig:Si_hmc}.
Figures shown in the main text are included here again for completeness.


\section{Mechanical properties}\label{sec:mechanical_properties}
Young's and shear moduli were computed using the strain tensor $C_{ijkl}$. 
First, a local geometry optimization was performed on the structural samples to obtain the relaxed atomic positions and lattice vectors.
The strain tensor was calculated as the derivative of the stress tensor $\sigma_{ij}$ with respect to the strain $\varepsilon_{kl}$, i.e.,
\begin{equation}
C_{ijkl} = \left. \frac{\partial \sigma_{ij}}{\partial \varepsilon_{kl}} \right|_{\varepsilon = 0}.
\end{equation}
Finite differences were used to calculate the derivatives and atomic positions were relaxed after straining the unit cell before the stress tensor was computed.

Since the investigated samples are largely isotropic, we can reduce $C_{ijkl}$ to $C_{ij}$ using Voigt notation, averaging redundant entries in the full tensor. 
The Young's and shear moduli are then computed as
\begin{equation}
  E = \frac{(C_{11} - C_{12}) \cdot (C_{11} + 2\,C_{12})}{C_{11} + C_{12}}
\end{equation}
and
\begin{equation}
  G = C_{44}
\end{equation}
respectively~\cite{gersten2001physics}.

\section{Structural features of the MEG dataset}\label{sec:meg_structure}

To validate the local structure of the generated multi-element glass (MEG) samples, we computed partial radial distribution functions (RDFs) and cumulative coordination numbers for six element--oxygen pairs: Si--O, P--O, Al--O, Li--O, Ti--O, and Ca--O.
These include all three network formers (Si, P, Al) and the three most abundant modifier cations (Li, Ti, Ca), which together account for the majority of cation sites in the dataset.
The remaining modifiers (K, Ba, Be, Zn) are too sparse for statistically reliable comparison.
The generated samples are taken from the Li-conditioned inverse design runs (targeting a Li molar fraction of 0.15 and Young's modulus in the range 20--160\,GPa), which produce structures with compositions that differ from the average training distribution.

\begin{figure}[!t]
  \centering
  \includegraphics[width=\textwidth]{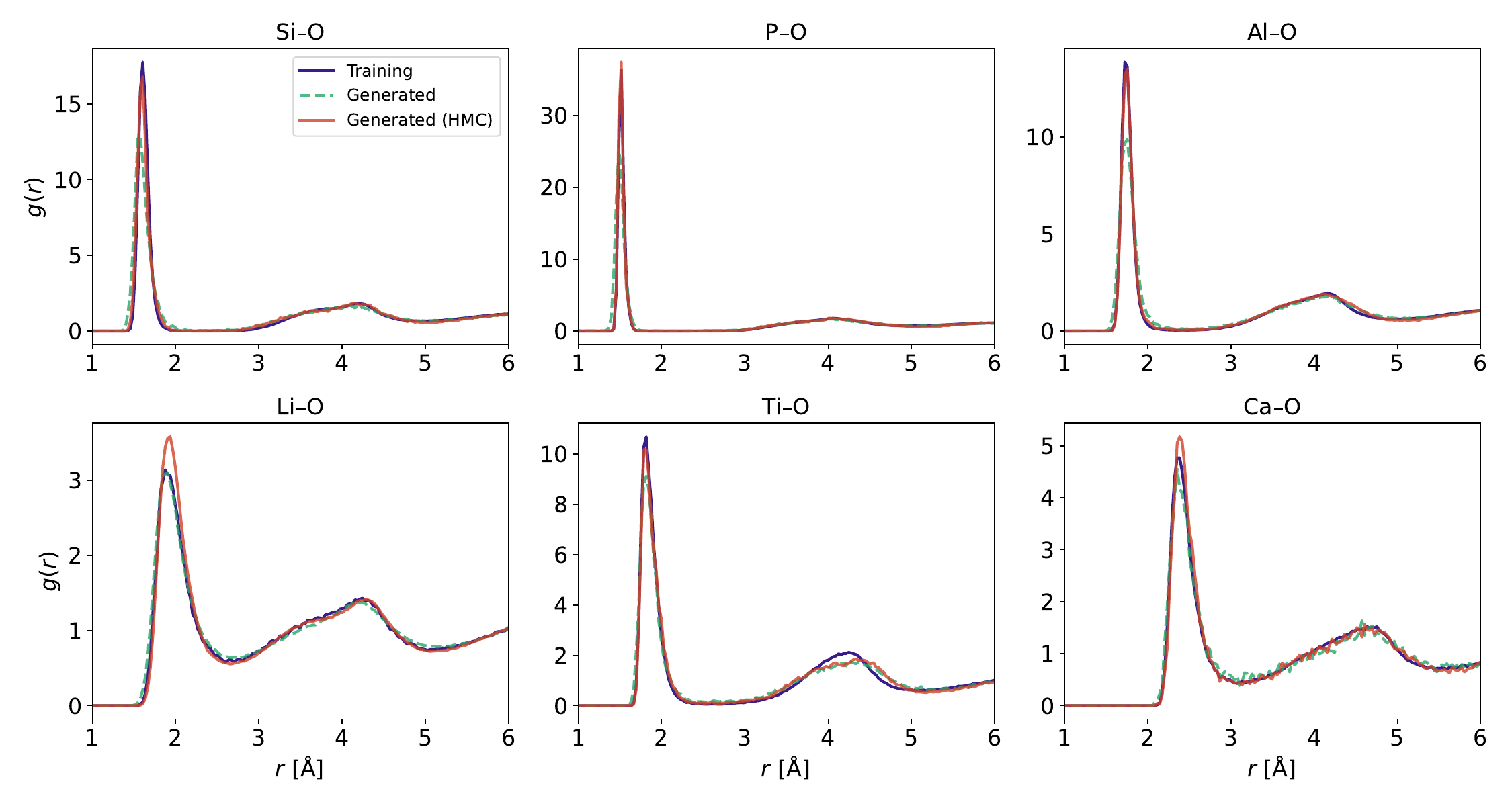}
  \caption{
    Partial radial distribution functions $g(r)$ for six element--oxygen pairs in the MEG system, comparing training data (solid blue), generated samples without HMC denoising (dashed green), and generated samples with HMC denoising (solid red).
  }
  \label{fig:meg_partial_rdf}
\end{figure}

Supplemental Figure~\ref{fig:meg_partial_rdf} shows the partial RDFs $g(r)$ comparing the training data with generated samples (with and without HMC refinement).
Standard denoising reproduces the correct first-shell peak positions for all six pairs, but the peaks are systematically lower and broader than the training data, indicating incomplete relaxation of the local bonding environments.
HMC refinement substantially improves the agreement, sharpening the first-shell peaks to closely match the training data. This is consistent with the improvements observed for the amorphous Si anneal dataset.
The Li--O pair is an exception: the HMC-refined samples show a slightly higher first peak than the training average, which reflects the elevated Li content in the conditioned generation.

\begin{figure}[!t]
  \centering
  \begin{subfigure}[b]{0.48\textwidth}
    \includegraphics[width=\textwidth]{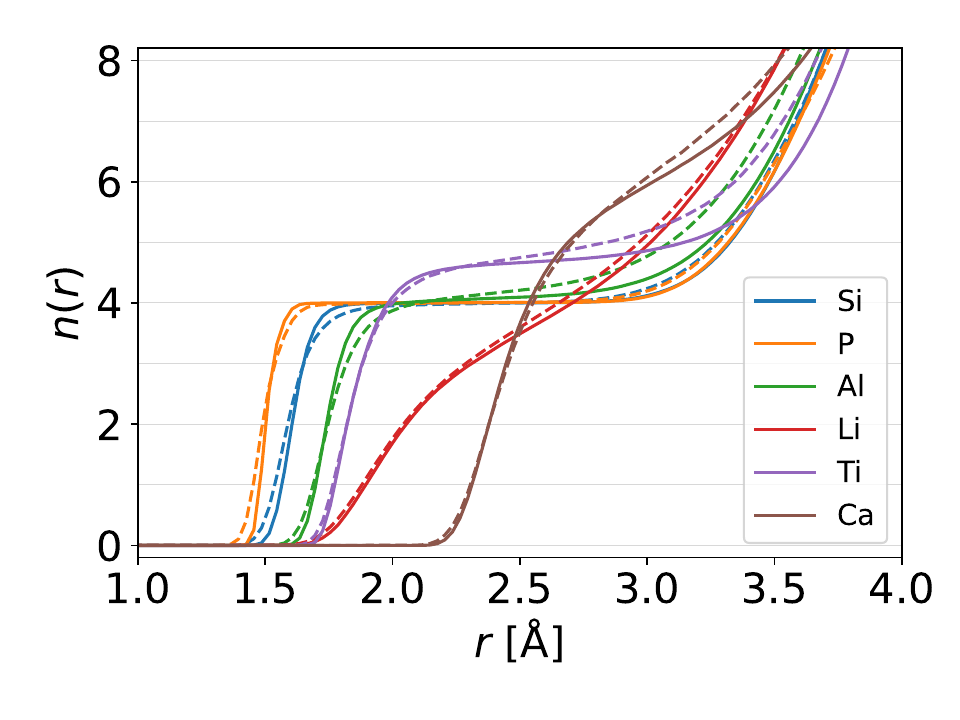}
    \caption{}
  \end{subfigure}\hfill
  \begin{subfigure}[b]{0.48\textwidth}
    \includegraphics[width=\textwidth]{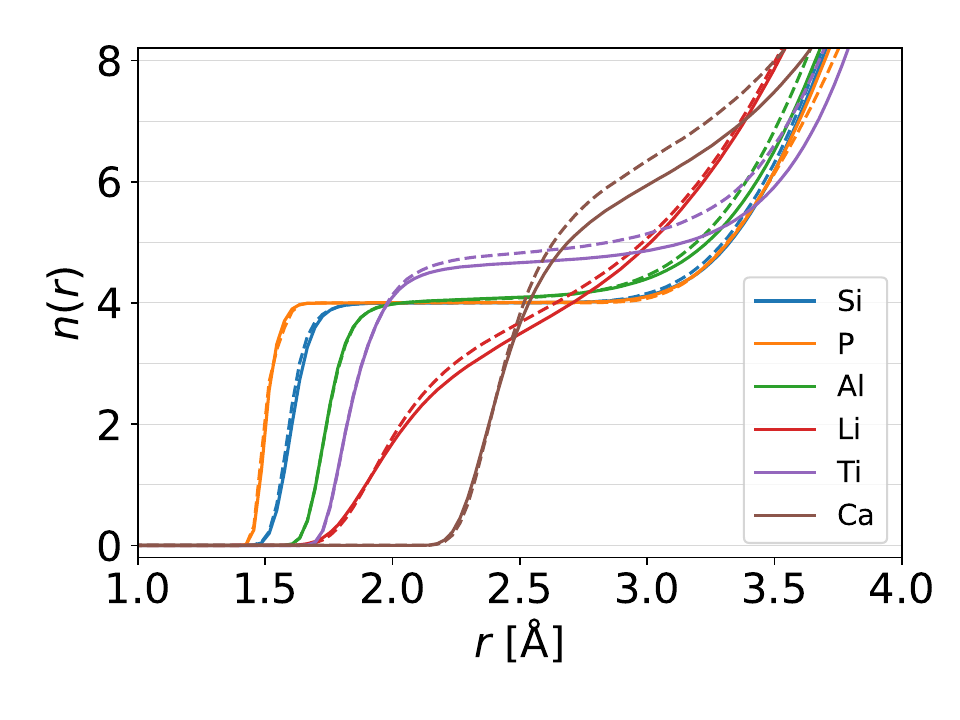}
    \caption{}
  \end{subfigure}
  \caption{
    Cumulative coordination number $n(r)$ of oxygen neighbors around each cation type, comparing training data (solid) with generated samples (dashed).
    (a)~Standard denoising.
    (b)~HMC denoising.
  }
  \label{fig:meg_cum_cn}
\end{figure}

Supplemental Figure~\ref{fig:meg_cum_cn} presents the cumulative coordination number $n(r)$, which counts the average number of oxygen neighbors within distance $r$ for each cation type.
The first-shell plateau values confirm that the generated structures reproduce the expected coordination numbers: ${\sim}4$ for Si, P, and Al, and higher values for the modifier cations.
The agreement is strongest in the first coordination shell, where the HMC-refined curves closely track the training data, while the standard denoising curves show small but visible offsets.
Beyond the first shell, minor deviations appear, reflecting the compositional differences between the conditioned generation and the training distribution.

\section{Property distributions of generated samples}
\begin{figure}[h]
  \centering
  \begin{subfigure}[b]{0.24\textwidth}
    \includegraphics[width=\textwidth]{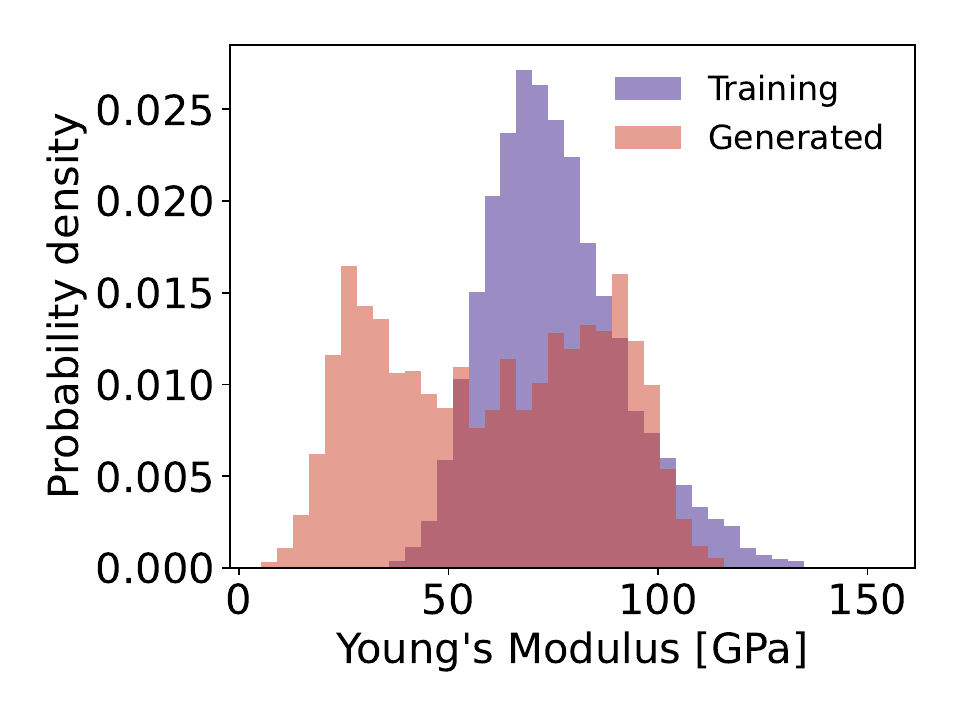}
    \caption{}
  \end{subfigure}\hfill
  \begin{subfigure}[b]{0.24\textwidth}
    \includegraphics[width=\textwidth]{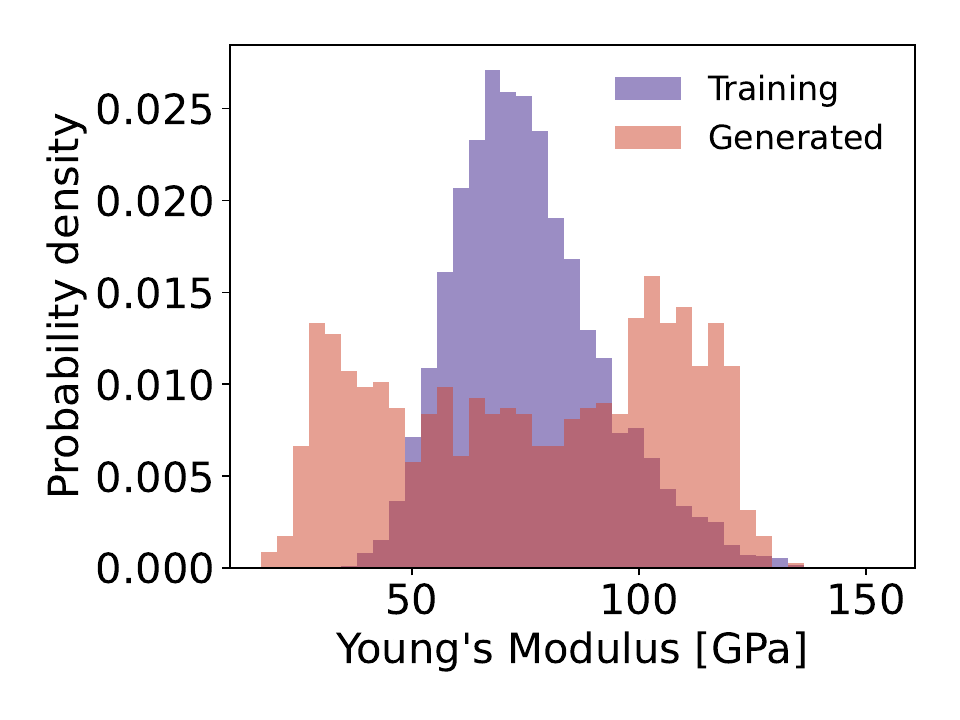}
    \caption{}
  \end{subfigure}\hfill
  \begin{subfigure}[b]{0.24\textwidth}
    \includegraphics[width=\textwidth]{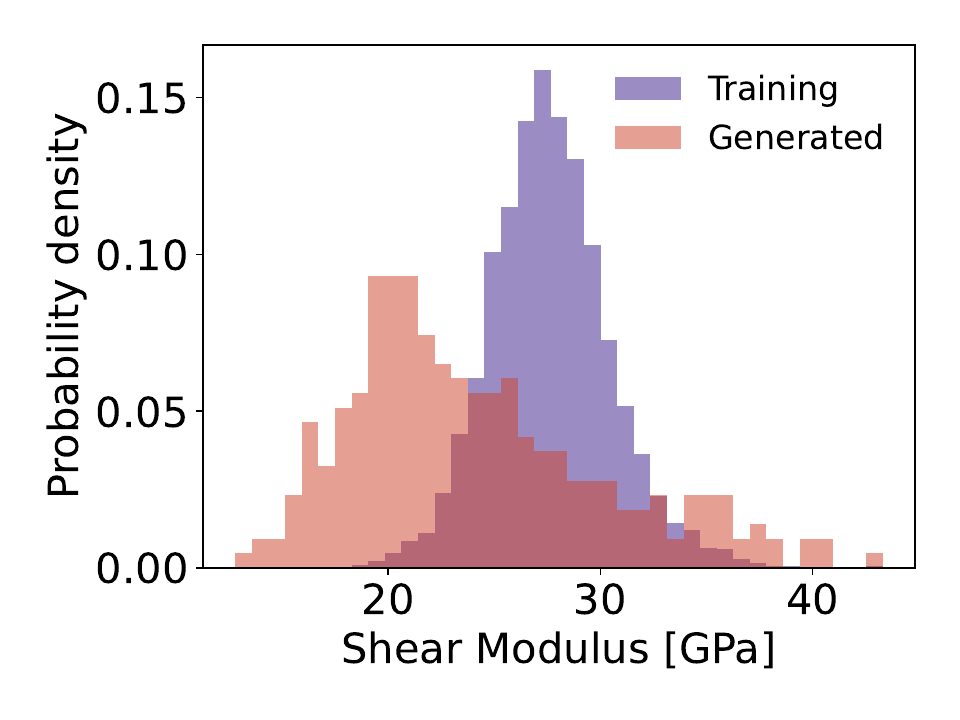}
    \caption{}
  \end{subfigure}\hfill
  \begin{subfigure}[b]{0.24\textwidth}
    \includegraphics[width=\textwidth]{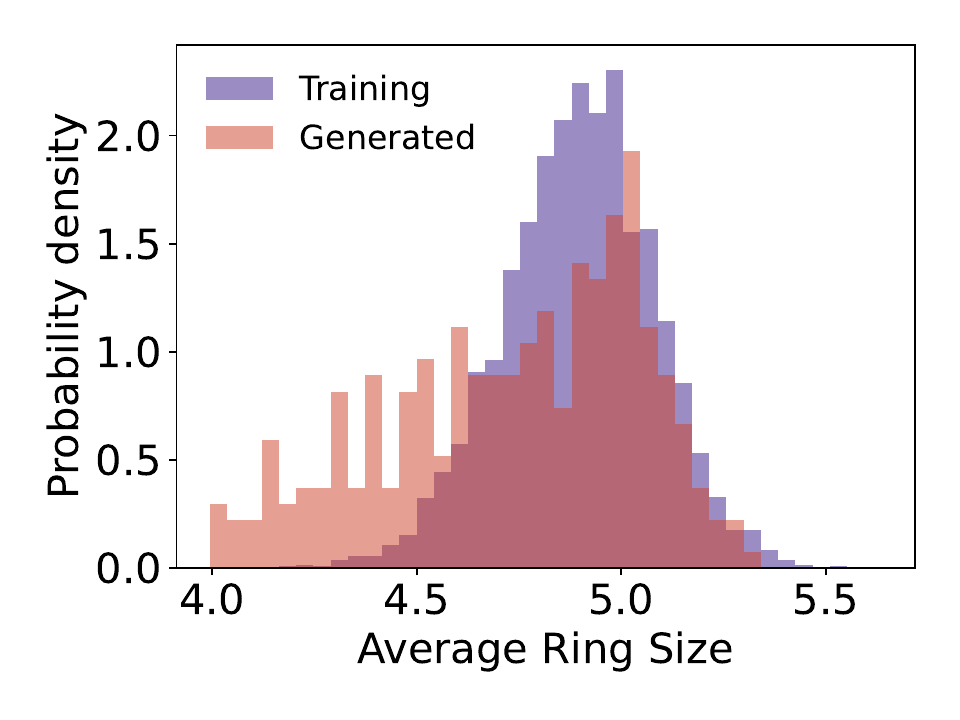}
    \caption{}
  \end{subfigure}
  \caption{
    Comparison of property distributions between training data and generated samples.
    (a) Young's modulus on the MEG dataset with standard denoising.
    (b) Young's modulus on the MEG dataset with HMC denoising.
    (c) Shear modulus on the amorphous \ce{SiO2} dataset.
    (d) Average ring size on the amorphous \ce{SiO2} dataset.
    In all panels, the generated distributions extend beyond the training range, particularly toward lower property values, indicating extrapolation capability of the model.
  }
  \label{fig:property_dist}
\end{figure}

Supplemental Figure~\ref{fig:property_dist} compares the distributions of target properties between the training data and the generated samples for both the MEG and amorphous \ce{SiO2} datasets. In all cases, the generated distributions extend beyond the training range, particularly toward lower property values.

\section{Generation validity statistics}\label{sec:valid_ratio}

Since AMDEN uses ghost atoms to control the density of generated structures, stoichiometric balance is not strictly guaranteed.
A generated sample is considered valid if its total formal charge is exactly zero, corresponding to an exact \ch{Si}:\ch{O} = 1:2 ratio for \ch{SiO2} and the appropriate cation-to-anion balance for the MEG system.
For the single-element amorphous Si datasets, all generated samples are valid by construction.
Table~\ref{tab:valid_ratio} summarizes the generation validity statistics across all datasets.

\begin{figure}[ht]
  \centering
  \begin{subfigure}[b]{0.48\textwidth}
    \includegraphics[width=\textwidth]{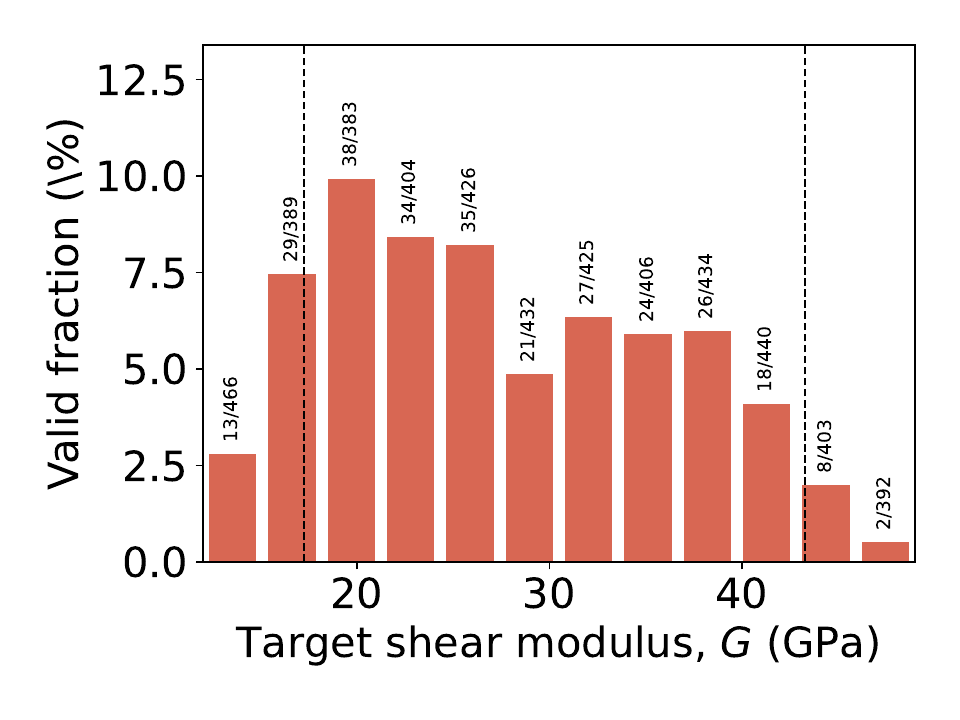}
    \caption{}
  \end{subfigure}\hfill
  \begin{subfigure}[b]{0.48\textwidth}
    \includegraphics[width=\textwidth]{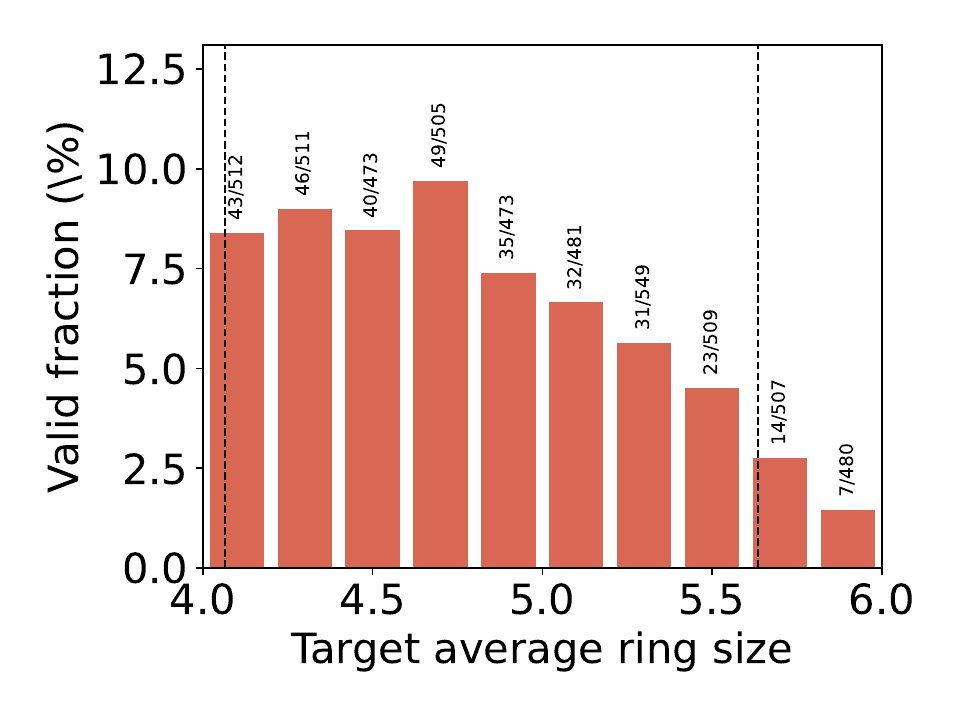}
    \caption{}
  \end{subfigure}
  \caption{
    Valid (stoichiometrically balanced) fraction of generated \ch{SiO2} samples as a function of target (a)~shear modulus and (b)~average ring size.
    Dashed vertical lines indicate the training data range.
    Each bar is annotated with the number of valid samples over the total in that bin.
  }
  \label{fig:valid_ratio_sio2}
\end{figure}

For the \ch{SiO2} dataset, the valid fraction is 5.5\% (275/5\,000) and 6.4\% (320/5\,000) for the shear modulus and average ring size conditioning, respectively.
As shown in Fig.~\ref{fig:valid_ratio_sio2}, the valid fraction depends on the target property value: it is highest within the training range and drops when the target falls outside the training distribution.
No samples fail during generation or property evaluation for either conditioning target.

\begin{figure}[!t]
  \centering
  \begin{subfigure}[b]{0.48\textwidth}
    \includegraphics[width=\textwidth]{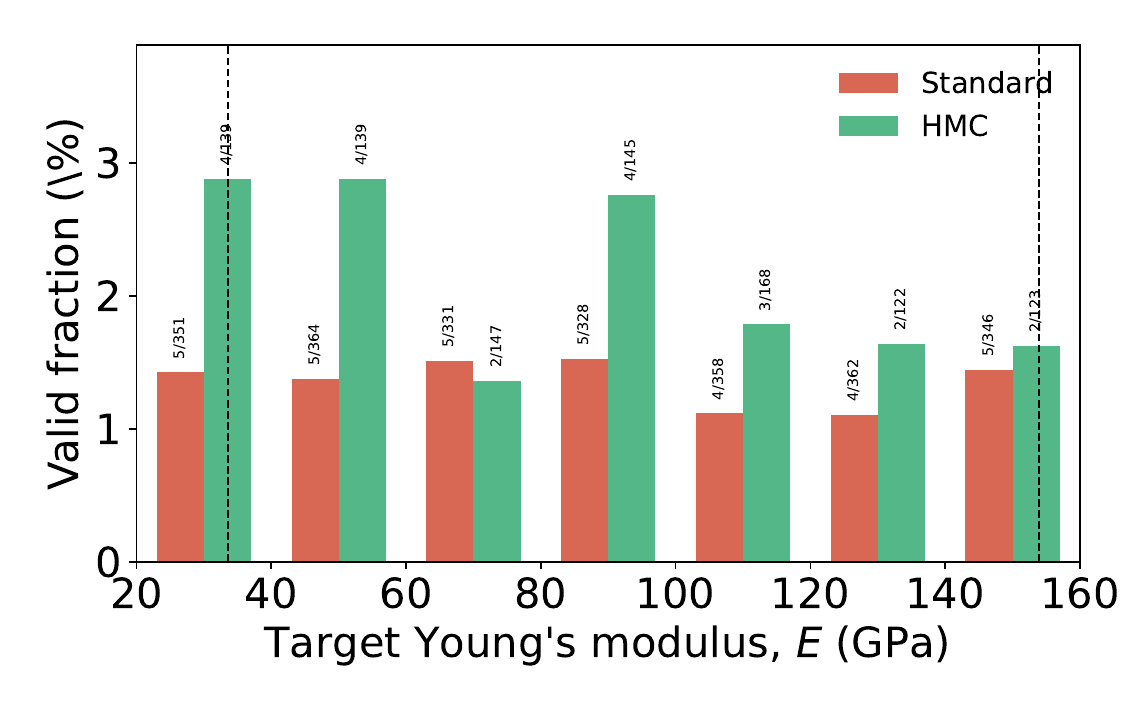}
    \caption{}
  \end{subfigure}\hfill
  \begin{subfigure}[b]{0.48\textwidth}
    \includegraphics[width=\textwidth]{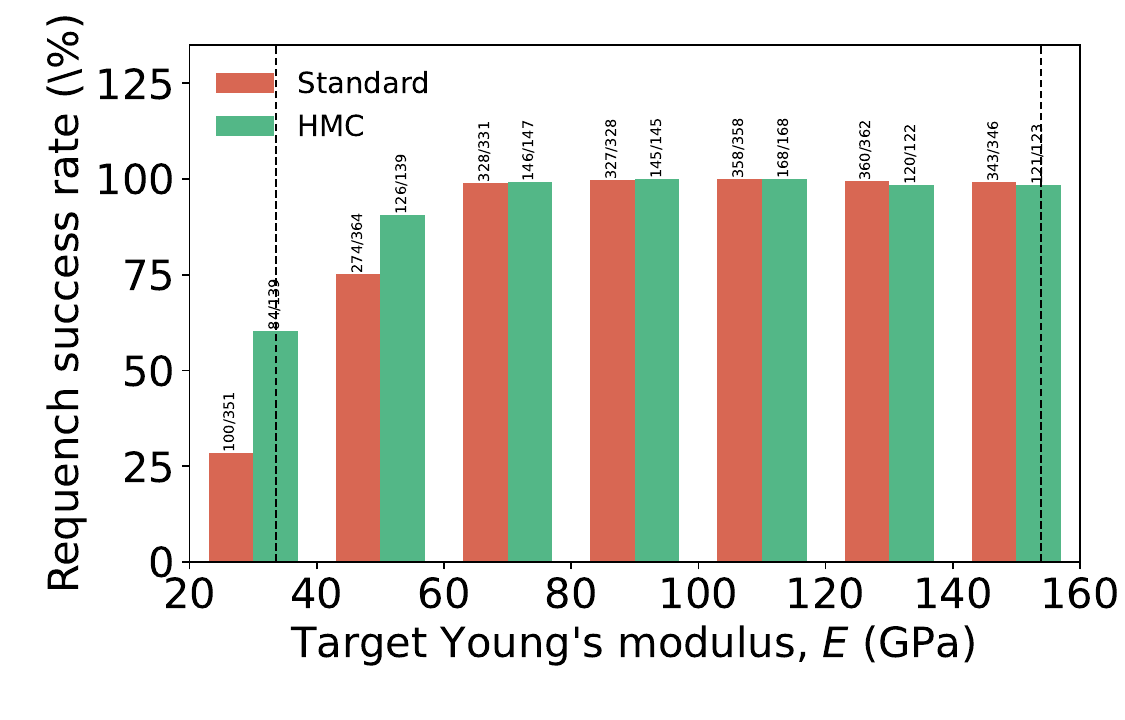}
    \caption{}
  \end{subfigure}
  \caption{
    MEG generation statistics as a function of target Young's modulus.
    (a)~Valid (stoichiometrically balanced) fraction.
    (b)~Requench success rate.
    Dashed vertical lines indicate the training data range.
    Each bar is annotated with the number of successful samples over the total in that bin.
  }
  \label{fig:valid_ratio_meg}
\end{figure}

For the MEG dataset, about 1--2\% of generated samples are strictly stoichiometrically balanced (Fig.~\ref{fig:valid_ratio_meg}a), reflecting the difficulty of achieving exact charge neutrality across 11 elements.
The number of HMC samples is smaller than for standard denoising due to the higher per-sample computational cost of HMC on this large system (${\sim}$800 atoms per sample), which also limits the number of stoichiometrically balanced HMC samples.
As discussed in the main text, the MEG results are not filtered by stoichiometric balance.
Among the MEG samples, 85.7\% (2\,090/2\,440) and 92.6\% (910/983) are successfully requenched for standard and HMC denoising, respectively.
As shown in Fig.~\ref{fig:valid_ratio_meg}b, the requench failures are concentrated in the extrapolative low-$E$ regime: for standard denoising, only 28.5\% of samples targeting $E \approx 30$\,GPa are successfully requenched, compared to $>$99\% for $E > 70$\,GPa.
HMC denoising substantially improves the requench success rate in the extrapolative low-$E$ regime (60.4\% vs.\ 28.5\% at $E \approx 30$\,GPa), while both methods achieve near-perfect success rates for targets within the training range.

\begin{table}[ht]
  \centering
  \caption{Generation validity statistics across all datasets.
  A sample is considered valid if its total formal charge is exactly zero.
  For amorphous Si, all samples are valid by construction.}
  \label{tab:valid_ratio}
  \begin{tabular}{llllrrl}
    \toprule
    Dataset & Denoising & Conditioning & Total & Valid & Valid (\%) \\
    \midrule
    a-Si    & --        & --           & --    & --    & 100.0 \\
    \ch{SiO2} & Standard & $G$       & 5\,000 & 275  & 5.5   \\
    \ch{SiO2} & Standard & Ring size  & 5\,000 & 320  & 6.4   \\
    MEG     & Standard  & $E$, $C_\text{Li}$ & 2\,440 & 33   & 1.4   \\
    MEG     & HMC       & $E$, $C_\text{Li}$ & 983    & 21   & 2.1   \\
    \bottomrule
  \end{tabular}
\end{table}

\section{Structural validation of generated \ch{SiO2} samples}\label{sec:sio2_validation}

To verify that the larger generated \ch{SiO2} cells used for inverse design (350--500 atoms, larger than the training range of 80--250 atoms) remain physically realistic at the medium-range level, we performed independent reference melt-quench MD simulations and compared the resulting ring size distributions.

The reference simulations were performed using the LAMMPS software~\cite{lammps} with the BKS interatomic potential~\cite{vanbeest1990force}, which is widely used for amorphous \ch{SiO2} and is independent of the Tersoff potential~\cite{munetoh2007interatomic} used for the training data.
For each conditioning target ($G$ and average ring size), the AMDEN-generated cells were melted in the NPT ensemble at 5\,000\,K for 500\,ps, cooled to 300\,K at a rate of 1\,K/ps, and finally equilibrated at 300\,K and zero pressure for 100\,ps to obtain the reference glass configurations.
The high-temperature melting step erases the initial atomic positions, so the resulting structures represent independent melt-quench references at the same enlarged cell size and composition.
Ring size distributions for both the AMDEN-generated and reference structures were computed according to the Guttman criterion~\cite{guttman1990ring}.

\begin{figure}[!t]
  \centering
  \begin{subfigure}[b]{0.48\textwidth}
    \includegraphics[width=\textwidth]{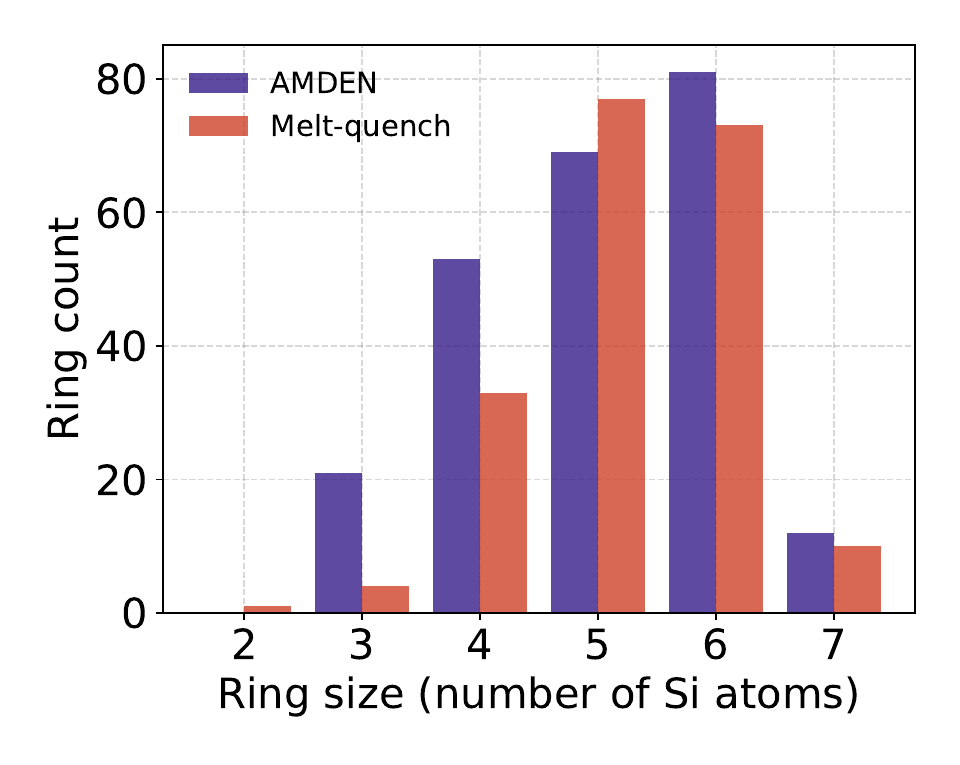}
    \caption{}
    \label{fig:sio2_ring_comparison_g}
  \end{subfigure}\hfill
  \begin{subfigure}[b]{0.48\textwidth}
    \includegraphics[width=\textwidth]{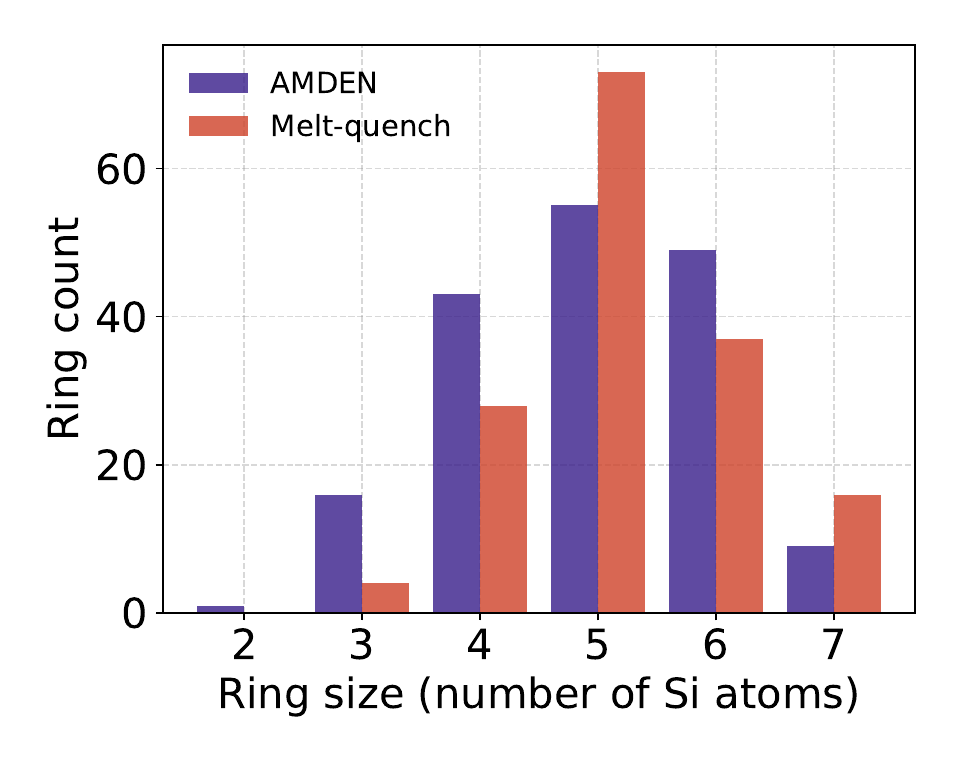}
    \caption{}
    \label{fig:sio2_ring_comparison_rsd}
  \end{subfigure}
  \caption{
    Ring size distributions of AMDEN-generated \ch{SiO2} samples (blue) compared with reference melt-quench MD simulations using the BKS potential (red).
    (a)~Generation conditioned on shear modulus $G$.
    (b)~Generation conditioned on average ring size.
    Ring sizes are reported as the number of \ch{Si} atoms in the ring, computed using the Guttman criterion.
  }
  \label{fig:sio2_ring_comparison}
\end{figure}

Supplemental Figure~\ref{fig:sio2_ring_comparison} compares the ring size distributions for both conditioning targets.
The two distributions agree on the peak location and overall range of ring sizes (3 to 7, concentrated around 4 to 6), consistent with established structural features of amorphous \ch{SiO2}, indicating that the larger generated cells remain physically realistic at the medium-range level.
Small but systematic differences are visible: the AMDEN distributions place more weight on smaller (3- and 4-membered) rings, reflecting the structural variation that AMDEN produces in response to the conditioning target.
This effect is most pronounced in the ring-size-conditioned case (Fig.~\ref{fig:sio2_ring_comparison_rsd}), where AMDEN shifts probability mass toward smaller rings to match the lower target ring sizes.
Together with the target--property parity already shown in Fig.~4 of the main text, these results indicate that AMDEN can bias the medium-range network topology in directions that are not naturally accessible to the melt-quench procedure while still producing physically realistic structures.

\begin{figure}[!t]
  \centering
  \includegraphics[width=0.9\textwidth]{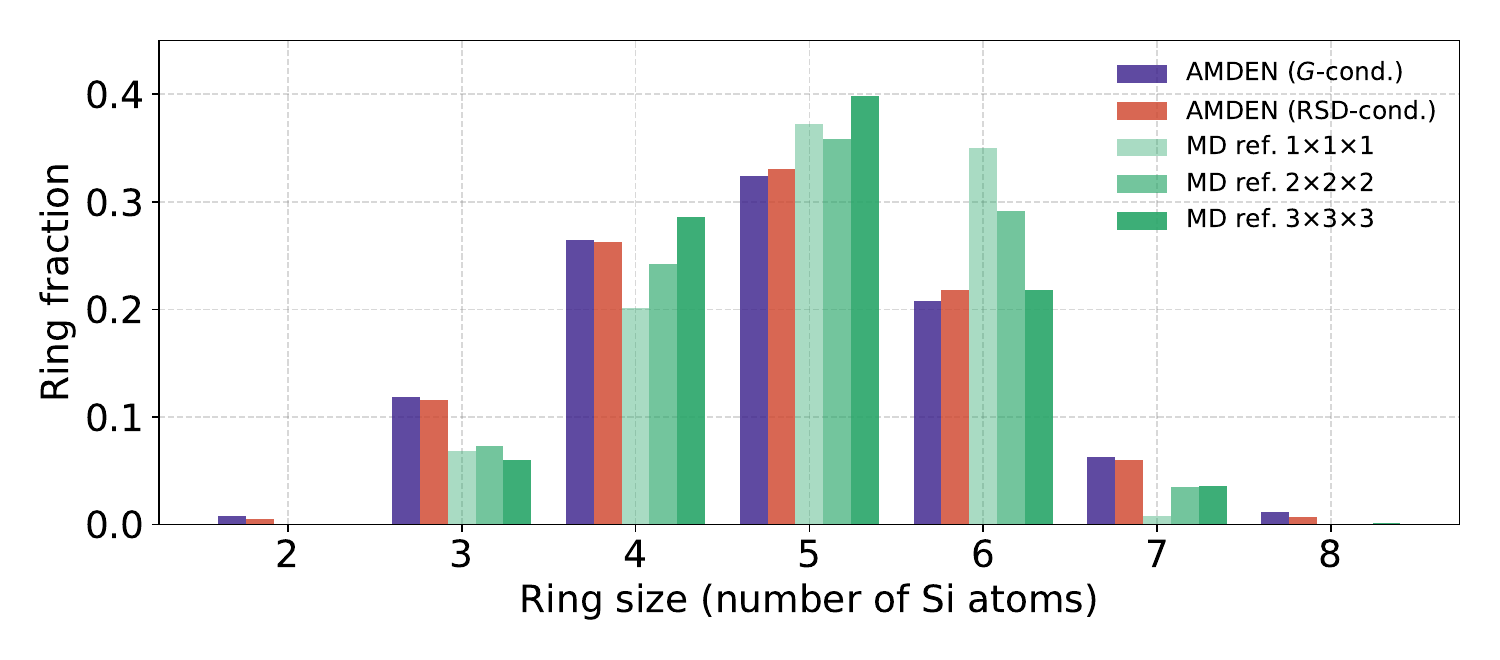}
  \caption{
    Ring fraction distributions of AMDEN-generated \ch{SiO2} samples for both conditioning targets (blue: $G$-conditioned; red: ring-size-conditioned), overlaid on reference Tersoff melt-quench MD simulations at three cubic supercell sizes ($1{\times}1{\times}1$, $2{\times}2{\times}2$, $3{\times}3{\times}3$).
    Ring sizes are reported as the number of \ch{Si} atoms in the ring, computed using the Guttman criterion.
    The reference distributions are stable across cell sizes, and the AMDEN distributions track them across the populated range with the same small shift toward smaller rings observed in Fig.~\ref{fig:sio2_ring_comparison}.
  }
  \label{fig:sio2_ring_multibox}
\end{figure}

To check that this comparison is not dominated by finite-size effects, we repeated the same melt-quench protocol with the Tersoff potential~\cite{munetoh2007interatomic} (the same potential used to generate the training data) at three cubic supercells of the base \ch{SiO2} cell ($1{\times}1{\times}1$, $2{\times}2{\times}2$, $3{\times}3{\times}3$), spanning the range of AMDEN-generated cell sizes (350--500 atoms).
Supplemental Figure~\ref{fig:sio2_ring_multibox} overlays the ring fraction distributions of the AMDEN-generated samples for both conditioning targets on the three reference distributions.
The three reference distributions are nearly superimposed across the populated 3- to 7-membered ring range, indicating that the reference ring statistics are converged with respect to cell size at the AMDEN scale.
The AMDEN distributions for both conditioning targets track the reference closely at the peak (5-membered rings) and across sizes 4 to 7, while displaying the same modest shift toward smaller (3- and 4-membered) rings already seen in Fig.~\ref{fig:sio2_ring_comparison}.
Two conclusions follow.
First, the medium-range agreement reported above persists across roughly an order of magnitude in cell volume and is therefore not an artifact of cell size.
Second, the shift toward smaller rings reflects a structural bias introduced by AMDEN's conditioning rather than a finite-size effect, reinforcing the interpretation that AMDEN can access medium-range topologies that the melt-quench procedure does not naturally produce.

\FloatBarrier
\bibliography{paper_abbreviated.bib}